\documentclass[symmetry,review,moreauthors,pdftex]{mdpi} 
\usepackage{amsmath,amscd,amssymb}
\usepackage{color,epsfig}
\usepackage{xfrac}
\usepackage{latexsym}
\usepackage{mathrsfs,bigints}
\usepackage{ulem}
\usepackage{slashed}
   
\newcommand{\dslash}{\partial \hskip -0.6em /}
\newcommand{\kslash}{k \hskip -0.5em /}
\newcommand{\vslash}{v \hskip -0.5em /}
\newcommand{\aslash}{a \hskip -0.5em /}
\newcommand{\pslash}{p \hskip -0.5em /}
\newcommand{\qslash}{q \hskip -0.5em /}
\newcommand{\nslash}{n \hskip -0.6em /}

\newcommand{\bD}{{\bf D}}
\newcommand{\bDp}{{\bf D}^{(\pi)}}

\newcommand{\zr}[1]{\mbox{\hspace*{#1em}}}
\newcommand{\ID}{\mbox{{\sf 1}\zr{-0.16}\rule{0.04em}{1.55ex}\zr{0.1}}}
\newcommand{\imu}{{\rm i}}
\newcommand{\bjlim}{\,{\stackrel{\scriptstyle{\rm Bj}}
{\textstyle\longrightarrow\,}}}
\newcommand{\tauom}{\vec{\tau}\hskip-0.3mm
\cdot\hskip-0.3mm\vec{\Omega}}

\newcommand{\xipl}{\vec{\xi}\hskip-0.6mm
+\hskip-0.6mm\lambda\hat{n}}
\newcommand{\ximl}{\vec{\xi}\hskip-0.6mm
-\hskip-0.6mm\lambda\hat{n}}

\allowdisplaybreaks

\firstpage{1} 
\pubvolume{xx}
\issuenum{1}
\articlenumber{5}
\pubyear{2020}
\copyrightyear{2020}
\history{Received: date; Accepted: date; Published: date}

\Title{Chiral Soliton Models and Nucleon Structure Functions}

\Author{Herbert Weigel $^{1}$\orcidA{0000-0002-2581-7717} 
and Ishmael Takyi $^{2}$\orcidA{0000-0002-1217-0889}}

\AuthorNames{Herbert Weigel and Ishmael Takyi}

\address{%
$^{1}$ \quad Institute of Theoretical Physics, Physics Department, Stellenbosch University, 
Matieland 7602, South Africa; weigel@sun.ac.za\\
$^{2}$ \quad Department of Mathematics, Kwame Nkrumah University of Science and Technology, 
Private Mail Bag, Kumasi, Ghana; ishmael.takyi@knust.edu.gh}

\abstract{We outline and review the computations of polarized and unpolarized nucleon structure 
functions within the bosonized Nambu-Jona-Lasinio chiral soliton model. We focus on a consistent 
regularization prescription for the Dirac sea contribution and present numerical results
from that formulation. We also reflect on previous calculations on quark distributions
in chiral quark soliton models and attempt to put them into perspective.}


\begin{document}

\section{Introduction}
 
In this mini-review we reflect on nucleon structure function calculations in chiral soliton models. 
This is an interesting topic not only because structure functions are of high empirical relevance 
but maybe even more so conceptually as of how much information about the nucleon structure can be 
retrieved from soliton models. In this spirit, this paper to quite an extend is a proof of concept 
review.

Solitons emerge in most non-linear field theories as classical solutions to the field equations. These
solutions have localized energy densities and can be attributed particle like properties. In the 
context of strong interactions, that govern the structure of hadrons, solitons of meson field
configurations are considered as baryons \cite{Witten:1979kh}.

Nucleon structure functions play an important role in deep inelastic scattering (DIS) that reveals 
the parton substructure of hadrons. In DIS leptons interact with partons by the exchange of a virtual 
gauge particle. Here we will mainly consider electrons that exchange a virtual photon with either a 
pion or a nucleon. The process is called deep inelastic as the produced hadrons are not detected. 
In a certain kinematical regime, the so-called Bjorken limit to be defined below, the DIS cross section 
can be parameterized as the product of the cross section for scattering off partons and distribution 
functions that measure the probabilities to find these partons inside the hadron. This is the 
factorization scheme~\cite{Collins:1989gx}. In this picture the structure functions are linear 
combinations of parton distribution functions.

DIS can also be explored without direct reference to partons by writing the cross section in terms of 
lepton and hadron components. The latter is the hadron matrix element of a current-current correlator 
and is parameterized by form factors. The structure functions are obtained from these form factors in 
a certain regime for the kinematic variables, again the Bjorken limit. The operator product 
expansion formally relates distribution and structure functions by expressing the hadron matrix elements 
of the current-current correlator as matrix elements of bilocal and bilinear quark operators in the 
Bjorken limit. The microscopic theory for the structure of hadrons is quantum-chromo-dynamics (QCD) which 
is the non-abelian gauge theory $SU(N_C)$, where $N_C=3$ is the number of color degrees of freedom. 
Though (perturbative) QCD only relates these functions at different energy scales and does so very 
successfully \cite{Abramowicz:2015mha} within the DGLAP formalism \cite{Gribov:1972ri}, neither 
structure nor distribution functions can be computed from first principles 
in QCD, except, maybe within the lattice formulation \cite{Lin:2017snn}\footnote{Another possibility is 
to apply QCD renormalization group equations to the empirical data at large energies and scale them 
down to the point at which the probability interpretation becomes inconsistent \cite{Gluck:1994uf}.}. 
Hence model calculations seem unavoidable for a theoretical approach to the structure functions that contain 
the information of the non-perturbative nature of hadrons. In such models it may or may not be possible 
to relate structure and distribution functions. For the quark model that we will employ, regularization 
stands in the way and we attempt to compute the structure functions directly from the current-current 
correlator.

Though chiral (soliton) models for baryons have so far not been derived from QCD, there is ample of 
motivation to explore nucleon properties in chiral models. The soliton approach goes back to the Skyrme 
model~\cite{Skyrme:1961vq} while the connection to QCD was later established by considering baryons in 
a generalized version of QCD with $N_C$ large \cite{Witten:1979kh}. Soon after those mainly combinatoric 
arguments for considering baryons in an effective meson theory, static baryons properties were
derived within the Skyrme model~\cite{Adkins:1983ya}. The soliton approach has ever since been 
very actively explored, {\it cf.} the reviews~\cite{Schwesinger:1988af}. The point of departure
for most of these models is an effective meson theory that reflects the major symmetries of 
QCD on the hadron level. In the low energy regime this is essentially the chiral symmetry with the
pions as would-be Goldstone bosons being the basic field degrees of freedom\footnote{On the other
end, the heavy quark effective symmetry has also been combined with the soliton picture. This
is outside the scope of this review. The interested reader may trace relevant publications from
 Ref.~[52] in the recent article~\cite{Liu:2019mxw}.}. Other mesons like $\omega$ and $\rho$ 
were then incorporated according to the rules of chiral symmetry. A major endeavor is to determine 
as many as possible model parameters from mesons to gain a high predictive power in 
the soliton sector, {\it i.e.} for baryon properties. Many of these properties have been reproduced 
in chiral soliton models to the accuracy that one expects from keeping the leading (and eventually 
next-to-leading) terms of a power expansion in $\frac{1}{N_C}$ when the actual value is $N_C=3$. 

Unfortunately, this is not the case for nucleon structure functions and very early on it was 
recognized that soliton models based on meson fields disagree with the parton model. Rather 
than leading to the parton model Callan-Gross relation between the unpolarized structure functions, 
the Skyrme model yields the Callan-Gross analog for boson constituents \cite{Chemtob:1987ut} when
evaluating current-current correlations that eventually lead to the structure functions. This 
problem is not unexpected as taking a purely meson model as point of departure implicitly relates 
local quark bilinears to the field degrees of freedom. On the other side, in QCD, the structure functions 
are related to bilocal quark bilinears and a successful exploration of these functions needs to trace 
the details of the bosonization procedure. To solve this fundamental problem of the soliton picture it 
is therefore compulsory to consider a model in which the bosonization is explicitly performed. Such
a model starts from a chirally symmetric quark self-interaction and introduces auxiliary boson fields
that make feasible the computation of the fermion path integral. Subsequently these boson fields 
take the role of the mesons in an effective theory. The Skyrme model problem is then approached by
formulating the current-current correlations before bosonization. For this purpose we will here 
consider bosonization~\cite{Ebert:1985kz} of the Nambu-Jona-Lasinio model \cite{Nambu:1961tp} that has 
well established soliton solutions \cite{Wakamatsu:1990ud,Alkofer:1994ph,Christov:1995vm}. The model by 
itself is not renormalizable and the regularization prescription is part of the model definition. 
Incorporating regularization is the major concern when computing structure functions in a bosonized 
chiral model. Essentially there are two approaches for computing nucleon structure functions. They differ 
conceptually but lead to similar results. The one that we will focus on here starts with a fully 
regularized action and extracts the structure functions from the absorptive part of the Compton tensor. 
We note that this approach is general enough to also predict the pion structure function \cite{Weigel:1999pc}.
We will take the position that we only identify the symmetries of QCD when adopting this model to describe 
hadrons. At this stage of the project we will not identify the quark degrees of freedom with those of QCD, 
which, for example, means that the current quark mass is a free parameter. The theoretical framework has 
been derived already some time ago \cite{Weigel:1999pc} while the numerical results arising from 
costly simulations have only been obtained recently \cite{Takyi:2019ahv}. As mentioned above, a 
major motivation for the soliton picture arises from generalizing QCD to a non-abelian gauge theory 
with large $N_C$. In this review we will make $N_C$ explicit in formulas, but actual calculations 
are performed with $N_C=3$.

Formal considerations of QCD relate DIS in the Bjorken limit to hadron matrix elements of bilocal 
bilinear quark operators. There are soliton model approaches that sandwich the quark operators from the 
self-consistent chiral soliton in those nucleon matrix elements and impose regularization 
{\it a posteriori}~\cite{Diakonov:1996sr,Pobylitsa:1998tk,Wakamatsu:1997en,Wakamatsu:1998rx,Pobylitsa:1996rs,Gamberg:1998vg,Schweitzer:2001sr,Ohnishi:2003mf,Wakamatsu:2003wg}.  We will comment on those approaches in Sect.\ref{sec:others}.

The study of structure functions in soliton models has, to quite some extend, been triggered by the so-called
{\it proton spin puzzle} \cite{Ashman:1987hv}: Data on the polarized structure function suggested that,
together with flavor symmetric relations, almost none of the nucleon spin was due to the spin of the
quark constituents. This picture emerges from the non-relativistic quark model in which the
nucleon spin equals the matrix element of the axial singlet current. It is actually this matrix
element that relates to the data and chiral soliton models indeed yield a small value, some
versions even predict zero \cite{Brodsky:1988ip}. See Ref.~\cite{Deur:2018roz} for a recent review
on the present understanding of the proton spin structure.

Earlier we have noted the relevance of structure and/or distribution functions for DIS. There are
other regimes of relevance. Let us first make explicit the factorization theorem for the cross-section 
for electron hadron scattering \cite{Collins:1989gx},
\begin{equation}
\sigma_{e}(x,Q^2)=\sum_{a}\int_x^1 d\xi\,f_a(\xi)\sigma_{ea}\left(\frac{x}{\xi},Q^2\right)\,.
\label{eq:factor}
\end{equation}
Here $f_a(\xi)$ is the distribution function for parton $a$ with momentum fraction $\xi$
in the hadron and $\sigma_{ea}$ is the (Born) cross-section for electron parton scattering.
Note that the sum over $a$ also includes different distributions for the same parton such as
polarized and unpolarized.
Furthermore $x$ and $Q^2$ are Lorentz invariant kinematical variables that will be defined in 
Section \ref{sec:DIS}. Essentially we are interested in the case where $x$ is fixed but $Q^2$
becomes large.  The $f_a(\xi)$ are equally important for the Drell-Yan process in which 
two hadrons ($A$ and $B$) scatter into a lepton-antilepton pair and other hadrons. That pair 
originates from a virtual gauge boson that is produced by quarks ($q$) and antiquarks~($\overline{q}$) 
within the hadrons. Without going into further detail this suggests that the scattering cross section 
is parameterized by the same distribution functions as DIS
\begin{equation}
\sigma\sim\sum_{a}\int_{x_A}^1 d\xi_q\,f_q(\xi_q)
\int_{x_B}^1 d\xi_{\overline{q}}\,f_{\overline{q}}(\xi_{\overline{q}})
\sigma^\prime(\xi_q,\xi_{\overline{q}},Q^2)\,,
\label{eq:factorDY}
\end{equation}
where $\sigma^\prime$ is the cross section for turning the quark-antiquark pair into a lepton-antileption 
pair by the exchange of a virtual gauge boson. For the detailed definition of the kinematic variables 
$x_A$ and $x_B$ for the two hadrons $A$ and $B$ we again refer to Ref.~\cite{Collins:1989gx}. Here we will 
not pursue the Drell-Yan process any further because it is not related to a current-current correlation 
matrix element of a single hadron. However, we will shortly come back to the Drell-Yan process in 
Section \ref{sec:others}.

The expansion, Eq.~(\ref{eq:factor}) indicates that different distributions $f_a(\xi)$ contribute with 
different inverse powers of $Q$ to the total cross section through $\sigma_{ea}$. Accordingly distributions are 
categorized by their {\it twist} which is extracted from the leading inverse power of $Q$ in $\sigma_{e}$. 
The definition of twist dwells in the operator product expansion and relates to the dimensionality and spin 
of the operators in that expansion. Here it is sufficient to mention that the leading contribution
(as $Q$ increases) has twist-2, distributions that contribute like $1/Q$ to the total cross section 
have twist-3 etc.\@ \cite{Jaffe:1996zw}.

The following section contains a brief recap of basic definitions in the context of structure functions. 
Section \ref{sec:model} describes the path from the self-interacting fermion theory to the bosonized chiral model
together with a review of the pion structure function calculation. This will be followed by the construction
of the soliton in that model in Section \ref{sec:soliton}. We explain the soliton model calculation of structure 
in Section \ref{sec:HTsol} and discuss the numerical results in Section \ref{sec:NR}. As mentioned, in 
Section \ref{sec:others} we will discuss related distribution function calculations in the chiral quark 
soliton model. Some concluding remarks are contained in Section \ref{sec:concl}.


\section{Framework of deep inelastic scattering}
\label{sec:DIS}
Deep inelastic scattering (DIS) is a major tool to explore the composition of the nucleon. In this
process electron scattering produces a virtual photon which then interacts with the charged 
components of the nucleon. To extract the structure functions, the scattering products need not 
be detected as they are summed over in the final scattering cross section.

\parbox[r]{7.5cm}{~\vskip0.4cm
The Feynman diagram to the right describes the kinematical set-up, where $k$ and $k^\prime$ 
are the momenta of the initial and final electrons, respectively, while $p$ is the momentum of
the incoming proton, typically taken in the rest frame. The set of final hadrons, $X$ is not 
detected and summed over, {\it cf.} Eq.~(\ref{eq:hten0}).}
\parbox[l]{1cm}{~}
\parbox[l]{6cm}{
\setlength{\unitlength}{3.5mm}
\begin{picture}(15.0,10.0)
\thicklines
\put(1,9){\vector(3,-1){3}}
\put(4,8){\line(3,-1){3}}
\put(2,7.3){\large $k$}
\put(7,7){\vector(3,1){3}}
\put(10,8){\line(3,1){3}}
\put(12,7.3){\large $k^\prime$}
\linethickness{1pt}
\bezier{50}(7,7)(8,6.5)(7,6)
\bezier{50}(7,6)(6,5.5)(7,5)
\bezier{50}(7,5)(8,4.5)(7,4)
\bezier{50}(7,4)(6,3.5)(7,3)
\bezier{50}(7,3)(8,2.5)(7,2)
\bezier{50}(7,2)(6,1.5)(7,1)
\put(7.7,5.5){\large $q=k-k^\prime$}
\put(7,1){\circle*{0.4}}
\linethickness{2pt}
\put(1,1){\vector(1,0){4}}
\put(4,1){\line(1,0){3}}
\put(2,1.8){\large $p$}
\linethickness{1pt}
\put(7,1,1){\vector(4,3){4}}
\put(7,1,1){\vector(2,1){4}}
\put(7,1,1){\vector(4,1){4}}
\put(7,1,1){\vector(1,0){4}}
\put(7,1,1){\vector(4,-1){4}}
\put(7,1,1){\vector(2,-1){4}}
\put(12,1){\huge $\Bigg\}$~~\large $X$}
\end{picture}}
\vskip1cm

The interaction vertex for the disintegration of the nucleon is the matrix element 
of the (electromagnetic) current $J_\mu(\xi)$. The cross-section contains the 
squared absolute value of this matrix element and we sum/integrate over all 
final states subject to energy momentum conservation. This defines the hadron
tensor for electron nucleon scattering
\begin{equation}
W_{\mu \nu}(p,q;s) = \frac{1}{4\pi} \sum_X \Big\langle p,s\Big|J_{\mu}(0)\Big|X\Big\rangle
\Big\langle X\Big|J^\dagger_{\mu}(0)\Big|p,s\Big\rangle
(2\pi)^4\delta^4(p+q-p_X)\,,
\label{eq:hten0}
\end{equation}
where $s$ denotes the nucleon spin. The nucleon momentum is $p$ and $q=k-k^\prime$ is the 
momentum of the virtual photon. As the interaction is inelastic we have $q_0>0$. This, together
with translational invariance, yields 
\begin{equation}
W_{\mu \nu}(p,q;s) = \frac{1}{4\pi} \int d^4\xi \, e^{iq\cdot \xi}\,
\Big\langle p,s\Big| [ J_{\mu}(\xi),J_{\nu}^{\dagger}(0)]
\Big| p,s \Big\rangle\, .
\label{eq:hten1}
\end{equation}
The interaction is space-like and it is customary to introduce $Q^2=-q^2>0$ as well as
$\nu=\frac{p\cdot q}{M_N}$ where $M_N$ is the nucleon mass. In the nucleon rest frame 
$\nu$ is the energy transferred from the electron to the virtual photon. Most prominent 
is the Bjorken variable 
\begin{equation}
x=\frac{Q^2}{2M_N\nu}\,,
\label{eq:Bj1}
\end{equation}
which in the parton model denotes the momentum fraction associated with a particular parton.
Since on-shell $p^2=M_N^2$, $Q^2$ and $x$ can be taken as the only dynamical Lorentz invariant
variables so that the hadron tensor has the form factor decomposition
\begin{align}
W_{\mu \nu}(p,q;s)
& =\left(-g_{\mu \nu} + \frac{q_{\mu} q_{\nu}}{q^2}\right) M_N W_{1} (x,Q^2)
+\left(p_{\mu} - q_{\mu}\frac{p\cdot q}{q^2}\right)
\left(p_{\nu} - q_{\nu}\frac{p\cdot q}{q^2}\right)
\frac{1}{M_N} W_{2} (x,Q^2)\cr
&\hspace{1cm}
+\imu\epsilon_{\mu \nu \lambda \sigma} \frac{q^{\lambda} M_N}{p\cdot q}
\left(\left[G_1(x,Q^2)+G_2(x,Q^2)\right]s^{\sigma}
-\frac{q\cdot s}{q\cdot p} p^{\sigma} G_2(x,Q^2) \right)
\label{eq:hten1a}
\end{align}
for parity conserving processes like electromagnetic scattering of photons.
The structure functions are the form factors in the so-called Bjorken scaling limit that 
takes $Q^2\to\infty$ with $x$ fixed. For the spin independent, unpolarized structure functions
$f_1(x)$ and $f_2(x)$ that is
\begin{equation}
M_N W_1(x,Q^2)\, \bjlim\,  f_1(x)
\quad {\rm and} \quad
\frac{p\cdot q}{M_N}\,W_2(x,Q^2)\, \bjlim\,  f_2(x)\, .
\label{eq:deff1f2}
\end{equation}
For the spin dependent, polarized structure functions no further scaling is 
involved and
\begin{equation}
G_1(x,Q^2)\, \bjlim\,  g_1(x)
\quad {\rm and} \quad
G_2(x,Q^2)\, \bjlim\,  g_2(x)\, .
\label{eq:defg1g2}
\end{equation}
Contracting the hadron tensor with projectors listed in Tab.~\ref{tab_1} extracts the 
pertinent structure functions. For the unpolarized structure functions these projectors 
directly lead to the Callan-Gross relation $f_2=2xf_1$. Observe also that these projectors
are to be combined with appropriate selections for the spin orientation of the nucleon state 
as indicated in the last row of Tab.~\ref{tab_1}.
\begin{table}
\renewcommand{\arraystretch}{1.3}
\caption{\label{tab_1}
Projection operators which extract the leading large $Q^2$ components
from the hadron tensor. The projectors given in the spin independent
cases presume the contraction of $W_{\rho\sigma}$ with
$S^{\mu\nu\rho\sigma}=g^{\mu\rho}g^{\nu\sigma}
+g^{\rho\nu}g^{\mu\sigma}-g^{\mu\nu}g^{\rho\sigma}$.
The last row denotes the required spin orientation of the nucleon.}
\begin{center}
\begin{tabular}{c|c|c|c}
$f_1$ & $f_2$ & $g_1$ & $g_T=g_1+g_2$ \\
\hline
$-\frac{1}{2}g^{\mu\nu}$&
$-xg^{\mu\nu}$&
$\frac{i}{2M_N}\epsilon^{\mu\nu\rho\sigma}
\frac{q_\rho p_\sigma}{q\cdot s}$ &
$\frac{-i}{2M_N}\epsilon^{\mu\nu\rho\sigma}
s_\rho p_\sigma$ \\
\hline
$\genfrac{}{}{-1pt}{}{\rm spin}{\rm independent}$&
$\genfrac{}{}{-1pt}{}{\rm spin}{\rm independent}$
& $\vec{s}\parallel\vec{q}$ &
$\vec{s}\perp\vec{q}$
\end{tabular}
\end{center}
\renewcommand{\arraystretch}{1.0}
\normalsize
\end{table}
Even though we employ the Bjorken limit to the form factors, that leading expansion may
still contribute with different (inverse) powers of $Q$ to the total cross section and
thus the structure functions may be assigned different (leading) twist.

Similarly to the commutator in the hadron tensor we consider the matrix element of 
the time-ordered current-current product
\begin{align}
T_{\mu \nu}(p,q;s) & = \imu \int d^4\xi\, e^{\imu q\cdot \xi}\,
\Big\langle p,s\Big| T\left( J_{\mu}(\xi)J_{\nu}^{\dagger}(0)\right)\Big| p,s \Big\rangle\cr
&=(2\pi)^3\sum_X\left\{\frac{\delta^3(\vec{p}_X-\vec{q}-\vec{p})}{p_X^0-q^0-p^0-\imu\epsilon}
\Big\langle p,s\Big|J_{\mu}(0)\Big|X\Big\rangle
\Big\langle X\Big|J^\dagger_{\mu}(0)\Big|p,s\Big\rangle
\right.\cr & \left. \hspace{2cm}
+\frac{\delta^3(\vec{p}_X+\vec{q}-\vec{p})}{p_X^0+q^0-p^0-\imu\epsilon}
\Big\langle p,s\Big|J_{\mu}(0)\Big|X\Big\rangle
\Big\langle X\Big|J^\dagger_{\mu}(0)\Big|p,s\Big\rangle\right\}\,.
\label{eq:comp1}
\end{align}
Cauchy's principal value prescription $\frac{1}{x\pm\imu\epsilon}=\mathcal{P}\left(\frac{1}{x}\right)
\mp\imu\pi\delta(x)$ shows that imaginary part of the first term is proportional to the hadron tensor as 
in Eq.~(\ref{eq:hten0}) while the second term does not have an imaginary part for the present
kinematical set-up. Hence we have
\begin{equation}
W_{\mu \nu}(p,q;s)
=\frac{1}{2\pi}{\sf Abs} T_{\mu\nu}\,,
\label{eq:comp2}
\end{equation}
where {\sf Abs} stands for absorptive part.
From the physics point of view, $T_{\mu\nu}$ is the forward amplitude for nucleon Compton 
scattering and the hadron tensor is its absorptive part.

This paves the way towards computing the structure functions in the bosonized quark model. 
The action for that model is obtained from a functional integral of a self-interacting 
quark model. Within that formulation matrix elements of time-ordered products are straightforward 
to compute. Subsequently Cutkosky's rules are applied to extract their absorptive parts.

\section{The Chiral Quark Model}
\label{sec:model}

We consider the simplest $SU(2)$ Nambu-Jona-Lasinio (NJL) model which contains a chirally symmetric 
quartic fermion interaction in the scalar and pseudoscalar bilinears. In Minkowski space the Lagrangian 
reads \cite{Nambu:1961tp},
\begin{equation}
\mathcal{L}_{\rm NJL} = \overline{q} \left( \imu \slashed{\partial} -m^{0} \right) q + 
\frac{G}{2}\left[\left(\overline{q}q\right)^{2}
+\left(\overline{q} \imu \gamma_{5}\vec{\tau\,}q\right)^{2} \right]\,.
\label{nambu} 
\end{equation} 
The field $q(x)$ denotes a spinor with two flavors (up, $u$ and down, $d$). There are no color interactions
but each spinor has $N_C$ color components.  Furthermore $m^{0}$ and $G$ are the current quark 
mass (average up and down quark mass) and the dimensionful coupling constant, respectively. The 
symmetry transformations are $q\,\longrightarrow\, q+\imu\vec{\epsilon}\cdot\vec{\tau}q$ for 
$m^0\propto\ID$ and $q\,\longrightarrow\, q+\imu\gamma_5\vec{\epsilon}_5\cdot\vec{\tau}q$ for $m^0=0$.

The effective bosonized action for the NJL model is constructed with the help of the auxiliary matrix 
field $M$ that has a quadratic potential and couples linearly to the quark bilinears $\overline{q}q$ and 
$\overline{q} \imu \gamma_{5}\vec{\tau\,}q$. Then the fermion part of the functional integral can be 
computed and its logarithm is an effective action which becomes a non-linear and non-local theory for 
$M$ \cite{Ebert:1985kz}. The entries of this matrix are identified with the fields of the 
low-lying mesons. The model (by invention) breaks chiral symmetry dynamically for sufficiently large $G$ 
and therefore the most import modes of $M$ are the pseudoscalar pions ($\pi^{\pm}$, $\pi^0$). The members 
of this isospin triplet would be Goldstone bosons in the chiral limit characterized by $m^{0}=0$.

At face value the effective action diverges and is not renormalizable. It is therefore mandatory to supplement 
it with a regularization prescription. It is standard to Wick-rotate to Euclidian space in which the effective 
action is complex. Apart from the cosmological constant contribution (which diverges quarticly but has 
no dynamical effect), the real part of this Euclidian action 
is quadratically divergent while the imaginary part is (conditionally) convergent. It is customary not to 
regularize the latter in order to properly reproduce the axial anomaly which can be analyzed by introducing 
photon fields, $\gamma$, (via minimal substitution in $\mathcal{L}_{\rm NJL}$) and studying the decay 
$\pi^0\to\gamma\gamma$. On the other hand the real part is subjected to standard regularization methods like 
proper-time \cite{Reinhardt:1989st} or Pauli-Villars \cite{Jaminon:1989ix}. Within the perturbative realm 
({\it i.e.} zero soliton sector) one can even work with a sharp momentum cut-off \cite{Klevansky:1992qe}.

We would like to avoid the Wick-rotation because we want any imaginary part in our calculation of 
the hadron tensor being solely due to the absorptive components that we will extract via Cutkosky's rules. 
There is indeed a procedure to identify the Minkowski space analogs of the real and imaginary parts of the 
Euclidian action \cite{Davidson:1994uv}. To this end we define Dirac operators 
\begin{align}
\imu \bD &= \imu\dslash - \left(S+\imu\gamma_5P\right)
+\vslash +\aslash\gamma_5
=:\imu\bDp+\vslash+\aslash\gamma_5\cr
\imu \bD_5 &= - \imu\dslash - \left(S-\imu\gamma_5P\right)
-\vslash+\aslash\gamma_5
=:\imu\bDp_5-\vslash+\aslash\gamma_5\, ,
\label{eq:defd5}
\end{align}
where $S=\frac{1}{2}(M+M^\dagger)$ and $P=\frac{1}{2}(M-M^\dagger)$. Furthermore $v_\mu$ and $a_\mu$ denote 
external (classical) source fields with respect which we will compute functional derivatives to explore 
correlation functions.  Finally we have also defined Dirac operators without those sources ($\bDp$ and $\bDp_5$) 
for later use. Wick-rotating $\bD_5$ produces the conjugate of the Wick-rotation of $\bD$ so that 
$\frac{1}{2}{\rm Tr}\, {\rm log}[\bD \bD_{5}]$ corresponds to the real 
part of the Euclidian action while its imaginary part is associated with 
$\frac{1}{2}{\rm Tr}\, {\rm log}[\bD(\bD_{5})^{-1}]$.
The introduction of $\bD_{5}$ comes at a price. Some of the Ward identities derived from the standard
Dirac operator $\bD$ do not hold anymore and rather occur with opposite signs \cite{Weigel:1999pc}. 
We will later cure that obstacle by a particular calculational procedure to extract the polarized 
structure functions. This procedure is part of the regularization scheme.
Even though the proper-time scheme has been very successfully applied for the solitons of the NJL model, 
we do not implement it here. This scheme induces an exponential dependence on the cut-off and it is unclear
how to implement the Bjorken limit. Rather we adopt a version of the Pauli-Villars scheme in which the 
cut-off essentially is additive to the quark mass and does not interfere with the Bjorken limit. With
all these preliminaries we are now in a position to write down the effective action for $M$:
\begin{align}
\mathcal{A}_{\rm NJL}&=\mathcal{A}_{\rm R}+\mathcal{A}_{\rm I}
+\frac{1}{4G} \int \mathrm{d}^{4} x\, {\rm tr} \left[m^{0}(M + M^{\dagger}) -MM^{\dagger} \right]
\label{act1} \\
\mathcal{A}_{\rm R}&=-\imu\frac{N_C}{2}
\sum_{i=0}^{2} c_{i} {\rm Tr}\, {\rm log}
\left[- \bD \bD_{5} +\Lambda_{i}^{2}-i\epsilon\right]\,,\cr
\mathcal{A}_{\rm I}&=-\imu\frac{N_{C}}{2}
{\rm Tr}\, {\rm log}
\left[-\bD \left(\bD_{5}\right)^{-1}-i\epsilon\right]\, .
\nonumber
\end{align} 
Here $\mathcal{A}_{\rm R}$ and $\mathcal{A}_{\rm I}$ are the Minkowski analogs of the real 
and imaginary parts of the Euclidian space effective action. Furthermore 'Tr' denotes the 
functional trace that includes space-time integration on top of summing over the discrete
Dirac and flavor indexes. The Pauli-Villars regularization scheme requires 
\begin{equation}
c_0=1\, ,\quad \Lambda_0=0\, ,\quad \sum_{i=0}^{2}c_i=0
\quad {\rm and}\quad \sum_{i=0}^2c_i\Lambda_i^2=0\,.
\label{pvcond}
\end{equation}
For simplicity we reduce the number of regulators by the limiting case $\Lambda_{1}=\Lambda_{2}=\Lambda$. 
For any quantity $Q(\Lambda)$ that is subject to regularization we then have
\begin{equation}
\sum_{i=0}^2 c_{i} Q(\Lambda_{i}^{2})= Q(0)-Q(\Lambda^{2})+\Lambda^{2} Q^{\prime} (\Lambda^{2} )\,,
\label{eq:LAMBDA}
\end{equation} 
where the prime denotes the derivative with respect to the argument. For notational simplicity we will 
usually write the formulas as on the left hand, understanding that the right hand side is implemented
in actual computations.

To analyze the model we need to find the ground state solution, $\langle M\rangle$. For symmetry 
reasons any non-zero solution can only be a (real) constant that is proportional to the unit matrix. 
We therefore substitute $\langle M\rangle=m$ in the so-called gap equation
\begin{equation}
\frac{1}{2G}\left(m-m^{0}\right)
=-4\imu N_{C} m\sum_{i=0}^{2}c_{i}
\int\frac{d^{4}k}{(2\pi)^{4}}
\left[-k^{2}+m^{2}+\Lambda_{i}^{2}-\imu\epsilon\right]^{-1}
\label{gap} 
\end{equation}  
that arises from $\frac{\delta A_{\rm NJL}}{\delta M}=0$. For sufficiently large coupling $G$ this 
equation has a solution with $m\gg m^0$ which obviously plays the role of a mass parameter when 
substituted for $M$ into $\bD$ (or $\bD_5$). It is therefore called the constituent quark mass. 

Any non-trivial vacuum solution signals dynamical symmetry breaking and applying a symmetry 
transformation onto that solution leads to (would-be) Goldstone bosons. In this case the relevant
transformation is chiral and the would-be Goldstone boson\footnote{We expect that boson to 
be massless only when the original theory has an exact chiral symmetry, $m^{(0)}=0$.} 
is the pseudoscalar iso-triplet pion $\vec{\pi}$. This field is most conveniently introduced 
via the non-linear realization
\begin{equation}
M=m U =m\,{\rm exp}\left[\imu \frac{g}{m}\vec{\pi}\cdot\vec{\tau}\right]
=m +\imu g \vec{\pi}\cdot\vec{\tau}+\mathcal{O}(\vec{\pi}^2) \,,
\label{xcircle}
\end{equation}
where $U$ is the chiral field while $g$ is the Yukawa coupling constant.
In the next step, we expand the effective action to quadratic order in the pion fields 
\begin{equation}
\mathcal{A}_{\rm NJL}=g^2\int \frac{d^4p}{(2\pi)^4}\,
{\vec{\widetilde\pi}}(p) \cdot {\vec{\widetilde\pi}}(-p)
\left[2N_C q^2\Pi(p^2)-\frac{1}{2G}\frac{m_0}{m}\right]
+\mathcal{O}\left(\vec{\pi}^4\right)\, ,
\label{A2}
\end{equation}
which has been written for the Fourier transform 
$\vec{\widetilde\pi}(p)=\bigintsss d^4x\, {\rm e}^{-\imu p\cdot\xi}\vec{\pi}(\xi)$.
The quadratic contribution contains the polarization function
\begin{align}
\Pi(p^2)&=\int_0^1 dx\, \Pi(p^2,x) \quad {\rm with}\quad
\Pi(p^2,x)=-i\sum_{i=0}^2 c_i\,
\frac{d^4k}{(2\pi)^4}\,
\left[-k^2-x(1-x)p^2+m^2+\Lambda_i^2-\imu\epsilon\right]^{-2}\,.
\label{specfct}
\end{align}
The factor in square brackets in Eq.~(\ref{A2}) times $g^2$ is the inverse pion propagator. 
Requiring this propagator to have a pole at the physical pion mass enforces
\begin{equation}
m^{0}=4 N_{C} G m m_{\pi}^{2} \Pi(m_{\pi}^{2})\,. 
\label{pion_mass}
\end{equation}
Furthermore the residue of that pole should be one thereby relating the Yukawa coupling 
constant $g$ to other model parameters,
\begin{equation}
\frac{1}{g^{2}} = 4 N_{C} 
\frac{\partial}{\partial m_{\pi}^{2}} \left[ m_{\pi}^{2} \Pi (m_{\pi}^{2}) \right]  \,.
\label{yukawa}
\end{equation}

We construct the axial current from the functional derivative with respect to the 
axial source $a_\mu$
$$
A_\mu(\xi)=\frac{\delta \mathcal{A}_{\rm NJL}}{\delta a^{\mu}(\xi)}\Big|_{v_\nu,a_\nu=0}\,.
$$
Expanding $A_\mu(\xi)$ to linear order in $\vec{\widetilde{\pi}}(p)$ yields the matrix element
($a$ and $b$ are flavor labels)
$$
\langle 0|A^{(a)}_\mu(\xi)|\widetilde{\pi}^{(b)}(p)\rangle\stackrel{!}{=}
\delta_{ab}f_\pi(p)p_\mu{\rm e}^{-\imu p\cdot\xi}
$$
from which we get the on-shell pion decay constant $f_\pi(0)=f_{\pi}=4N_{c} m g \Pi(m_{\pi}^{2})$. 
Taking this together with Eqs.~(\ref{pion_mass}) and (\ref{yukawa}) gives three equations for four model 
parameters ($g$, $\Lambda$, $G$ and $m^{0}$) after inserting the empirical data $f_{\pi}=93{\rm MeV}$ 
and $m_{\pi}=138 {\rm MeV}$. This leaves one parameter, say $G$, undetermined. We employ the 
gap equation~(\ref{gap}) to express that undetermined parameter as a function of the constituent 
quark mass $m$ which we take as the sole variable from now on. After all, we have quite some intuition 
about $m$ and expect it to be somewhere around $400{\rm MeV}$. This procedure is reflected by the
first three columns of Tab.~\ref{tab:model} in the proceeding Section. In this calculation the 
current quark mass is only about one third of what is obtained within proper-time regularization 
scheme~\cite{Alkofer:1994ph}. This significant difference again suggests that quarks fields of the 
model are merely some effective degrees of freedom, with little or no relation to fundamental particles.

To apprehend the nucleon structure function calculation let us have a short look at 
DIS off pions which is characterized by a single structure function, $F(x)$,
\begin{equation}
\frac{1}{2\pi} {\sf Abs}\, T_{\mu\nu}(p,q) \quad \bjlim \quad
F(x)\left[-g_{\mu\nu}+\frac{q_\mu q_\nu}{q^2}
-\frac{1}{q^2}\left(p_\mu-\frac{q_\mu}{2x}\right)
\left(p_\nu-\frac{q_\nu}{2x}\right)\right]\, ,
\label{disp1}
\end{equation}
where the Bjorken limit defined after Eq.~(\ref{eq:hten1a}) has been indicated. In order to 
compute the Compton amplitude (\ref{disp1}) we calculate the time-ordered product
\begin{equation}
T\left(J_\mu(\xi) J_\nu(0)\right)=
\frac{\delta^2}{\delta v^\mu(\xi)\delta v^\nu(0)}
\mathcal{A}_{\rm NJL}\Bigg|_{v_\mu=0}
\label{disp2}
\end{equation}
from the action, $\mathcal{A}_{\rm NJL}$ in Eq.~(\ref{eq:defd5}) with $a_\mu=0$ and the 
substitution $v_\mu\,\to\,v_\mu Q$, where $\mathcal{Q}=\frac{1}{3}{\rm diag}(2,-1)$ is the 
quark charge matrix. In principle
we would have to fully expand $\mathcal{A}_{\rm NJL}$ to quadratic order in both the 
photon vector source, $v_\mu$ and the pion field $\vec{\widetilde{\pi}}(p)$. Fortunately
there is some simplification in expanding $\bD\bD_5$. Contributions to this product
that are quadratic in either of the two fields add Feynman diagrams to the 
Compton amplitude that depend only on one of the two momenta. These, kind of local diagrams,
do not have an absorptive component. It is thus sufficient to consider
\begin{equation}
-\bD\bD_5=\partial^2+m^2
+g\gamma_5\left[\dslash,\vec{\pi}\cdot\vec{\tau}\right]
-\imu\left(\dslash\vslash\mathcal{Q}+\vslash\mathcal{Q}\dslash\right)
+\imu g\gamma_5\left[\vec{\pi}\cdot\vec{\tau},\vslash\mathcal{Q}\right]+\ldots\, .
\label{disp3}
\end{equation}
Even with this simplification, the expansion of the logarithm in $\mathcal{A}_{\rm R}$ 
($\mathcal{A}_{\rm I}$ does not contribute) has some un-wanted terms with the flavor 
trace ${\rm tr}[\vec{\pi}\mathcal{Q}\vec{\pi}\mathcal{Q}]$ that would lead to different 
structure functions for the charged and un-charged pions. Fortunately these terms cancel 
in the Bjorken limit. Even when omitting terms which are suppressed in this limit
or eventually do not contribute to the absorptive part, the pion Compton amplitude 
is still quite cumbersome to compute \cite{Weigel:1999pc}
\begin{align}
&\int d^4\xi\, {\rm e}^{\imu q\cdot\xi}\,
\langle \pi(p) |\frac{\delta^2}{\delta v^\mu(\xi)\delta v^\nu(0)}
\mathcal{A}_{\rm NJL}\Bigg|_{v_\mu=0}|\pi(p) \rangle
\nonumber \\* & \hspace{0.5cm}
=\frac{5g^2N_C}{9}\sum_{i=0}^2c_i\int \frac{d\,^4k}{(2\pi)^4}
\frac{1}{-k^2+m^2+\Lambda_i^2-\imu\epsilon}\,
\frac{1}{\left[-(k-p)^2+m^2+\Lambda_i^2-\imu\epsilon\right]^2}
\nonumber \\ &\hspace{2cm} \times
\Bigg\{\frac{-(k-p)^2+m^2+\Lambda_i^2}
{-(k+q-p)^2+m^2+\Lambda_i^2-\imu\epsilon}
{\rm tr}\left(\pslash\gamma^\mu\qslash\gamma^\nu
+\pslash\gamma^\nu\qslash\gamma^\mu\right)
\nonumber \\* & \hspace{2.8cm}
-\frac{-(k-p)^2+m^2+\Lambda_i^2}
{-(k-q-p)^2+m^2+\Lambda_i^2-\imu\epsilon}
{\rm tr}\left(\pslash\gamma^\mu\qslash\gamma^\nu
+\pslash\gamma^\nu\qslash\gamma^\mu\right)
\nonumber \\* & \hspace{2.8cm}
+2m^2\Bigg[\frac{{\rm tr}\left([\kslash-\pslash]\gamma^\nu
\qslash\gamma^\mu\right)}
{-(k-q-p)^2+m^2+\Lambda_i^2-\imu\epsilon}
- \frac{{\rm tr}\left([\kslash-\pslash]
\gamma^\mu\qslash\gamma^\nu\right)}
{-(k+q-p)^2+m^2+\Lambda_i^2-\imu\epsilon}\Bigg]\Bigg\}\,.
\label{disp4}
\end{align}
The last two terms are products of four propagators as expected from an
expansion up to fourth order. The first two terms only have three propagators 
and are represented by diagrams with a pion and a photon at a single 
vertex. This interaction stems from the last term in Eq.~(\ref{disp3}).
As in Eq.~(\ref{eq:comp1}) the absorptive part is extracted by putting all 
intermediate propagators on-shell according to Cutkosky's rule
\begin{align}
\frac{1}{-k^2+m^2+\Lambda_i^2-\imu\epsilon}\quad
&\longrightarrow \quad -2\imu\pi\delta(k^2-m^2-\Lambda_i^2)\cr
\frac{1}{-(k\pm q-p)^2+m^2+\Lambda_i^2-\imu\epsilon}\quad
&\longrightarrow \quad -\frac{\imu \pi}{q^-}
\delta\left(q^+\pm(k-p)^+\right)\, .
\label{disp5}
\end{align}
In the second substitution we introduced light-cone coordinates (the full definition is given
in Section~\ref{ssec:boost}) because they render the implementation of the Bjorken limit quite 
transparent: $q^-\to\infty$ and $q^+\to -x p^+=\frac{xm_\pi}{\sqrt{2}}$
(in the pion rest frame). These coordinates bring in the factor $\frac{1}{q^{-}}$ when extracting 
the absorptive part. {\it A posteriori} this justifies the omission of all terms in Eq.~(\ref{disp4}) 
that did not contain a factor $\qslash$ in the numerator\footnote{The full calculation 
also produces terms involving $(\kslash-\pslash\pm\qslash)^2$ in Eq.~(\ref{disp4}). 
With Eq.~(\ref{disp5}) it is obvious that they do not contribute to the absorptive 
part even though they have a finite Bjorken limit.}. After taking the traces in color and spinor 
spaces the structure function can be read off from $T_{11}+T_{22}\bjlim2F(x)$ using, {\it e.g.},
$$
\gamma^1\qslash\gamma^1\bjlim \frac{1}{2}q^{-}\gamma^1\gamma^{+}\gamma^1
=-\frac{1}{2}q^{-}\gamma^{+}\,.
$$
We find
\begin{align}
F(x)&=-\frac{5\imu}{18}(4N_Cg^2) \sum_{i=0}^2c_i \int \frac{d^4k}{(2\pi)^4}\,
\frac{2\pi \delta(k^2-m^2-\Lambda_i^2)}{\left[-(k-p)^2+m^2+\Lambda_i^2-\imu\epsilon\right]^2}\cr
&\hspace{1cm}\times
\bigg\{\left[-(k-p)^2+m^2+\Lambda_i^2\right]\left[
\delta(k^{+}-p^{+}-q^{+})-\delta(k^{+}-p^{+}+q^{+})\right]p^{+}\cr
&\hspace{2cm}
+m^2\left[\delta(k^{+}-p^{+}+q^{+})-\delta(k^{+}-p^{+}-q^{+})\right](k^{+}-p^{+})\bigg\}\,.
\label{eq:Fpi1}
\end{align}
The $\delta$-functions straightforwardly produce the $k^{+}$ integrals fixing this variable
to either $m_\pi(1-x)$ or $m_\pi(1+x)$. The integral over $k^{-}$ can then be computed via
the $\delta$-function in first numerator. The result from these two integrals is
\begin{align}
F(x)&=\frac{5}{18}(4N_Cg^2) \sum_{i=0}^2c_i \int \frac{d^2k_\perp}{(2\pi)^3}\,
\left\{\frac{M^2_i(x)\theta(x)\theta(1-x)}{x(1-x)\left[M^2_i(x)-m_\pi^2\right]^2}+
\big(x\,\longleftrightarrow\,-x\big)\right\}\,,
\label{eq:Fpi2}
\end{align}
where $M^2_i(x)=\frac{1}{x(1-x)}\left[m^2+\Lambda_i^2+k_\perp^2\right]$. This expression for the pion 
structure function was earlier obtained using light-cone wave-functions~\cite{Frederico:1994dx,Davidson:1994uv}.
In the chiral limit, $m_\pi=0$, this structure function is just a constant on the interval $-1\le x\le1$.
It is interesting to note that the light-cone coordinate momentum variables can also be integrated in the 
pion polarization function, Eq.~(\ref{specfct}) leading to the same $k_\perp$ integral allowing the compact 
expression
\begin{equation}
F(x)=\frac{5}{9} (4N_C g^2)
\frac{\partial}{\partial p^2}\left[p^2\Pi(p^2,x)\right]\Bigg|_{p^2=m_\pi^2}\, .
\label{disp8}
\end{equation}
At this point one important aspect has not been considered. As it stands, Eq.~(\ref{disp8})
is the pion structure function at the scale at which the NJL-model is supposed to 
approximate QCD. Stated otherwise, the structure function computed from Eq.~(\ref{disp8})
approximates the QCD result at a (presumably) low renormalization scale. To allow a 
comparison with data, the QCD evolution equations must be applied to the model prediction.
At that stage, the low renormalization scale enters as a new parameter that is tuned
to optimize the agreement with the data at the higher energy scale of the experiments. This
calculation has been carried out in Ref.~\cite{RuizArriola:2002bp}. Here we will not
further elaborate on QCD evolution but will get back to it in Section \ref{sec:NR} in
the context of the nucleon structure functions.

The main lesson learned from this pion structure function study is that the calculation 
simplifies significantly when identifying the propagators that carry the momentum which is large 
in the Bjorken limit and ignoring the others (the many terms not shown in Eq.~(\ref{disp4})) 
and/or simplifying them by approximating them with free quark propagators.

\section{Self-consistent soliton}
\label{sec:soliton}

The soliton is a static meson configuration that minimizes the bosonized action. To construct
this configuration we define a Dirac Hamiltonian $h$ via the Dirac operators in Eq.~(\ref{eq:defd5})
\begin{equation}
\imu\mathbf{D}^{(\pi)}=\beta(\imu\partial_t-h) \quad {\rm and}\quad
\imu\mathbf{D}^{(\pi)}_5=(-\imu\partial_t-h)\beta\, .
\label{defh}
\end{equation}
Its diagonalization
\begin{equation}
h\Psi_\alpha = \epsilon_\alpha \Psi_\alpha \, ,
\label{diagh}
\end{equation}
yields eigenvalues $\epsilon_\alpha$ and eigen-spinors 
$\Psi_\alpha=\sum_{\beta}V_{\alpha\beta}\Psi^{(0)}_\alpha$ 
as linear combinations of the free Dirac spinors $\Psi^{(0)}_\alpha$ in a spherical basis.

When constraining the meson configuration 
to the chiral circle, {\it i.e.} parameterizing $M=mU$ with only $U$ being 
dynamical, the so-called hedgehog configuration \cite{Pa46} minimizes the action 
in the unit baryon number sector\footnote{We refer to the earlier review 
articles \cite{Alkofer:1994ph,Christov:1995vm} for obstacles and their solutions 
for hedgehog configurations away from the chiral circle.}. This, together with 
(the assumption of) spherical symmetry suggest the {\it ansatz}
\begin{equation}
h=\vec{\alpha}\cdot\vec{p} +\beta\, m\, U_5(\vec{r}) 
\qquad {\rm where}\qquad 
U_5(\vec{r})={\rm exp} \left[\imu \hat{r}\cdot\vec{\tau}\,
\gamma_5 \Theta(r)\right]\, .
\label{hedgehog}
\end{equation}
The radial profile function  $\Theta(r)$ is called the chiral angle. The hedgehog 
configuration, Eq.~(\ref{hedgehog}) is invariant under so-called grand spin 
transformations that combine flavor and coordinate rotations. Accordingly, the 
Dirac and flavor components of the eigenfunctions $\Psi_\alpha$ are products of 
radial functions and grand spin eigenfunctions. The latter are products of 
spherical harmonic functions, spinors and iso-spinors. Final discretization is 
accomplished by imposing boundary conditions on the radial functions at a distance 
$D$ much larger than typical extensions of the chiral angle \cite{Kahana:1984be}. 
Different boundary conditions are equivalent in the limit $D\to\infty$, however, at 
large but finite $D$ a certain choice may be preferable depending on which quantity 
is to be computed~\cite{Alkofer:1994gd}. All possible boundary conditions require 
that there is no flux through the sphere at $D$.

Once the structure of the spinors is established, particular profile functions can be 
considered. For profiles with $\Theta(0)=-\pi$ and $\lim_{r\to\infty}\Theta(r)=0$ the 
diagonalization, Eq.~(\ref{diagh}) yields a distinct, strongly bound level, (eigenvalue
$\epsilon_{\rm v}$, eigen-spinor $\Psi_{\rm v}$) in the grand spin zero channel. This 
level is referred to as the valence quark level \cite{Alkofer:1994ph}: the wider the 
chiral angle, the more strongly bound is this distinct level. Its (explicit) occupation 
ensures unit baryon number. 

The functional trace in $\mathcal{A}_R$ ($\mathcal{A}_I$ vanishes for static 
configurations) is computed as an integral over the time interval $T$ and a discrete 
sum over the basis levels defined by Eq.~(\ref{diagh}). In the limit $T\to\infty$ 
the vacuum contribution to the static energy is then extracted from 
$\mathcal{A}_R\to-TE_{\rm vac}$. Collecting pieces, we obtain
the total energy functional as \cite{Alkofer:1994ph,Christov:1995vm}
\begin{equation}
E_{\rm tot}[\Theta]=
\frac{N_C}{2}\left[1+{\rm sign}(\epsilon_{\rm v})\right]
\epsilon_{\rm v}
-\frac{N_C}{2}\sum_{i=0}^2 c_i \sum_\alpha
\left\{\sqrt{\epsilon_\alpha^2+\Lambda_i^2}
-\sqrt{\epsilon_\alpha^{(0)2}+\Lambda_i^2}
\right\}
+m_\pi^2f_\pi^2\int d^3r \, \left[1-{\rm cos}(\Theta)\right]\, .
\label{etot}
\end{equation}
Here we have also subtracted the vacuum energy associated with the non-dynamical meson field 
configuration $\Theta\equiv0$ (denoted by the superscript on the energy eigenvalues) that is 
often called the cosmological constant contribution. This subtraction will also 
play an important role for the unpolarized isoscalar structure function as it enters
via the momentum sum rule. Obviously, the soliton energy is linear in $N_C$ as 
ascertained for baryon masses in QCD \cite{Witten:1979kh}.

The soliton profile is then obtained as the profile function $\Theta(r)$ that minimizes the total 
energy $E_{\rm tot}$ self-consistently subject to the above mentioned boundary conditions on 
$\Theta(r)$. The energy eigenvalues $\epsilon_\alpha$ are functionals of the chiral angle through 
the diagonalization in Eq.~(\ref{diagh}). Hence the minimization of $E_{\rm tot}[\Theta]$ involves
$$
\frac{\delta \epsilon_\alpha}{\delta \Theta(r)}
=m\int d^3r^\prime\, \Psi^\dagger_\alpha(\vec{r\,}^\prime)\beta
\left[-\sin\Theta(r^\prime)+\imu \hat{r}^\prime\cdot\vec{\tau}\cos\Theta(r^\prime)\right]
\Psi_\alpha(\vec{r\,}^\prime)\delta(r-r^\prime)\,,
$$
by the chain rule. Self-consistency arises as the wave-functions in this functional derivative
emerge from diagonalizing an operator that contains $\Theta(r)$. Though this Hartree-type 
problem is quite elaborate, it has been established some time ago \cite{Reinhardt:1988fz}
and ever been refined \cite{Alkofer:1994ph,Christov:1995vm}. The two main contributions 
to $E_{\rm tot}[\Theta]$ act in opposite directions: the binding of the distinct level
is attractive while the Dirac sea piece (partially) compensates for this reduction.
As the binding of the valence level increases with the constituent quark mass $m$, 
the soliton is kinematically stable against decaying into $N_C$ unbound quarks 
for $m\gtrsim400{\rm MeV}$, {\it cf.} Tab.~\ref{tab:model}.

This soliton represents an object which has unit baryon number but neither good quantum 
numbers for spin and flavor (isospin). Such quantum numbers are generated by canonically 
quantizing the time-dependent collective coordinates $A(t)$ that parameterize the 
spin-flavor orientation of the soliton via
\begin{equation}
U_5(\vec{r},t)=A(t)U_5(\vec{r})A^\dagger(t)\,,
\label{collq0}
\end{equation}
where $U_5(\vec{r})$ is the self-consistent static configuration from Eq.~(\ref{hedgehog}). 
For a rigidly rotating soliton the Dirac operator becomes, after transforming to 
the flavor rotating frame \cite{Reinhardt:1989st},
\begin{equation}
\imu\mathbf{D}^{(\pi)}=A\beta\left(\imu\partial_t - 
\frac{1}{2}\vec{\Omega}\cdot\vec{\tau}-h\right)A^\dagger
\quad{\rm and}\quad
\imu\mathbf{D}^{(\pi)}_5=A\left(-\imu\partial_t + 
\frac{1}{2}\vec{\Omega}\cdot\vec{\tau} -h\right)\beta A^\dagger\, .
\label{collq1}
\end{equation}
Actual computations involve an expansion with respect to the angular velocities $\vec{\Omega}$ 
that are defined by that time derivative of the collective coordinates as
\begin{equation}
A^\dagger \frac{d}{dt} A = \frac{\imu}{2}\vec{\Omega}\cdot\vec{\tau}\,.
\label{collq2}
\end{equation}
According to the canonical quantization rules the angular velocities are replaced by the spin operator 
\begin{equation}
\vec{\Omega}\longrightarrow \frac{1}{\alpha^2}\, \vec{J}\, .
\label{collq3}
\end{equation}
The constant of proportionality is the moment of inertia 
\begin{align}
\alpha^2&=\frac{N_C}{4}\left[1+{\rm sign}(\epsilon_{\rm v})\right]
\sum_{\beta\ne{\rm v}} \frac{|\langle{\rm v}|\tau_3|\beta\rangle|^2}
{\epsilon_{\beta}-\epsilon_{\rm v}}\cr
&\hspace{0.5cm}
+\frac{N_C}{8}\sum_{\alpha\ne\beta}\sum_{i=0}^2c_i
\frac{|\langle\alpha|\tau_3|\beta\rangle|^2}{\epsilon_\alpha^2-\epsilon_\beta^2}
\left\{\frac{\epsilon_\alpha^2+\epsilon_\alpha\epsilon_\beta+2\Lambda_i^2}
{\sqrt{\epsilon_\alpha^2+\Lambda_i^2}}
-\frac{\epsilon_\beta^2+\epsilon_\alpha\epsilon_\beta+2\Lambda_i^2}
{\sqrt{\epsilon_\beta^2+\Lambda_i^2}}\right\}\,,
\label{mominert}\end{align}
expressed by introducing eigenstates $|\alpha\rangle$ of $h$; {\it i.e.}
$\Psi_\alpha(\vec{r})=\langle \vec{r}|\alpha\rangle$. The moment of inertia is $\mathcal{O}(N_C)$ 
and is extracted as twice the constant of proportionality of the $\mathcal{O}(\vec{\Omega}^2)$ term 
in the Lagrange function ($\mathcal{A}/T$). With Eq.~(\ref{collq3}) the expansion in $\vec{\Omega}$ 
is thus equivalent to the one in $\frac{1}{N_C}$. After quantizing the collective coordinates 
the Hamilton operator is that of a rigid rotor leading to the energy formula
\begin{equation}
E(j)=E_{\rm tot}+\frac{1}{2\alpha^2}\,j(j+1)\,,
\label{eq:rotor}
\end{equation}
with spin eigenvalues $j=\frac{1}{2}$ for the nucleon and $j=\frac{3}{2}$ for the 
$\Delta$-resonance. Note that this energy formula contains a piece linear in $N_C$ and one linear 
in $\frac{1}{N_C}$. The contribution $\mathcal{O}(N_C^0)$, which is the vacuum polarization energy 
from the meson fluctuations, is generally omitted in soliton models. There is no robust calculation 
of this vacuum polarization energy because these models are not renormalizable. Estimates indicate 
that the $\mathcal{O}(N_C^0)$ component significantly reduces the energy \cite{Meier:1996ng}. Since this 
part does not depend on the baryon quantum numbers, it is customary to only consider mass differences, 
in particular, the $\Delta$-nucleon mass difference $\Delta M=\frac{3}{2\alpha^2}$. The results shown
in Tab.~\ref{tab:model} suggest that $m\approx400{\rm MeV}$ reproduces the experimental value 
of $293{\rm MeV}$ reasonably well.

The nucleon wave-function becomes a (Wigner D) function of the collective coordinates. A useful relation 
in computing matrix elements of nucleon states $|N\rangle$ is \cite{Adkins:1983ya}
\begin{equation}
\langle N |D_{ab}|N\rangle=-\frac{4}{3}\langle N | I_a J_b | N\rangle
\qquad {\rm with}\qquad
D_{ab}=\frac{1}{2}{\rm tr} \left(A^\dagger\tau_a A\tau_b\right)\,.
\label{collq4}
\end{equation}
Here $I_a$ and $J_b$ are iso- and spin operators, respectively. The above matrix element
arises from the operator identity $I_a=-D_{ab}J_b$ which by itself reflects the invariance
of the hedgehog configuration under combined isospin and coordinate rotations.

As an example for the computation of a static nucleon property we consider the vacuum contribution 
to the axial charge, $g_a$, of the nucleon because in Section \ref{sec:HTsol} it will be 
paradigmatic for how sum rules for structure functions emerge in this model and its treatment. 
In the first step we require the spatial components of 
the axial current as a function of the collective coordinates $A$. This is achieved by expanding the 
regularized action to leading order in the axial source $\aslash$ with $a^0=0$
\begin{align}
\mathcal{A}_{\rm NJL}&=-\imu \frac{N_C}{2}\sum_{i=0}^2 c_i
{\rm Tr}\,{\rm log}\left\{\beta\left(\partial_t^2+h^2\right)\beta
+\beta\left(\imu \partial_t+h\right)\aslash\gamma_5
+\aslash\gamma_5\left(-\imu \partial_t+h\right)\beta
+\Lambda_i^2-\imu\epsilon\right\}\cr
&=-\imu \frac{N_C}{2}\sum_{i=0}^2 c_i
{\rm Tr}\,{\rm log}\left\{\partial_t^2+h^2+\left\{h,\aslash\gamma_5\beta\right\}
+\Lambda_i^2-\imu\epsilon\right\} \,.
\label{eq:ga1}
\end{align}
The next simplification is that we only need the (space) integral of that current
and therefore may take $\aslash \gamma_5\beta=
-\vec{a}^{(a)}\cdot\vec{\alpha}\gamma_5\frac{\tau_a}{2}
=-\vec{a}^{(a)}\cdot\vec{\Sigma}\frac{\tau_a}{2}$ with 
constant $\vec{a}^{(a)}$ to compute
\begin{align}
\frac{\partial \mathcal{A}_{\rm NJL}}{\partial \vec{a}^{(a)}}\Big|_{\vec{a}^{(a)}=0}&=
\imu \frac{N_C}{2}\sum_{i=0}^2 c_i {\rm Tr}\left\{
\left(\partial_t^2+h^2+\Lambda_i^2-\imu\epsilon\right)^{-1}
\left\{h,\vec{\Sigma}\frac{\tau_a}{2}\right\}\right\}\cr
&=\imu \frac{N_C}{2}\sum_{i=0}^2 c_i T\int \frac{d\omega}{2\pi} {\rm Tr}^\prime
\left\{\left(-\omega^2+h^2+\Lambda_i^2-\imu\epsilon\right)^{-1}
\left\{h,\vec{\Sigma}\frac{\tau_a}{2}\right\}\right\}\,.
\label{eq:ga2}
\end{align}
As for any path integral, the limit $T\to\infty$ extracts the vacuum (Dirac sea) component. 
In the next step we want to evaluate the remaining trace ${\rm Tr}^\prime$ using the eigenvalues 
$\epsilon_\alpha$ and the eigenstates $|\alpha\rangle$ of $h$. Substituting the rotating hedgehog 
configuration from Eq.~(\ref{collq0}) and using the cyclic property of the trace yields
\begin{align}
\frac{\partial \mathcal{A}_{\rm NJL}}{\partial \vec{a}^{(a)}}\Big|_{\vec{a}^{(a)}=0}&=
\imu \frac{N_C}{2}\sum_{i=0}^2 c_i T\int \frac{d\omega}{2\pi} \sum_\alpha
\left\{\left(-\omega^2+\epsilon_\alpha^2+\Lambda_i^2-\imu\epsilon\right)^{-1}
2\epsilon_\alpha\left\langle \alpha\left|\vec{\Sigma}A^\dagger \frac{\tau_a}{2} A
\right|\alpha\right\rangle\right\}\cr
&=-\frac{N_C}{2} T D_{ab}\sum_{i=0}^2 c_i \sum_\alpha
\frac{\epsilon_\alpha}{\sqrt{\epsilon_\alpha^2+\Lambda_i^2}}
\left\langle \alpha\left|\vec{\Sigma}\frac{\tau_b}{2}\right|\alpha\right\rangle \,,
\label{eq:ga3}
\end{align}
where the frequency integral has been computed by contour integration. The vacuum contribution 
to the axial charge is then obtained as the proton matrix element
\begin{equation}
g^{\rm (s)}_a=\lim_{T\to\infty}\frac{1}{T}\left\langle P \left|2
\frac{\partial \mathcal{A}_{\rm NJL}}{\partial a_z^{(3)}}\Bigg|_{\vec{a}^{(3)}=0}\right|P\right\rangle
=\frac{N_C}{6}\sum_{i=0}^2 c_i \sum_\alpha\frac{\epsilon_\alpha}{\sqrt{\epsilon_\alpha^2+\Lambda_i^2}}
\langle \alpha|\Sigma_3\tau_3|\alpha\rangle \,,
\label{eq:ga4}
\end{equation}
with spin projection $J_3=+\frac{1}{2}$. In addition we have the contribution from the valence quark
that we get via a similar derivative after ''gauging'' the valence level
\begin{equation}
\frac{N_C}{2}\left[1+{\rm sign}(\epsilon_{\rm v})\right]\epsilon_{\rm v}
\,\longrightarrow\,
\frac{N_C}{2}\left[1+{\rm sign}(\epsilon_{\rm v})\right] D_{ab}
\left\langle {\rm v}\left|h+\vec{a}^{(a)}\cdot\vec{\Sigma}\frac{\tau_b}{2}\right|{\rm v}\right\rangle
\label{eq:ga5}
\end{equation}
so that
$g^{\rm (v)}_a=-\frac{N_C}{6}\left[1+{\rm sign}(\epsilon_{\rm v})\right]
\langle {\rm v}|\Sigma_3\tau_3|{\rm v}\rangle$. In total we have
\begin{equation}
g_a=g^{\rm (v)}_a+g^{\rm (s)}_a=-\frac{N_C}{6}\left[1+{\rm sign}(\epsilon_{\rm v})\right]
\langle {\rm v}|\Sigma_3\tau_3|{\rm v}\rangle
+\frac{N_C}{6}\sum_{i=0}^2 c_i \sum_\alpha\frac{\epsilon_\alpha}{\sqrt{\epsilon_\alpha^2+\Lambda_i^2}}
\langle \alpha|\Sigma_3\tau_3|\alpha\rangle \,.
\label{eq:ga6}
\end{equation}
It is illuminating to make the single cut-off regularization from Eq.~(\ref{eq:LAMBDA}) explicit
\begin{equation}
g_a=-\frac{N_C}{6}\left[1+{\rm sign}(\epsilon_{\rm v})\right]
\langle {\rm v}|\Sigma_3\tau_3|{\rm v}\rangle
+\frac{N_C}{6}\sum_\alpha \left[{\rm sign}(\epsilon_\alpha)
-\epsilon_\alpha\frac{\epsilon_\alpha^2+\frac{3}{2}\Lambda^2}
{\left(\epsilon_\alpha^2+\Lambda^2\right)^{\frac{3}{2}}}\right]
\langle \alpha|\Sigma_3\tau_3|\alpha\rangle \,.
\label{eq:ga7}
\end{equation}
The strongly bound valence level is also included in the sum over $\alpha$. As the binding of 
that level is increased, for example by increasing the constituent quark mass $m$ in the 
self-consistent construction, the corresponding energy eigenvalue eventually changes sign. The 
particular combination of valence and sea contributions ensures that $g_a$ is continuous as the 
terms with ${\rm sign}(\epsilon_{\rm v})$ cancel. This feature is universal for any quantity; 
there is no discontinuity as the sign of the valence energy eigenvalue changes\footnote{Taking
the "chemical potential" to be zero is a choice anyhow.}. This is also true for the energy, 
Eq.~(\ref{etot}) and the moment of inertia, Eq.~(\ref{mominert}). Essentially this occurs 
by construction as the prefactor $\frac{1}{2}\left[1+{\rm sign}(\epsilon_{\rm v})\right]$ is 
introduced to ensure unit baryon number\footnote{In analogy to $g_a$ the baryon number is 
obtained from a functional derivative with respect to constant $v_0$. The vacuum contribution 
stems from $\mathcal{A}_{\rm I}$ and is not regularized. As mentioned earlier 
$\mathcal{A}_{\rm I}$ is conditionally convergent in the sense that the sum over $\alpha$ must 
be taken over a symmetric interval.}
$$
B=\frac{1}{2}\left[1+{\rm sign}(\epsilon_{\rm v})\right]
-\frac{1}{2}\sum_\alpha {\rm sign}(\epsilon_\alpha)\,.
$$
Stated otherwise, when the valence level is so strongly bound that its energy eigenvalue
is negative, the baryon number is carried by the polarized Dirac sea (vacuum). This is 
an implicitly assumed feature of topological chiral soliton models like the Skyrme model 
because the topological current is the leading term in the gradient expansion for the vacuum 
contribution of the baryon current in chiral quark models \cite{Alkofer:1992wy}.

\begin{table}[t]
\centerline{
\begin{tabular}{c|cc|cc|cc}
$m [{\rm MeV}]$  & $m^0 [{\rm MeV}]$ & $\Lambda [{\rm GeV}]$ & $E_{\rm tot} [{\rm MeV}]$ &
$\alpha^2 [1/{\rm GeV}]$ & $\Delta M [{\rm MeV}]$ & $g_a$\cr
\hline
350 & 7.9  & 0.77 & 1267 & 8.65 & 173 & 0.85\cr
400 & 8.4  & 0.74 & 1269 & 5.89 & 255 & 0.80\cr
450 & 8.5  & 0.73 & 1257 & 4.82 & 311 & 0.77
\end{tabular}}
\caption{\label{tab:model} Model parameters and results. See the main text for their 
definitions.}
\end{table}

In Tab.~\ref{tab:model} we also list the model predictions for $g_a$. They are about 30\% below the 
empirical value of $1.26$ \cite{Barnett:1996hr}. Note, however, that only the leading $\frac{1}{N_C}$ 
result is given. It has been asserted that, because of the time-ordering prescription in the path integral for 
bosonization, subleading contributions can significantly increase the model prediction \cite{Christov:1995vm}.
These contributions are, unfortunately, not without further problems. For example, they violate
PCAC: In soliton models a partially conserved axial current (PCAC) results from the field equation for 
the soliton. This equation contains only the leading order in $\frac{1}{N_C}$ and any subleading piece 
in the axial current is not covered. Altering the field equation accordingly \cite{Alkofer:1993pv} does 
not produce a stable soliton when the subleading Dirac sea contribution\footnote{Early studies 
\cite{Wakamatsu:1993nq} omitted that part.} is properly included \cite{Christov:1995vm}.

\section{Hadron tensor for the nucleon as soliton}
\label{sec:HTsol}
We now get to a central topic of this short review: extracting the nucleon structure functions 
from the hadron tensor in the soliton background while realizing regularization from the onset 
of the action, Eq.~(\ref{act1}). Here we will consider mainly the example of the leading 
$\frac{1}{N_C}$ component of the longitudinal polarized structure function, $g_1$. For this 
example we will also explain how sum rules are established in the fully regularized formulation. 
For further details on other structure functions, that are obtained using quite a similar 
procedure, we refer to to original literature \cite{Weigel:1999pc,Takyi:2019ahv}.

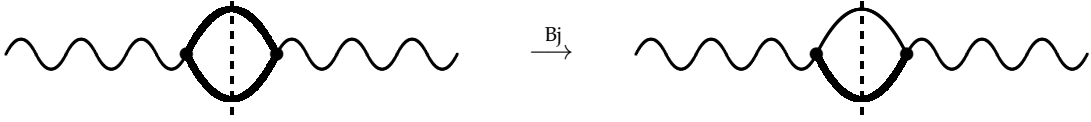
\begin{figure}[t]
\parbox[l]{0.5cm}{~}
\setlength{\unitlength}{2mm}
\begin{picture}(34.0,8.0)
\linethickness{1pt}
\bezier{40}(0,4)(1,6)(2,4)
\bezier{40}(2,4)(3,2)(4,4)
\bezier{40}(4,4)(5,6)(6,4)
\bezier{40}(6,4)(7,2)(8,4)
\bezier{40}(8,4)(9,6)(10,4)
\bezier{40}(10,4)(11,2)(12,4)
\put(12,4){\circle*{0.8}}
\linethickness{2pt}
\bezier{200}(12,4)(15,10)(18,4)
\bezier{200}(12,4)(15,-2)(18,4)
\put(18,4){\circle*{0.8}}
\linethickness{1pt}
\bezier{40}(18,4)(19,6)(20,4)
\bezier{40}(20,4)(21,2)(22,4)
\bezier{40}(22,4)(23,6)(24,4)
\bezier{40}(24,4)(25,2)(26,4)
\bezier{40}(26,4)(27,6)(28,4)
\bezier{40}(28,4)(29,2)(30,4)
\linethickness{1.4pt}
\multiput(15,0)(0,1){8}{\line(0,1){0.5}}
\end{picture}
\parbox[t]{1.4cm}{\vspace{-1.2cm}$\bjlim$}
\begin{picture}(32.0,8.0)
\linethickness{1pt}
\bezier{40}(0,4)(1,6)(2,4)
\bezier{40}(2,4)(3,2)(4,4)
\bezier{40}(4,4)(5,6)(6,4)
\bezier{40}(6,4)(7,2)(8,4)
\bezier{40}(8,4)(9,6)(10,4)
\bezier{40}(10,4)(11,2)(12,4)
\put(12,4){\circle*{0.8}}
\bezier{200}(12,4)(15,10)(18,4)
\linethickness{2pt}
\bezier{200}(12,4)(15,-2)(18,4)
\put(18,4){\circle*{0.8}}
\linethickness{1pt}
\bezier{40}(18,4)(19,6)(20,4)
\bezier{40}(20,4)(21,2)(22,4)
\bezier{40}(22,4)(23,6)(24,4)
\bezier{40}(24,4)(25,2)(26,4)
\bezier{40}(26,4)(27,6)(28,4)
\bezier{40}(28,4)(29,2)(30,4)
\linethickness{1.2pt}
\multiput(15,0)(0,1){8}{\line(0,1){0.5}}
\end{picture}
\caption{\label{fig2}Two photon coupling to fermion loop. Thick lines are the full
fermion propagators ${\bDp}^{-1}$ (or ${\bDp}_5^{-1}$) without any perturbation expansion. 
The thin line in the loop represents a free (massless) fermion propagator, $\dslash^{-1}$.
Dashed lines denote Cutkosky cuts as discussed after Eq.~(\ref{nuc3}).}
\end{figure}

Similar to the pion structure function in 
Sect.~\ref{sec:model} we start from the Compton tensor, Eq.~(\ref{disp2}). However, this time 
we have to account for the non-perturbative nature of the solitonic meson fields, and may not
approximate $\bDp$ except for the $\frac{1}{N_C}$ expansion.
As mentioned in that earlier Section, isospin violating contributions may arise that only 
cancel once the Bjorken limit is assumed. Can we anticipate this type of cancellations for the 
soliton configuration at an earlier stage and thus simplify the calculation (somewhat)? As a matter 
of fact the appearance of these terms is indeed an artifact of the simultaneous expansion in the pion 
and photon fields, Eq.~(\ref{disp3}). We might equally well have expanded only in the photon field 
first (taking the charge matrix $\mathcal{Q}$ as part of $\vslash$, for simplicity)
\begin{equation}
-{\rm Tr}\,\left\{\left(-\bDp\bDp_5+\Lambda_i^2\right)^{-1}
\left[\left(\bDp\vslash+\vslash\bDp_5\right)
\left(-\bDp\bDp_5+\Lambda_i^2\right)^{-1}
\left(\bDp\vslash+\vslash\bDp_5\right)\right]\right\}\, .
\label{simple1}
\end{equation}
Here square brackets have been introduced to mark those factors that are sensitive to the large photon 
momentum. Due to the cyclic properties of the trace this is merely a choice but it must contain all vertices
with $\vslash$. In momentum space the propagator inside the square brackets behaves like $1/Q^2$ when
assuming the Bjorken limit. In particular this implies that 
\begin{align}
\Big[\ldots\Big] &\bjlim
\left(\bDp\vslash+\vslash\bDp_5\right)
\left(-\bDp\bDp_5\right)^{-1}
\left(\bDp\vslash+\vslash\bDp_5\right)\cr
&\bjlim -\bDp\vslash\left(\bDp_5\right)^{-1}\vslash
-\vslash\left(\bDp\right)^{-1}\vslash\bDp_5\, .
\label{simple2}
\end{align}
Terms with either the unit or the $\left(-\bDp\bDp_5\right)^{-1}$ operators between two 
vector sources have been omitted because they will either not depend on the photon momentum, 
{\it cf.} the discussion before Eq.~(\ref{disp3}), or are additionally suppressed by factors 
of $\frac{1}{Q^2}$. The above replacement tells us that in the Bjorken limit the propagator 
through which the large photon momentum runs will not contain the cut-offs $\Lambda_i$. In 
particular there will be no contributions which behave like $\frac{Q^2}{\Lambda_i^2}$; thereby 
the proper scaling behavior is manifest. In other regularization schemes, like {\it e.g.} 
proper-time, wherein the cut-off is not additive to the loop momenta, the absence of such scaling 
violating contributions is not apparent. Previously, Eq.~(\ref{simple1}), we expanded the 
operator in powers of the pion field leading to complicated three and four vertex quark loops. 
Now we see that the Bjorken limit enforces the cancellations among those diagrams that we 
observed for the pion structure function. The expression (\ref{simple2}) simplifies even further 
by noting that the quark propagator between the two photon insertions carries the large photon 
momentum and should hence be approximated by the free massless propagator, 
\begin{align}
\Big[\ldots\Big] &\longrightarrow
\bDp\vslash\left(\dslash\right)^{-1}\vslash
-\vslash\left(\dslash\right)^{-1}\vslash\bDp_5\, .
\label{simple3}
\end{align}
The transition from the expression (\ref{simple1}) to (\ref{simple3}) is illustrated in Fig.~\ref{fig2}.
Substituting this simplification into Eq.~(\ref{simple1}) leads to
\begin{equation}
{\rm Tr}\,\left\{\left[\left({\bDp}\right)^{-1}-\left({\bDp_5}\right)^{-1}\right]
\vslash\left(\dslash\right)^{-1}\vslash\right\}\,+\quad
\mbox{reguarlization terms}\,.
\label{simple3a}
\end{equation}
Essentially we only include small and moderate momenta from the loop integrals for one of the two 
propagators, keeping in mind that the sum of the momenta in the propagators is subject to the Bjorken 
limit. The integration regime in which that large momentum is distributed (approximately) equally 
among the two propagators does not contribute in the Bjorken limit \cite{Weigel:1999pc}.

Having simplified the construction of the Compton tensor with the soliton background
in the Bjorken limit we see that it will be sufficient to differentiate (bringing back the 
charge matrix $\mathcal{Q}$)
\begin{equation}
\mathcal{A}_{\Lambda,{\rm R}}^{(2,v)}=
-\imu\frac{N_C}{4}\sum_{i=0}^2c_i
{\rm Tr}\,\left\{\left(-\bDp\bDp_5+\Lambda_i^2\right)^{-1}
\left[\mathcal{Q}^2\vslash\left(\dslash\right)^{-1}\vslash\bDp_5
-\bDp(\vslash\left(\dslash\right)^{-1}\vslash)_5
\mathcal{Q}^2\right]\right\}
\label{simple6}
\end{equation}
with respect to the vector sources. As already mentioned after Eq.~(\ref{eq:defd5})
the operator $\bD_5$, which was introduced to accomplish regularization, produces
an unconventional Ward identity because, in contrast to $\bD$, this $\gamma_5$-odd 
operator has a relative minus sign between the derivative operator $\imu\dslash$ and
the axial vector source $\aslash\gamma_5$. To correct this regularization artifact in a way
consistent with the Bjorken sum rule \cite{Bjorken:1966jh} for the nucleon axial charge, 
$g_a$, this relative sign must also be reflected in the Dirac decomposition of
$(\vslash\left(\dslash\right)^{-1}\vslash)_5=v^\mu\frac{\partial^\rho}{\partial^2}v^\nu
(\gamma_\mu\gamma_\rho\gamma_\nu)_5$:
\begin{equation}
\gamma_\mu\gamma_\rho\gamma_\nu
=S_{\mu\rho\nu\sigma}\gamma^\sigma
-\imu\epsilon_{\mu\rho\nu\sigma}\gamma^\sigma\gamma^5
\quad {\rm while} \quad
(\gamma_\mu\gamma_\rho\gamma_\nu)_5
=S_{\mu\rho\nu\sigma}\gamma^\sigma+
\imu\epsilon_{\mu\rho\nu\sigma}\gamma^\sigma\gamma^5\, ,
\label{defsign}
\end{equation}
where $S_{\mu\rho\nu\sigma}=g_{\mu\rho}g_{\nu\sigma}
+g_{\rho\nu}g_{\mu\sigma}-g_{\mu\nu}g_{\rho\sigma}$.
We recall that the $\bD_5$ model, which is not physical, has been solely introduced as a device 
to allow for a regularization which maintains the anomaly structure of the underlying theory by 
regularizing $\mathcal{A}_{\rm R}$ and $\mathcal{A}_{\rm I}$ differently. Hence it is not at all 
surprising that further specification of this regularization prescription is demanded in order to 
formulate a fully consistent model. We stress that this issue is not specific to the Pauli-Villars 
scheme. All schemes that regularize the sum, ${\rm log}\,(\bD)+{\rm log}\,(\bD_5)$ but not the 
difference, ${\rm log}\,(\bD)-{\rm log}\,(\bD_5)$ will require the specification~(\ref{defsign}). 
Since only the polarized, {\it i.e.} spin dependent, structure functions are effected, this issue has 
not shown up when computing the pion structure function.

For the imaginary part of the action the expression analogous to
Eq.~(\ref{simple6}) reads
\begin{align}
\mathcal{A}_{\Lambda,{\rm I}}^{(2,v)}&=
-\imu\frac{N_C}{4}
{\rm Tr}\,\left\{\left(-\bDp\bDp_5\right)^{-1}
\left[\mathcal{Q}^2\vslash\left(\dslash\right)^{-1}\vslash\bDp_5
+\bDp(\vslash\left(\dslash\right)^{-1}\vslash)_5
\mathcal{Q}^2\right]\right\}\cr
&=\imu\frac{N_C}{4}{\rm Tr}\,\left\{\left(\bDp\right)^{-1} 
\mathcal{Q}^2\vslash\left(\dslash\right)^{-1}\vslash
+\left(\bDp_5\right)^{-1} (\vslash\left(\dslash\right)^{-1}\vslash)_5 \mathcal{Q}^2\right\}\,.
\label{simple7}
\end{align}
Again, it is understood that the large photon momentum runs only through the operators in square brackets 
in the first equation. Note that in the unregularized case ($\Lambda_i\equiv0$) the contributions associated 
with $\bD_5$ would cancel in the sum of Eqs.~(\ref{simple6}) and~(\ref{simple7}) leaving
\begin{equation}
\mathcal{A}^{(2,v)}=
\mathcal{A}_{\Lambda,{\rm R}}^{(2,v)}+\mathcal{A}_{\Lambda,{\rm I}}^{(2,v)}=
\imu\frac{N_C}{2}
{\rm Tr}\,\left\{\left(\bDp\right)^{-1}
\left[\mathcal{Q}^2\vslash\left(\dslash\right)^{-1}
\vslash\right]\right\}\,+\,\mbox{regularization terms}\, ,
\label{simple8}
\end{equation}
and the adjustment, Eq.~(\ref{defsign}) would not be efficacious.
Expanding this expression to quadratic order in the pseudoscalar field $P$ produces 
the standard `handbag' diagram with the propagators connecting the quark-pion and
quark-photon vertices \cite{Frederico:1994dx}. In particular, there are no isospin
violating terms of the form ${\rm tr}\left(P\vslash P\vslash\right)$.

In the next step we will detail the calculation of the leading $\frac{1}{N_C}$ contribution from 
the polarized vacuum (Dirac sea) to the nucleon structure functions. The contribution of the 
distinct valence level, will later be added as for $g_a$ in Eqs.~(\ref{eq:ga5}) and~(\ref{eq:ga6}). 
For the NJL soliton model this valence quark contribution has been thoroughly discussed in 
Refs.~\cite{Weigel:1996kw,Weigel:1996jh}. In other models, like the MIT bag model~\cite{Chodos:1974je}, 
the calculation is quite similar \cite{Jaffe:1974nj}.

The above discussion and definition of the structure functions (form factors) in the 
hadron tensor was based on translational invariance. To apply it to a soliton configuration 
we need to restore translational invariance. This is accomplished by introducing a collective 
coordinate, $\vec{R}$, which describes the position of a soliton (nucleon) \cite{Gervais:1975yg} 
with its momentum $\vec{p}\,$ conjugate to this collective coordinate\footnote{This procedure 
is common to all soliton models when {\it e.g.} computing form factors \cite{Braaten:1986md}.}, 
{\it i.e.} $\langle \vec{R}|\vec{p}\,\rangle= \sqrt{2E}\,{\rm exp}\left(\imu\vec{R}\cdot\vec{p}\right)$.
Here $E=\sqrt{\vec{p}\,^2+M_N^2}$ denotes the nucleon energy. The Compton amplitude is then 
obtained by taking the pertinent matrix element and averaging over the position of the soliton,
\begin{align}
T_{\mu\nu}^{ab} &= 2\imu M_N
\int d^4\xi\, \int d^3R\, e^{\imu q\cdot \xi}\,
\Big\langle p,s\Big| T\left\{J_{\mu}^{a}(\xi-R)
J_{\nu}^{b \dagger}(-R)\right\}\Big| p,s \Big\rangle\cr
&= 2\imu  M_N
\int d^4\xi_1\, \int d^3\xi_2\, e^{\imu q\cdot(\xi_1-\xi_2)}
\Big\langle s\Big| T\left\{J_{\mu}^{a}(\xi_1)
J_{\nu}^{b \dagger}(\xi_2)\right\}\Big| s \Big\rangle\, .
\label{nuc1}
\end{align}
Here we have made use of the fact that the initial and final nucleon states not only have identical 
momenta but are actually considered in the rest frame. For simplicity we will treat $\xi_2$ and $R$ 
as four-vectors noting that their temporal components vanish, $\xi_2^0=R^0=0$. 
The spin-isospin matrix elements will be evaluated in the space of the collective coordinates $A$, 
which have been introduced in Eq.~(\ref{collq1}).

To see how Cutkosky's rule works in the soliton sector it instructive to briefly (and only formally) 
consider the leading $\frac{1}{N_C}$ contribution in the unregularized case
\begin{align}
T\left\{J_\mu(\xi_1)J_\nu(\xi_2)\right\}
&=\imu\frac{N_C}{2}\, {\rm Tr}\,
\Bigg\{\left(-\bDp\right)^{-1}\mathcal{Q}^2
\Big[\gamma_\mu\delta^4\left(\hat{x}-\xi_1\right)
\left(\dslash\right)^{-1}
\gamma_\nu\delta^4\left(\hat{x}-\xi_2\right) \cr
& \hspace{4cm}
\,+\gamma_\nu\delta^4\left(\hat{x}-\xi_2\right)
\left(\dslash\right)^{-1}
\gamma_\mu\delta^4\left(\hat{x}-\xi_1\right)\Big]\Bigg\}\, .
\label{nuc2}
\end{align}
Here $\hat{x}$ refers to the position operator. The above functional trace is computed by using 
a plane-wave basis for the operator in the square brackets while the matrix elements of $\bDp$ 
are evaluated employing the eigenfunctions $\Psi_\alpha$ of the Dirac Hamiltonian (\ref{hedgehog}):
\begin{align}
T_{\mu\nu}(q)&=
-M_NN_C\int\frac{d\omega}{2\pi} \sum_\alpha
\int d^4\xi_1\,\int d^3\xi_2\, \int \frac{d^4k}{(2\pi)^4}\,
{\rm e}^{\imu\xi_1^0(q^0+k^0)}\,
{\rm e}^{-\imu(\vec{\xi}_1-\vec{\xi}_2)\cdot(\vec{q}+\vec{k})}\,
\frac{1}{k^2+\imu\epsilon}\cr
& \hspace{3.0cm}\times
\frac{\omega+\epsilon_\alpha}
{\omega^2-\epsilon^2_\alpha+\imu\epsilon}
\Big\langle N \Big|
\Bigg\{\overline{\Psi}_\alpha(\vec{\xi}_1)\mathcal{Q}_A^2
\gamma_\mu \kslash\gamma_\nu\Psi_\alpha(\vec{\xi}_2)\,
{\rm e}^{\imu\xi_1^0\omega} \cr
& \hspace{5.5cm}
-\,\overline{\Psi}_\alpha(\vec{\xi}_2)\mathcal{Q}_A^2
\gamma_\nu \kslash\gamma_\nu\Psi_\alpha(\vec{\xi}_1)\,
{\rm e}^{-\imu\xi_1^0\omega}\Bigg\}\Big| N\Big\rangle
+\mathcal{O}\left(\frac{1}{N_C}\right)\, .\qquad
\label{nuc3}
\end{align}
The dependence on the collective coordinates is contained in $\mathcal{Q}_A=A^\dagger\mathcal{Q}A$.
We clearly recognize the two propagators, one in the massless plane wave basis and
the other in the soliton background. Cutkosky's rule produces respective $\delta$-functions 
$-2\pi\imu\delta(k^2)$ and $-2\pi\imu\delta(\omega^2-\epsilon^2_\alpha)$. We perform the 
frequency integral, write $t=\xi_1^0$ and employ the prescription from Eq.~(\ref{defsign})
so that the hadron tensor becomes
\begin{align}
W_{\mu\nu}&=
M_NN_C\sum_\alpha {\rm sign}(\epsilon_\alpha)
\int dt\, \int d^3\xi_1\,\int d^3\xi_2\,
\int \frac{d^4k}{(2\pi)^4}\, {\rm e}^{\imu t(q^0+k^0)}\,
{\rm e}^{-\imu(\vec{\xi}_1-\vec{\xi}_2)\cdot(\vec{q}+\vec{k})}\,
\delta\left(k^2\right)k^\rho \cr
& \hspace{0.5cm}\times\hspace{-0.1cm}
\Big\langle N \Big|S_{\mu\rho\nu\sigma}
\Bigg\{\overline{\Psi}_\alpha(\vec{\xi}_1)
\mathcal{Q}_A^2\gamma^\sigma\Psi_\alpha(\vec{\xi}_2)\,
{\rm e}^{\imu\epsilon_\alpha t}
-\overline{\Psi}_\alpha(\vec{\xi}_2)
\mathcal{Q}_A^2\gamma^\sigma\Psi_\alpha(\vec{\xi}_1)\,
{\rm e}^{-\imu\epsilon_\alpha t}\Bigg\}\cr
& \hspace{0.7cm}
-\imu\epsilon_{\mu\rho\nu\sigma}
\Bigg\{\overline{\Psi}_\alpha(\vec{\xi}_1)
\mathcal{Q}_A^2\gamma^\sigma\gamma_5\Psi_\alpha(\vec{\xi}_2)\,
{\rm e}^{\imu\epsilon_\alpha t}
+\overline{\Psi}_\alpha(\vec{\xi}_2)
\mathcal{Q}_A^2\gamma^\sigma\gamma_5\Psi_\alpha(\vec{\xi}_1)\,
{\rm e}^{-\imu\epsilon_\alpha t}\Bigg\}
\Big| N\Big\rangle +\mathcal{O}\left(\frac{1}{N_C}\right).\quad
\hspace{0.5cm}
\label{nuc5}
\end{align}
In the above we have four contributions, two for each the unpolarized 
($S_{\mu\rho\nu\sigma}$) and polarized ($\epsilon_{\mu\rho\nu\sigma}$) components.
One of the two components propagates from $\xi_1$ to $\xi_2$ and the other 
in the opposite direction. Typically they are denoted particle and anti-particle
distributions. Note, however, that in the present case $\epsilon_\alpha$ may have
either sign so that both particle and anti-particles spinors contribute in all 
terms.

In deriving Eq.~(\ref{nuc5}) only the pole from $\omega=+\epsilon_\alpha$ contributed. 
That will be different when regularization is accounted for. We display the result 
without further derivation as the calculation for the fully regularized scenario goes 
along the same lines as above
\begin{align}
T_{\mu\nu}(q)&=-M_N\frac{N_C}{2}\int \frac{d\omega}{2\pi}
\sum_\alpha \int dt\, \int d^3\xi_1\,\int d^3\xi_2\,
\int \frac{d^4k}{(2\pi)^4}\,
{\rm e}^{\imu(q_0+k_0)t}\,
{\rm e}^{-\imu({\vec q}+{\vec k})\cdot({\vec \xi}_1-{\vec \xi}_2)}\,
\frac{1}{k^2+\imu\epsilon}\cr
& \hspace{-1cm}\times \Big\langle N\Big|
\Bigg\{\left[{\rm e}^{\imu\omega t}
\Psi_\alpha^\dagger({\vec \xi}_1)\beta \mathcal{Q}_A^2\gamma_\mu\kslash
\gamma_\nu \Psi_\alpha({\vec \xi}_2)
-{\rm e}^{-\imu\omega t}
\Psi_\alpha^\dagger({\vec \xi}_2)\beta \mathcal{Q}_A^2\gamma_\nu\kslash
\gamma_\mu \Psi_\alpha({\vec \xi}_1)\right]f_\alpha^+(\omega)
\label{treg} \\ &\hspace{-0.6cm}
+\left[{\rm e}^{\imu\omega t}
\Psi_\alpha^\dagger({\vec \xi}_1)\mathcal{Q}_A^2
(\gamma_\mu\kslash\gamma_\nu)_5\beta
\Psi_\alpha({\vec \xi}_2)-{\rm e}^{-\imu\omega t}
\Psi_\alpha^\dagger({\vec \xi}_2)\mathcal{Q}_A^2
(\gamma_\nu\kslash\gamma_\mu)_5\beta
\Psi_\alpha({\vec \xi}_1)\right]f_\alpha^-(\omega)
\Bigg\}\Big| N\Big\rangle +\mathcal{O}\left(\frac{1}{N_C}\right),
\nonumber
\end{align}
with the spectral functions
\begin{align}
f_\alpha^\pm(\omega)=\sum_{i=0}^2 c_i \frac{\omega\pm\epsilon_\alpha}
{\omega^2-\epsilon_\alpha^2-\Lambda_i^2+\imu\epsilon}
\pm\frac{\omega\pm\epsilon_\alpha}
{\omega^2-\epsilon_\alpha^2+\imu\epsilon}\, .
\label{sfct}
\end{align}
The first term in these spectral functions arises from the regularized real
part, and the second from the unregularized imaginary part. Without regularization
$f_\alpha^+(\omega)\sim\frac{2(\omega+\epsilon_\alpha)}
{\omega^2-\epsilon_\alpha^2+i\epsilon}$ and $f_\alpha^-(\omega)\sim0$ so that 
Eq.~(\ref{nuc3}) would be recovered.

Before applying Cutkosky's rule we integrate over the time variable which is 
distinct from the spatial coordinates because the soliton is static. This integral
yields $2\pi\delta(q_0+k_0\pm\omega)$ which we subsequently use to integrate $k_0$. 
Then the $\delta$-function for the absorptive part of the Compton amplitude is
$\delta((q_0\pm\omega)^2-|\vec{k}|^2)$. To perform the spatial integrals we define 
the Fourier transform of the single particle wave-functions\footnote{The 
single particle wave-functions are parity eigenfunctions so that spatial reflections
can be compensated by factors of $\beta$.} as 
\begin{equation}
\widetilde{\Psi}_\alpha(\vec{p})=\int \frac{d^3\xi}{4\pi}\,
\Psi_\alpha(\vec{\xi})\, {\rm e}^{\imu\vec{\xi}\cdot\vec{p}}
\label{ftrans}
\end{equation}
and get
\begin{align}
W_{\mu\nu}(q)&=\imu M_N\frac{N_C}{\pi}\int \frac{d\omega}{2\pi}
\sum_\alpha \int d^3k\cr
&\hspace{1cm}\times
 \Big\langle N\Big|\Bigg\{\Big[
\widetilde{\Psi}_\alpha^\dagger(\vec{q}+\vec{k})\mathcal{Q}_A^2
\beta\gamma_\mu\kslash\gamma_\nu
\widetilde{\Psi}_\alpha(\vec{q}+\vec{k})\delta(|\vec{k}|^2-(q_0+\omega)^2)\cr
& \hspace{2cm}
-\, \widetilde{\Psi}_\alpha^\dagger(\vec{q}+\vec{k})\mathcal{Q}_A^2
\gamma_\nu\kslash\gamma_\mu\beta
\widetilde{\Psi}_\alpha(\vec{q}+\vec{k})\delta(|\vec{k}|^2-(q_0-\omega)^2)\Big]
f_\alpha^+(\omega)\Big|_{\rm p}\cr
& \hspace{1.5cm}
+\,\Big[\widetilde{\Psi}_\alpha^\dagger(\vec{q}+\vec{k})\mathcal{Q}_A^2
(\gamma_\mu\kslash\gamma_\nu)_5\beta
\widetilde{\Psi}_\alpha(\vec{q}+\vec{k})\delta(|\vec{k}|^2-(q_0+\omega)^2)\cr
& \hspace{2cm}
-\, \widetilde{\Psi}_\alpha^\dagger(\vec{q}+\vec{k})\mathcal{Q}_A^2
\beta(\gamma_\nu\kslash\gamma_\mu)_5
\widetilde{\Psi}_\alpha(\vec{q}+\vec{k})\delta(|\vec{k}|^2-(q_0-\omega)^2)
\Big]f_\alpha^-(\omega)\Big|_{\rm p}\Bigg\}\Big| N\Big\rangle\, ,\qquad
\label{eq:wten0}
\end{align}
where, again, we only wrote the leading $\frac{1}{N_C}$ term. An example for the pole extraction is
\begin{align}
\left( \sum_{i=0}^2c_i \frac{1}{\omega^2-\epsilon_\alpha^2
-\Lambda_i^2+\imu\epsilon} \right)_{\rm p} & =  \sum_{i=0}^2 c_i
\frac{-\imu \pi}{ \omega_\alpha } \left[ \delta \left( \omega + \omega_\alpha \right)
+ \delta \left( \omega -\omega_\alpha \right) \right]\,,
\quad {\rm with}\quad 
\omega_\alpha = \sqrt{\epsilon_\alpha^2 + \Lambda_i^2}\,.
\label{eq:pol1} \end{align}
To get an expression that looks like a bilocal and bilinear distribution function we 
shift the integration variable $\vec{p}=\vec{q}+\vec{k}$ and recognize that the 
single particle wave-functions will have support only for small $\vec{p}$, as compared
to the large momenta in $\vec{q}$. This allows us to replace $\kslash$ by $-\qslash$ in 
the Bjorken limit (recall that $k_0=-q_0\mp\omega$ from the $t$ integral) for the 
Dirac matrices sandwiched between the spinors. Furthermore
\begin{align*}
|\vec{k}|^2-(q_0\pm\omega)^2&=|\vec{p}-\vec{q}|^2-(q_0\pm\omega)^2
=\vec{p}^2-2\vec{p}\cdot\hat{n}|\vec{q}|+|\vec{q}|^2-(q_0\pm\omega)^2\cr
&\bjlim 
-2|\vec{q}|\left[\vec{p}\cdot\hat{n}-(M_Nx\mp\omega)\right]\,.
\end{align*}
Here $\hat{n}$ is the unit vector in the direction of the spatial photon 
momentum $\vec{q}$. Then
\begin{align}
W_{\mu\nu}(q)&=\imu M_N\frac{N_C}{2\pi}\int \frac{d\omega}{2\pi}
\sum_\alpha \int d^3p\cr
&\hspace{1cm}\times
 \Big\langle N\Big|\Bigg\{\Big[
\widetilde{\Psi}_\alpha^\dagger(\vec{p})\mathcal{Q}_A^2
\beta\gamma_\mu\nslash\gamma_\nu
\widetilde{\Psi}_\alpha(\vec{p})\delta(\vec{p}\cdot\hat{n}-(M_Nx-\omega))\cr
& \hspace{2cm}
-\, \widetilde{\Psi}_\alpha^\dagger(\vec{p})\mathcal{Q}_A^2
\gamma_\nu\nslash\gamma_\mu\beta
\widetilde{\Psi}_\alpha(\vec{p})\delta(\vec{p}\cdot\hat{n}-(M_Nx+\omega))\Big]
f_\alpha^+(\omega)\Big|_{\rm p}\cr
& \hspace{1.5cm}
+\,\Big[\widetilde{\Psi}_\alpha^\dagger(\vec{p})\mathcal{Q}_A^2
(\gamma_\mu\nslash\gamma_\nu)_5\beta
\widetilde{\Psi}_\alpha(\vec{p})\delta(\vec{p}\cdot\hat{n}-(M_Nx-\omega))\cr
& \hspace{2cm}
-\, \widetilde{\Psi}_\alpha^\dagger(\vec{p})\mathcal{Q}_A^2
\beta(\gamma_\nu\nslash\gamma_\mu)_5
\widetilde{\Psi}_\alpha(\vec{p})\delta(\vec{p}\cdot\hat{n}-(M_Nx+\omega))
\Big]f_\alpha^-(\omega)\Big|_{\rm p}\Bigg\}\Big| N\Big\rangle\, ,\qquad
\label{eq:wten1}
\end{align}
where $n^\mu=(1,\hat{n})^\mu$ is a light-like vector. Eq.~(\ref{eq:wten1}) is well 
suited for our numerical simulations in Section~\ref{sec:NR}, in particular when treating 
the $\delta$-functions by averaging the directions of $\hat{n}$ \cite{Diakonov:1996sr}. 
However, the similarity with distribution functions is more apparent when returning to 
coordinate space and writing the $\delta$-functions as integrals of exponential functions
\begin{align}
W_{\mu\nu}^{(s)}(q)&=\imu M_N\frac{N_C}{4}\int \frac{d\omega}{2\pi}
\sum_\alpha \int d^3\xi \int \frac{d\lambda}{2\pi}\,{\rm e}^{\imu M_nx\lambda}\cr
& \hspace{1cm} \times
 \Big\langle N\Big|\Bigg\{
\Big[\overline\Psi_\alpha(\vec{\xi})\mathcal{Q}_A^2
\gamma_\mu\nslash\gamma_\nu\Psi_\alpha(\vec{\xi}+\lambda\hat{n}){\rm e}^{-\imu\lambda\omega}\cr
&\hspace{4cm}
-\overline\Psi_\alpha(\vec{\xi})\mathcal{Q}_A^2
\gamma_\nu\nslash\gamma_\mu\Psi_\alpha(\vec{\xi}-\lambda\hat{n}){\rm e}^{\imu\lambda\omega}
\Big] f_\alpha^+(\omega)\Big|_{\rm p}\cr
& \hspace{2cm}
+\Big[\overline\Psi_\alpha(\vec{\xi})\mathcal{Q}_A^2
(\gamma_\mu\nslash\gamma_\nu)_5\Psi_\alpha(\vec{\xi}-\lambda\hat{n}){\rm e}^{-\imu\lambda\omega}\cr
&\hspace{4cm}
-\overline\Psi_\alpha(\vec{\xi})\mathcal{Q}_A^2
(\gamma_\nu\nslash\gamma_\mu)_5\Psi_\alpha(\vec{\xi}+\lambda\hat{n}){\rm e}^{\imu\lambda\omega}
\Big] f_\alpha^-(\omega)\Big|_{\rm p}
\Bigg\}\Big| N\Big\rangle\, ,\qquad
\label{eq:wten2}
\end{align}
where we have added the superscript on $W_{\mu\nu}^{(s)}(q)$ to clarify that Eq.~(\ref{eq:wten2}) 
represents the vacuum (Dirac sea) component only. The valence component is most conveniently 
obtained by restricting the sum to $\alpha={\rm v}$ and omitting regularization 
\begin{align}
W_{\mu\nu}^{(v)}(q)&=\imu\left[1+{\rm sign}(\epsilon_{\rm v})\right] 
M_N\frac{N_C}{4}\int d^3\xi \int \frac{d\lambda}{2\pi}\,{\rm e}^{\imu M_nx\lambda}
 \Big\langle N\Big|\Bigg\{
\Big[\overline\Psi_{\rm v}(\vec{\xi})\mathcal{Q}_A^2
\gamma_\nu\nslash\gamma_\mu\Psi_{\rm v}(\vec{\xi}-\lambda\hat{n}){\rm e}^{\imu\lambda\epsilon_{\rm v}}
\cr &\hspace{6.0cm}
-\overline\Psi_{\rm v}(\vec{\xi})\mathcal{Q}_A^2
\gamma_\mu\nslash\gamma_\nu\Psi_{\rm v}(\vec{\xi}+\lambda\hat{n}){\rm e}^{-\imu\lambda\epsilon_{\rm v}}
\Big]\Bigg\}\Big| N\Big\rangle\,.
\label{eq:wten2v}
\end{align}

Eqs.~(\ref{eq:wten2}) and~(\ref{eq:wten2v}) indeed have the form of bilocal and bilinear quark 
distributions. However these are the distributions for the quarks in the chiral model interacting
self-consistently with the soliton. So far, no connection to distributions in QCD has been 
incorporated; our calculation is solely based on the electromagnetic interaction within the 
chiral model.  Several features needed consideration to arrive at an expression of 
the form of distributions. Most importantly and, of course, not surprisingly the Bjorken limit
was implemented. In addition the one of the two propagators that occur in the Compton amplitude
is taken to be that of a free massless fermion, while the other contains all information about
the soliton that resembles the nucleon. Again, this separation is an indirect consequence of
the Bjorken limit. Furthermore we made use of the fact that the (momentum space) quark wave-functions
only have support at momenta that are tiny compared to the momentum of the exchanged virtual
photon. Finally we stress that the appearance of single distribution functions in Eq.~(\ref{eq:wten2}) 
is kind of deceptive as the spectral functions $f_\alpha^{(\pm)}(\omega)$ pick up more than 
a single pole.

In Section \ref{sec:soliton} we have computed that axial charge, $g_a$, of the nucleon. It is the prime 
example to see how sum rules work in the presence of regularization. The Bjorken sum rule \cite{Bjorken:1966jh} 
relates that charge to the $x$-integral of the isovector combination of longitudinal polarized nucleon 
structure functions $g_1(x)$ for proton and neutron. These functions are obtained from the anti-symmetric 
component of the hadron tensor
\begin{align}
W_{\mu\nu}^{(s,\rm A)}&=
-M_N\frac{N_C}{2}\epsilon_{\mu\rho\nu\sigma}{\rm n}^\rho
\int \frac{d\omega}{2\pi} \sum_\alpha \int d^3\xi
\int \frac{d\lambda}{2\pi}\, {\rm e}^{\imu M_Nx\lambda}
\left(\sum_{i=0}^2c_i\frac{\omega+\epsilon_\alpha}
{\omega^2-\epsilon_\alpha^2-\Lambda_i^2+\imu\epsilon}\right)_{\rm p}\cr
&\hspace{0.5cm}\times \Big\langle N\Big|
\overline{\Psi}_\alpha(\vec{\xi})\mathcal{Q}^2_A\gamma^\sigma\gamma_5
\Psi_\alpha(\xipl){\rm e}^{-\imu\omega\lambda}
+\overline{\Psi}_\alpha(\vec{\xi})\mathcal{Q}^2_A\gamma^\sigma\gamma_5
\Psi_\alpha(\ximl)
{\rm e}^{\imu\omega\lambda}\Big| N\Big\rangle\, .
\label{wmna}
\end{align}
The spectral function is fully regularized because it originates from 
$$
f_\alpha^+(\omega)-f_\alpha^-(-\omega)
=\sum_{i=0}^2c_i\frac{\omega+\epsilon_\alpha-(-\omega-\epsilon_\alpha)}
{\omega^2-\epsilon_\alpha^2-\Lambda_i^2+\imu\epsilon}
+\frac{\omega+\epsilon_\alpha+(-\omega-\epsilon_\alpha)}
{\omega^2-\epsilon_\alpha^2+\imu\epsilon}
=2\sum_{i=0}^2c_i\frac{\omega+\epsilon_\alpha}
{\omega^2-\epsilon_\alpha^2-\Lambda_i^2+\imu\epsilon}\,.
$$
Here the prescription from Eq.~(\ref{defsign}) has had a major impact. Without this specification 
the relative sign between the spectral functions  would have been positive resulting 
in the spectral function $(\omega+\epsilon_\alpha)/ (\omega^2-\epsilon_\alpha^2+\imu\epsilon)$.
In that case $W_{\mu\nu}^{(s,\rm A)}$ would have to be associated with unregularized
imaginary part of the action which is not compatible with the sum rules. The reason is 
that the leading order $\left({\rm in~}\frac{1}{N_C}\right)$ contribution to the axial 
charges stems from the regularized real part of the action.

Taking $\hat{n}=\hat{e}_3$ and the projection operator given in Tab.~\ref{tab_1} we find for 
the Dirac sea component of the longitudinal polarized structure function
\begin{align}
g^{(s)}_1(x)&=-\imu\frac{M_NN_C}{36}
\Big\langle N\Big| I_3 \Big| N\Big\rangle
\int \frac{d\omega}{2\pi} \sum_\alpha \int d^3\xi
\int \frac{d\lambda}{2\pi}\, {\rm e}^{iM_Nx\lambda}
\left(\sum_{i=0}^2c_i\frac{\omega+\epsilon_\alpha}
{\omega^2-\epsilon_\alpha^2-\Lambda_i^2+i\epsilon}\right)_{\rm p}\cr
& \hspace{0.5cm}\times
\left[\Psi^\dagger_\alpha(\vec{\xi})\tau_3
\left(1-\alpha_3\right)\gamma_5
\Psi_\alpha(\xi+\lambda\hat{e}_3)
{\rm e}^{-\imu\omega\lambda}
+\Psi^\dagger_\alpha(\vec{\xi})\tau_3
\left(1-\alpha_3\right)\gamma_5
\Psi_\alpha(\xi-\lambda\hat{e}_3)
{\rm e}^{\imu\omega\lambda}\right]\, ,
\label{g1x}
\end{align}
where we have substituted the matrix element (\ref{collq4}) of the collective coordinates, $A$, 
sandwiched between nucleon states. To establish a sum rule we first note that $0\le x<\infty$. The 
upper bound is not unity because the soliton breaks translational invariance. Eventually that will 
be accounted for by boosting the soliton to the infinite momentum frame \cite{Gamberg:1997qk}, as 
will be discussed in Subsection~\ref{ssec:boost}. Furthermore the two terms in Eq.~(\ref{g1x}) 
are related by $\lambda\,\leftrightarrow\,-\lambda$ which allows us to integrate only one of them 
but over $-\infty<x<\infty$ thereby producing $\frac{2\pi}{M_N}\delta(\lambda)$. From parity 
conservation we have $\int d^3\xi\, \Psi^\dagger_\alpha(\vec{\xi})\tau_3\gamma_5\Psi_\alpha(\xi)=0$
and the poles are straightforwardly extracted as
\begin{align}
\left(\frac{\omega+\epsilon_\alpha}
{\omega^2-\epsilon_\alpha^2-\Lambda^2+\imu\epsilon}\right)_{\rm p}
&=-\frac{\imu\pi \epsilon_\alpha}{\sqrt{\epsilon_\alpha^2+\Lambda^2}}\,
\left[\delta\left(\omega+\sqrt{\epsilon_\alpha^2+\Lambda^2}\right)
+\delta\left(\omega-\sqrt{\epsilon_\alpha^2+\Lambda^2}\right)\right]\cr
&\hspace{1cm}-\imu\pi\left[\delta\left(\omega+\sqrt{\epsilon_\alpha^2+\Lambda^2}\right)
-\delta\left(\omega-\sqrt{\epsilon_\alpha^2+\Lambda^2}\right)\right]\, .\qquad
\label{poles}
\end{align}
Because of $\delta(\lambda)$ there is no other dependence on $\omega$ in Eq.~(\ref{g1x}) and thus
the second square bracket in Eq.~(\ref{poles}) vanishes when integrating 
$\int_{-\infty}^\infty dx\, g^{(s)}_1(x)$. Therefore the vacuum contribution to the Bjorken sum rule 
($p$ and $n$ are proton and neutron, respectively)
\begin{equation}
\int_0^\infty dx \left(g_1^{(s,\rm p)}(x)-g_1^{(s,\rm n)}(x)\right)
=\frac{1}{6}g^{(s)}_{\rm A}
\label{bjsum}
\end{equation}
is immediately verified from Eq.~(\ref{eq:ga4}) after taking care of the isospin matrix elements 
of the nucleon. Adding the valence level component to this sum rule is a trivial simplification of 
the calculation leading to Eq.~(\ref{bjsum}).

The above example for the verification of a sum rule is (almost) general. The symmetries under 
$\lambda\,\leftrightarrow\,-\lambda$ extend the $x$ integral over whole real axis, not only the positive 
half-line. That integral then produces $\delta(\lambda)$ which turns the bilocal matrix elements into 
the expectation values that occur in the expressions for the static properties that occur in the 
particular sum rule. Then the sum rule is verified level by level, {\it i.e.} separately for each 
term in $\sum_{\alpha}$. The one exception is the momentum sum rule which involves the isoscalar 
component of the unpolarized structure function $f_1(x)$. When adapting the calculation of the 
Bjorken sum rule to the unpolarized structure function $f_1(x)$, the integral $\int dx\, x f_1(x)$ 
produces the fermion part of the classical soliton energy in Eq.~(\ref{etot}). However, there is 
an additional contribution proportional to\footnote{The factor $x$ under the integral is written 
as a derivative with respect to $\lambda$. Integrating by parts and averaging over angles turns this 
into the expectation value of $\vec{\alpha}\cdot\vec{\partial}$.}
$$
\left[1+{\rm sign}(\epsilon_{\rm v})\right]
\int d^3\xi\,\Psi^\dagger_{\rm v}(\vec{\xi})\vec{\alpha}\cdot\vec{\partial}\Psi_{\rm v}(\vec{\xi})
-\sum_{i=0}^2c_i\sum_\alpha \frac{\epsilon_\alpha}{\sqrt{\epsilon_\alpha^2+\Lambda_i^2}}
\int d^3\xi\,\Psi^\dagger_\alpha(\vec{\xi})\vec{\alpha}\cdot\vec{\partial}\Psi_\alpha(\vec{\xi})
$$
and the sum rule is only verified when this piece vanishes. One shows that this is indeed the case
by recognizing that 
$$
\vec{\alpha}\cdot\vec{\partial}\propto \left[\vec{\xi}\cdot\vec{\partial},h\right]
-m\left(\vec{\xi}\cdot\vec{\partial}U_5(\xi)\right)
$$
so that the matrix elements in that un-wanted contribution are those of the dilatation operator acting 
on the soliton. In turn the above sum is the change in energy obtained when squeezing or stretching the 
soliton infinitesimally. As the soliton minimizes the energy, this change must indeed be 
zero \cite{Weigel:1999pc,Diakonov:1996sr}\footnote{There is a (numerically negligible) complication
due to chiral symmetry breaking: for $m_\pi\ne0$ the dilatation term in the sum over the 
quark levels is not exactly zero but compensates for the local integral in Eq.~(\ref{etot}). Numerically 
more concerning is the fact that the nucleon mass has $\mathcal{O}\left(\frac{1}{N_C}\right)$ corrections, 
Eq.~(\ref{eq:rotor}), which are not contained in this structure function. We also note that 
the sum rule actually yields $\frac{E_{\rm tot}}{M_N}-1$, which does not vanish as we have defined
the hadron tensor to contain the physical mass parameter. Nevertheless this sum rule is perfectly 
suited to test the numerical simulation.}. We must thus keep in mind that the momentum
sum rule only works when summing all levels.

For completeness (and an attempt to frighten the reader) we display the Bjorken limit of the hadron tensor 
including the next to leading order term for the expansion in $\frac{1}{N_C}$,
\begin{align}
W_{\mu\nu}^{(s)}&\bjlim\imu M_N\frac{N_C}{4}
\int \frac{d\omega}{2\pi}\sum_\alpha \int d^3 \xi
\int \frac{d\lambda}{2\pi}\, {\rm e}^{\imu M_Nx\lambda}\,
\Big\langle N\Big|
\label{eq:wten3} \\
&\hspace{-2mm}\times
\Bigg\{\Big[\overline{\Psi}_\alpha({\vec\xi})\mathcal{Q}_A^2
\gamma_\mu\nslash\gamma_\nu\Psi_\alpha(\xipl)
{\rm e}^{-\imu\lambda\omega}
-\overline{\Psi}_\alpha({\vec\xi})\mathcal{Q}_A^2
\gamma_\nu\nslash\gamma_\mu
\Psi_\alpha(\ximl)
{\rm e}^{\imu\lambda\omega}\Big]
f_\alpha^+(\omega)\Big|_{\rm p}\cr
& +\Big[\overline{\Psi}_\alpha({\vec\xi})\mathcal{Q}_A^2
(\gamma_\mu\nslash\gamma_\nu)_5
\Psi_\alpha(\ximl)
{\rm e}^{-\imu\lambda\omega}
-\overline{\Psi}_\alpha({\vec\xi})\mathcal{Q}_A^2
(\gamma_\nu\nslash\gamma_\mu)_5
\Psi_\alpha(\xipl)
{\rm e}^{\imu\lambda\omega}\Big]
f_\alpha^-(\omega)\Big|_{\rm p}\cr
&+\frac{\imu\lambda}{4}\Big[\overline{\Psi}_\alpha({\vec\xi})
\tauom \mathcal{Q}_A^2 \gamma_\mu\nslash\gamma_\nu
\Psi_\alpha(\xipl) {\rm e}^{-\imu\lambda\omega}
+\overline{\Psi}_\alpha({\vec\xi})
\mathcal{Q}_A^2\tauom\gamma_\nu\nslash\gamma_\mu
\Psi_\alpha(\ximl) {\rm e}^{\imu\lambda\omega}\Big]
f_\alpha^+(\omega)\Big|_{\rm p}\cr
&+\frac{\imu\lambda}{4}\Big[\overline{\Psi}_\alpha({\vec\xi})
\tauom \mathcal{Q}_A^2(\gamma_\mu\nslash\gamma_\nu)_5
\Psi_\alpha(\ximl) {\rm e}^{-\imu\lambda\omega}
+\overline{\Psi}_\alpha({\vec\xi})\mathcal{Q}_A^2\tauom
(\gamma_\nu\nslash\gamma_\mu)_5
\Psi_\alpha(\xipl)
{\rm e}^{\imu\lambda\omega}\Big]
f_\alpha^-(\omega)\Big|_{\rm p}\cr
&+\sum_\beta\langle\alpha|\tauom|\beta\rangle 
\Bigg(\Big[\overline{\Psi}_\beta({\vec\xi})\mathcal{Q}_A^2
\gamma_\mu\nslash\gamma_\nu\Psi_\alpha(\xipl)
{\rm e}^{-\imu\lambda\omega}
-\overline{\Psi}_\beta({\vec\xi})\mathcal{Q}_A^2
\gamma_\nu\nslash\gamma_\mu
\Psi_\alpha(\ximl)
{\rm e}^{\imu\lambda\omega}\Big]
g_{\alpha\beta}^+(\omega)\Big|_{\rm p}\cr
&+\Big[\overline{\Psi}_\beta({\vec\xi})\mathcal{Q}_A^2
(\gamma_\mu\nslash\gamma_\nu)_5
\Psi_\alpha(\ximl)
{\rm e}^{-\imu\lambda\omega}
-\overline{\Psi}_\beta({\vec\xi})\mathcal{Q}_A^2
(\gamma_\nu\nslash\gamma_\mu)_5
\Psi_\alpha(\xipl)
{\rm e}^{\imu\lambda\omega}\Big]
g_{\alpha\beta}^-(\omega)\Big|_{\rm p}
\Bigg)\Bigg\}\Big| N \Big\rangle\, ,
\nonumber
\end{align}
with the spectral functions
\begin{equation}
g_{\alpha\beta}^\pm(\omega)=\sum_{i=0}^2c_i
\frac{(\omega\pm\epsilon_\alpha)
(\omega\pm\epsilon_\beta)+\Lambda_i^2}
{(\omega^2-\epsilon_\alpha^2-\Lambda_i^2+\imu\epsilon)
(\omega^2-\epsilon_\beta^2-\Lambda_i^2+\imu\epsilon)}
\pm
\frac{(\omega\pm\epsilon_\alpha)
(\omega\pm\epsilon_\beta)}
{(\omega^2-\epsilon_\alpha^2+\imu\epsilon)
(\omega^2-\epsilon_\beta^2+\imu\epsilon)}\,.
\label{blsfct}
\end{equation}
The subleading $\frac{1}{N_C}$ terms contain the angular velocity, Eq.~(\ref{collq2}).
They arise from the expansions (after transforming 
$|\omega,\beta\rangle\,\to\,A|\omega,\beta\rangle$)
\begin{equation}
\langle\omega,\alpha|\left(\bDp\right)^{-1}|\omega,\beta\rangle
=\frac{\delta_{\alpha\beta}}{\omega-\epsilon_\alpha}
+\frac{1}{\omega-\epsilon_\alpha}\,
\langle \alpha|\frac{1}{2}\tauom|\beta\rangle\,
\frac{1}{\omega-\epsilon_\beta} +
{\cal O}\left(\vec{\Omega}\,^2\right)
\label{nuc4}
\end{equation}
and
\begin{equation}
\langle t, \vec{\xi}| A(\hat{t}) | \omega,\alpha\rangle
= A(t)\, {\rm e}^{-\imu\omega t} \Psi_\alpha(\vec{\xi})
=A(0)\left[1+\frac{\imu t}{2}\,\tauom\right]
{\rm e}^{-\imu\omega t} \Psi_\alpha(\vec{\xi})+
{\cal O}\left(\vec{\Omega}\,^2\right)\, .
\label{nuc4a}
\end{equation}
The explicit appearance of the time variable is treated in the context of the Fourier transform 
$t{\rm e}^{\imu q_0t}=-\imu \frac{\partial}{\partial q_0}{\rm e}^{\imu q_0t}$ while
(in the nucleon rest frame) $x=-\frac{-q_0^2+\vec{q}}{2M_Nq_0}$ allows us to write 
$\frac{\partial}{\partial q_0}=\frac{\partial x}{\partial q_0}\frac{\partial}{\partial x}$
with
\begin{align*}
\frac{\partial x}{\partial q_0}=-\frac{1}{M_N}-\frac{q_0x}{q_0^2}\bjlim-\frac{1}{M_N}\,.
\end{align*}
This clarifies that the factors of $\imu\lambda$ in Eq.~(\ref{eq:wten3}) originated from the explicit 
appearance of the time variable via the derivative with respect to the Bjorken variable $x$.

Again, Eq.~(\ref{eq:wten3}) is the vacuum contribution. The valence part is most 
easily obtained by substituting the cranked valence level wave-function
\begin{align}
\Psi_{\rm v}^{(\rm rot)} (\vec{r},t) & = \Biggl\{ \Psi_{\rm v}(\vec{r}) 
+\frac{1}{2} \sum_{\alpha \neq \rm v}
\Psi_\alpha (\vec{r}) \frac{ \langle \alpha | \vec{\tau} \cdot \vec{\Omega}
| \rm v \rangle }{ \epsilon_{\rm v}-\epsilon_\alpha}\Biggr\}
\label{cranked_valence_level}
\end{align}
into Eq.~(\ref{eq:wten2v}) and taking care of the bilocal dependence on time as in Eq.~(\ref{nuc4a}).

In this chapter we have reviewed the formal derivation of the hadron tensor for a chiral quark soliton 
model starting from the electromagnetic coupling before bosonization and making ample use of the 
Bjorken limit.  We have detailed the case of the longitudinal polarized structure function to illuminate 
the calculational principle and verify the relevant sum rule. Detailed formulas for other structure 
functions are derived and presented in Refs.~\cite{Weigel:1999pc,Takyi:2019ahv}.

\section{Numerical results}
\label{sec:NR}
The results discussed in this section are mostly taken from Ref.~\cite{Takyi:2019ahv}. There are
several steps until we can perform a sensible comparison with experimental data. First we numerically
simulate the analytic results from the previous section. This produces structure functions that we call 
rest frame (RF) structure functions. We will present the results for the RF structure functions in the 
following two subsections. These structure functions have the unwanted feature that they do not vanish for
$|x|>1$.  We will therefore briefly describe a formalism \cite{Gamberg:1997qk} to boost the soliton to the 
infinite momentum frame (IMF). In the IMF the structure functions indeed vanish for $|x|>1$. That formalism 
is essentially adapted from a similar study \cite{Jaffe:1980qx} of the MIT bag model in $D=1+1$. This 
adoption is made possible as in the Bjorken limit it suffices to restore Lorentz invariance in direction 
of the (large) photon momentum only. Once support of the structure functions is confined to $|x|<1$ we 
can apply the DGLAP evolution formalism and compare with available data in the last subsection. We will 
only apply a first order evolution as a proof of concept; after all, the model is not constructed for 
high precision predictions.

We obtain the RF structure functions from the momentum space representation of Eq.~(\ref{eq:wten1}) and 
the momentum space analog of Eq.~(\ref{eq:wten3}). This momentum space computation is the most costly part 
of the simulation because we explicitly perform the Fourier transformation, Eq.~(\ref{ftrans}), on the 
eigen-spinors of the self-consistent soliton. Large momenta on a dense grid are needed to maintain the 
normalization of the spinors (and thus the sum rules). A typical simulation takes several CPU days/weeks 
on a standard desktop PC. In related work~\cite{Diakonov:1996sr,Wakamatsu:1997en} the expansion coefficients 
defined after Eq.~(\ref{diagh}) were directly used. Since they are discrete, some smoothing procedure was
needed in that treatment of the momentum space wave-functions.

We will refrain from presenting lengthy formulas as, {\it e.g.} the extremely bulky expressions involving the 
Fourier transforms of the radial functions in $\Psi_\alpha$ \cite{Takyi:2019ahv}. Rather we focus on explaining the 
treatment of the pole terms without going into too much details. This treatment is non-trivial and interferes with 
regularization, the central topic of this review and therefore deserves closer consideration. Below we therefore 
describe some key ingredients that are relevant for all our calculations.

As in Ref.~\cite{Diakonov:1996sr} we treat the Dirac $\delta$-functions in Eq.~(\ref{eq:wten1}) by 
averaging over $\hat{n}$, that is, we replace these $\delta$-functions by
\begin{equation}
\frac{1}{4\pi}\int d\Omega_{\hat{n}}\delta\left(E+\vec{p}\cdot\vec{n}\right)
=\frac{1}{2|\vec{p}|}\theta\left(|\vec{p}|-|E|\right)
\label{eq:average}
\end{equation}
and generalizations thereof that contain additional factors of $\hat{n}$ under the solid angle integral. We 
then need to evaluate expressions like (in sums over single particle levels $\alpha$, but 
omitting that index)
\begin{equation}
\int \frac{d\omega}{2\pi}\int \frac{d\lambda}{2\pi}\sum_{i=0}^2c_i\left(\frac{\omega+\epsilon}
{\omega^2-\epsilon^2-\Lambda_i^2+i\epsilon}\right)_{\rm p}
\int d^3p\,\widetilde{\Psi}^\dagger(\vec{p})\widetilde{\Psi}(\vec{p})
{\rm e}^{\imu(M_Nx-\hat{n}\cdot\vec{p})\lambda}{\rm e}^{\imu\omega\lambda}\,.
\label{eq:p1}
\end{equation}
Defining $\omega_i=\sqrt{\epsilon^2+\Lambda_i^2}$ we have (see also Eq.~(\ref{eq:pol1}))
\begin{equation}
\int \frac{d\omega}{2\pi}\sum_{i=0}^2c_i\left(\frac{\omega+\epsilon}
{\omega^2-\epsilon^2-\Lambda_i^2+i\epsilon}\right)_{\rm p}{\rm e}^{\imu\omega\lambda}
=-\frac{\imu}{2}\sum_{i=0}^2\frac{c_i}{|\omega_i|}
\left[\left(\epsilon+\omega_i\right){\rm e}^{\imu\omega_i\lambda}
+\left(\epsilon-\omega_i\right){\rm e}^{-\imu\omega_i\lambda}\right]
\label{eq:p2}
\end{equation}
and therefore
\begin{align}
&\int\frac{d\lambda}{2\pi} {\rm e}^{\imu(M_Nx-\hat{n}\cdot\vec{p})\lambda}
\int \frac{d\omega}{2\pi}\sum_{i=0}^2c_i\left(\frac{\omega+\epsilon}
{\omega^2-\epsilon^2-\Lambda_i^2+i\epsilon}\right)_{\rm p}
{\rm e}^{\imu\omega\lambda}\cr
&\hspace{2cm}
=-\frac{\imu}{2}\sum_{i=0}^2\frac{c_i}{|\omega_i|}
\left[\left(\epsilon+\omega_i\right)\delta(M_Nx-\hat{n}\cdot\vec{p}+\omega_i)
+\left(\epsilon+\omega_i\right)\delta(M_Nx-\hat{n}\cdot\vec{p}-\omega_i)\right]\cr
&\quad \longrightarrow\quad
-\frac{\imu}{4|\vec{p}|}\sum_{i=0}^2\frac{c_i}{|\omega_i|}\left[
\left(\epsilon+\omega_i\right)\theta(|\vec{p}|-|M_Nx+\omega_i|)
+\left(\epsilon-\omega_i\right)\theta(|\vec{p}|-|M_Nx-\omega_i|)\right]\,,
\label{eq:p3}
\end{align}
where the arrow denotes the averaging procedure from Eq.~(\ref{eq:average}). Note that, 
due to the step function, the cut-off also appears as the lower boundary of the momentum 
integral and we treat these boundaries according to the single cut-off prescription, 
Eq.~(\ref{eq:LAMBDA})\footnote{Here $f(p,\omega_i)$ contains angular matrix elements like 
$\int d\Omega_{\vec{p}}\widetilde{\Psi}^\dagger(\vec{p})\widetilde{\Psi}(\vec{p})$
or $\int d\Omega_{\vec{p}}\widetilde{\Psi}^\dagger(\vec{p})
\vec{\alpha}\cdot\vec{p\,}\widetilde{\Psi}(\vec{p})$
multiplied by powers of $\omega_i$.}
\begin{align}
\sum_{i=0}^2c_i\int^\infty_{|M_Nx+\omega_i|} p dp\, f(p,\omega_i)
&=\int^\infty_{|M_Nx+\epsilon|} p dp\, f(p,\epsilon)
-\int^\infty_{|M_Nx+\sqrt{\epsilon^2+\Lambda^2}|} p dp\, f(p,\sqrt{\epsilon^2+\Lambda^2})\cr
&\qquad
+\Lambda^2\int^\infty_{|M_Nx+\sqrt{\epsilon^2+\Lambda^2}|} p dp\, 
\frac{\partial}{\partial\Lambda^2}f(p,\sqrt{\epsilon^2+\Lambda^2})\cr
&\qquad
-\frac{\Lambda^2}{2\sqrt{\epsilon^2+\Lambda^2}}
\left[pf(p,\sqrt{\epsilon^2+\Lambda^2})\right]_{p=|M_Nx+\sqrt{\epsilon^2+\Lambda^2}|}\,.
\label{eq:p4}
\end{align}

\subsection{Unpolarized structure functions}

We will not present detailed formulas, except for some leading terms of the 
$\frac{1}{N_C}$ expansion. We refer the reader to Ref.~\cite{Takyi:2019ahv}
for more details (even though some factors of $\pi$ were not written there). As an example we 
present the expression for the isoscalar component of the unpolarized RF structure function 
\begin{align}
\left[ f_1^{s} (x)\right]_{I=0}^{\mp} & = \frac{5M_NN_{c}}{72\pi} \sum_\alpha 
\sum_{i=0}^{2} c_i  \int_{\vert M_N  x_\alpha^{\pm} \vert}^{\infty} p\, d p\,  
\int d \Omega_p \Biggl\{ \pm  \widetilde{\Psi}_\alpha^\dagger (\vec{p}) \widetilde{\Psi}_\alpha (\vec{p})
 \cr & \hspace{6cm}
- \frac{\epsilon_\alpha}{\sqrt{\epsilon_\alpha^2+\Lambda_i^2}}
\frac {M_N x_\alpha^{\pm}}{p} \widetilde{\Psi}_\alpha^{\dagger} (\vec{p}) 
\hat{p} \cdot \vec{\alpha} \widetilde{\Psi}_\alpha (\vec{p}) \Biggr\}, 
\hspace{1cm}
\label{vacuum_cont_f1}
\end{align} 
where   
\begin{equation}
M_N x_\alpha^{\pm}  = M_N x \pm \sqrt{\epsilon_\alpha^2 + \Lambda_i^2}. \label{def_of_Mx}
\end{equation} 
The superscripts $\mp$ denote the positive (negative) frequency components which are typically 
referred to as quark and antiquark distribution. They arise from the two poles (for a particular 
$\omega_\alpha$) of the $\delta$-function in Eq.~(\ref{eq:pol1}) and materialize in the 
$\pm\omega$ terms in Eq.~(\ref{eq:wten1}). The total Dirac sea contribution to the isoscalar 
unpolarized structure function is the sum
\begin{equation}
[f_1^{s}(x)]_{I=0} = [f_1^{s}(x)]_{I=0}^{-} + [f_1^{s}(x)]_{I=0}^{+}\,.
\label{eq:f1s}
\end{equation}
On first sight it seems as if the first term under the integral in Eq.~(\ref{vacuum_cont_f1}) would
not be subject to regularization. This is not the case, as the momentum integral is computed 
according to Eq.~(\ref{eq:p4}).
Since the isoscalar unpolarized structure functions are related to the classical energy of the soliton 
by the momentum sum rule, we must still subtract the analog of this calculation that is obtained by 
substituting spinor wave-functions for $\Theta=0$. We have numerically checked the sum rule and
achieved agreement better than 1\%. In view of the many elaborate elements of the simulation, this is 
more than satisfactory. We get the valence contribution from substituting Eq.~(\ref{cranked_valence_level}) 
into the unregularized expression. This then adds
\begin{align}
\left[ f_1^{\rm v}(x) \right]_{I=0}^{\mp}  & = -\frac{5 M_N N_c}{72\pi}   
\left[1+{\rm sign}(\epsilon_{\rm v})\right]
\int_{M_N \vert x_{\rm v}^{\pm} \vert}^{\infty} p\, d p \int d \Omega_p 
\Biggl\{ \pm  \widetilde{\Psi}_{\rm v}^{\dagger} (\vec{p}) \widetilde{\Psi}_{\rm v} 
(\vec{p}) - \frac{M_N x_{\rm v}^{\pm}}{p}  \widetilde{\Psi}_{\rm v}^{\dagger} 
(\vec{p}) \hat{p} \cdot \vec{\alpha} \widetilde{\Psi}_{\rm v} (\vec{p})\Biggr\}, 
\label{valence_cont_f1}
\end{align}
to the positive and negative frequency components of the isoscalar unpolarized structure function. In this 
case there is no need to subtract the $\Theta=0$ counterpart because this level is not occupied in the baryon 
number zero sector.

\begin{figure}[t]
\centerline{
\includegraphics[width=14cm,height=5cm]{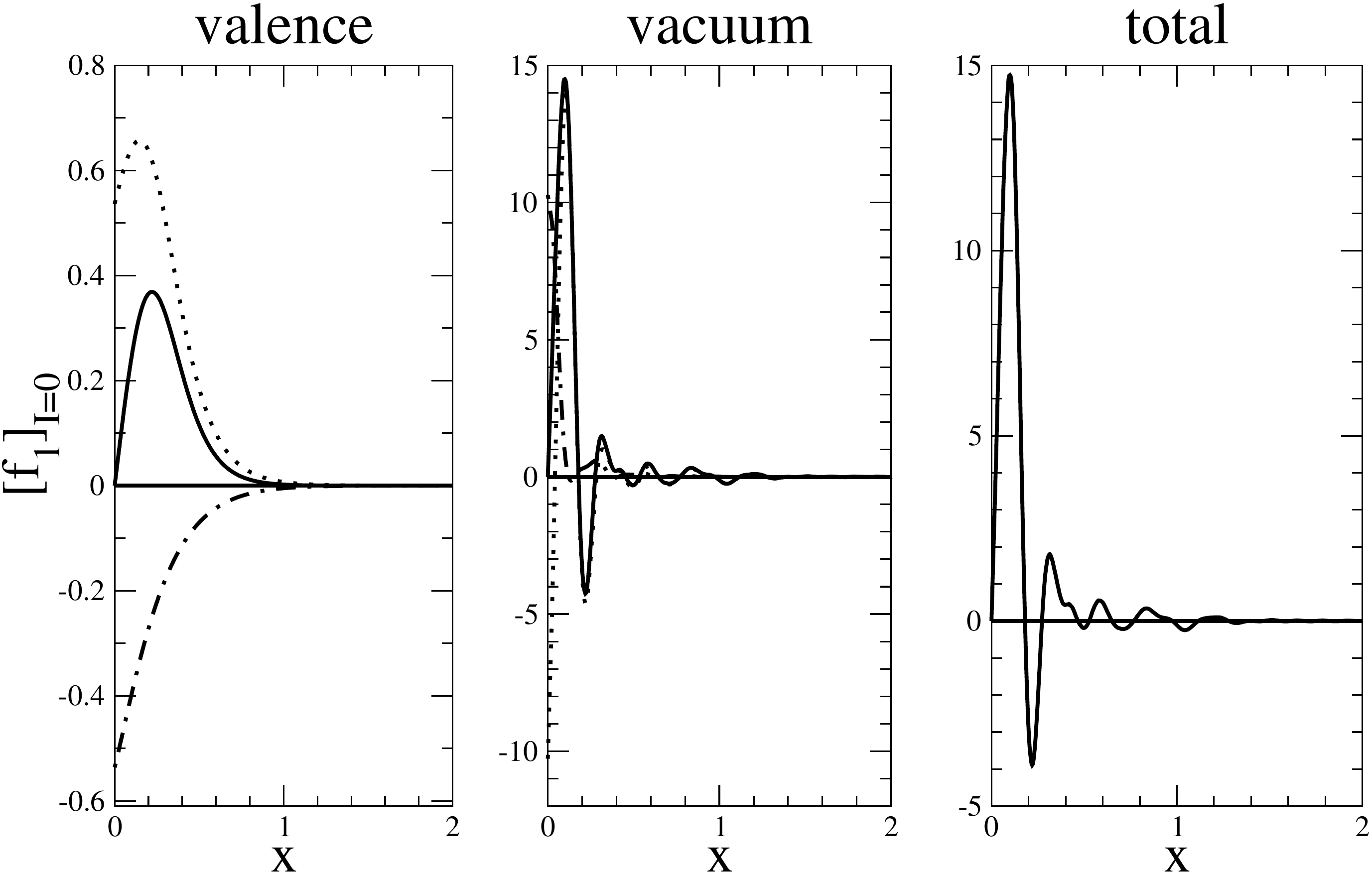}}
\caption{Model prediction (with $m=400 {\rm MeV}$) for the isoscalar unpolarized structure function 
in the nucleon rest frame. Dotted and dotted-dashed lines refer to the positive 
and negative frequency contributions, respectively.}
\label{f1a}
\end{figure}
In Fig.~\ref{f1a} we present typical numerical results. While the valence contribution is smooth,
the vacuum part exhibits large peaks at small $x$. We consider this as an artifact of the 
$\Theta=0$ subtraction, which actually has no dynamical justification other than setting the 
zero energy scale. But this is merely a consistency condition on the sum rule which is only an
integral over the structure function. This may actually be too strong a condition and we will
comment on that in the conclusion.

\begin{figure}
\centerline{
\includegraphics[width=14cm,height=5cm]{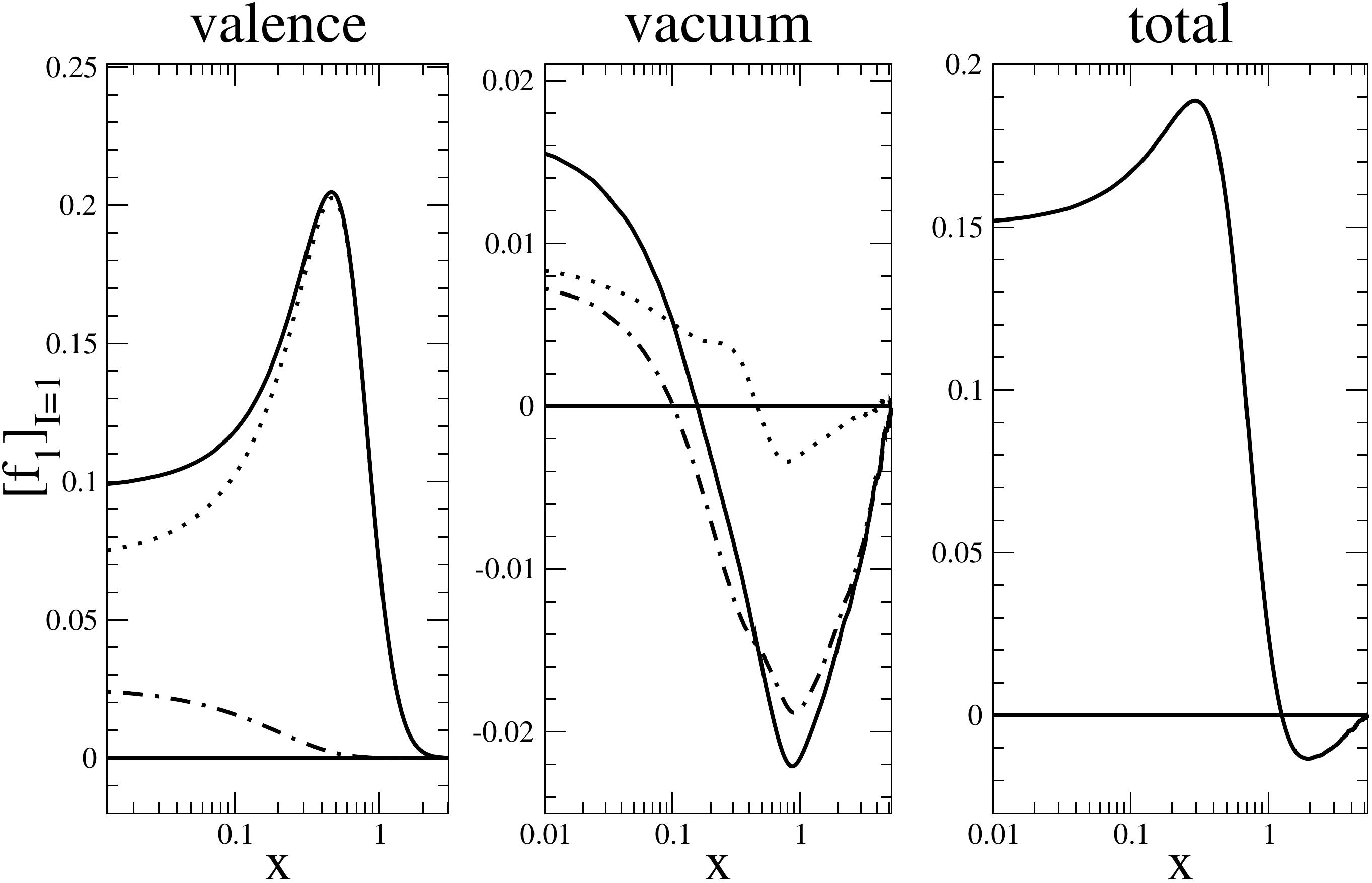}}
\caption{Same as Fig.~\ref{f1a} for the isovector unpolarized structure function.
Observe the logarithmic scale for the Bjorken variable $x$.}
\label{f1b}
\end{figure}

Fig.~\ref{f1b} shows the isovector counterpart which is subleading in $\frac{1}{N_C}$ and does 
not have any (artificial) $\Theta=0$ subtraction. Obviously this structure function is dominated by 
the valence level contribution while the vacuum part is almost negligible.

\begin{figure}
\centerline{
\includegraphics[width=14cm,height=5cm]{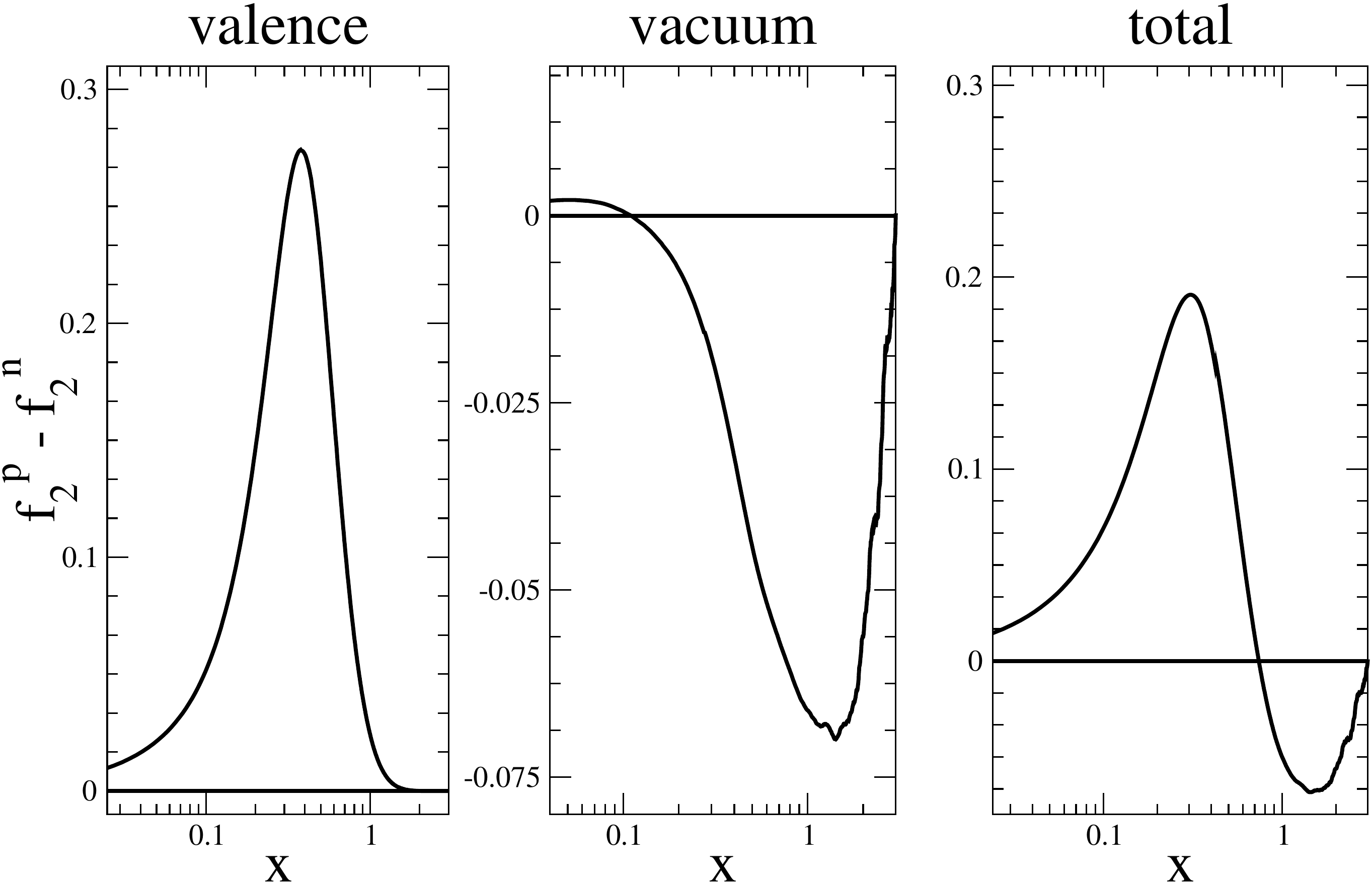}}
\caption{Model prediction of the unpolarized structure function $f_{2}^{p}(x)-f_{2}^{n}(x)$ 
for the constituent quark mass of $m=400{\rm MeV}$.}
\label{f2}
\end{figure}

In Fig.~\ref{f2} we display the numerical results for the unpolarized structure function that enters the 
Gottfried sum rule, {\it i.e.} $f_{2}^{p}(x)-f_{2}^{n}(x) =2x \left[f_{1}^{p}(x)-f_{1}^{n}(x)\right]$, as 
the Callan-Gross relation holds in the rest frame, {\it cf.} Tab.~\ref{tab_1}. At large $x$ the vacuum 
contribution turns slightly negative. Though the valence contribution is generally dominant, the small 
negative piece persists in the total contribution of this structure function. In Tab.~\ref{tab_2}, we 
compare our model prediction for the Gottfried sum rule, 
\begin{equation}
\mathcal{S}_{G}= \int_{0}^{\infty}\frac{d x}{x}\,\left(f_{2}^{p}-f_{2}^{n}\right)\,,
\label{eq:SG}
\end{equation}
for various constituent quark masses to that of the value extracted from data by the NM 
Collaboration~\cite{Arneodo:1994sh}. The agreement is astonishingly good. The integral is 
almost completely saturated by the valence level contribution.

\begin{table}
\caption{The Gottfried sum rule for various values of $m$. The subscripts 'v' and 's' denote the 
valence and vacuum contributions, respectively. The fourth column contains their sums.}
\label{tab_2}
\begin{center}
\begin{tabular}{|c| c| c| c|c|}
\hline
$m\,[{\rm MeV}]$  & $[\mathcal{S}_{G}]_{{\rm v}}$  & $[\mathcal{S}_{G}]_{{\rm s}}$
& $\mathcal{S}_{G}$ & emp.\@ value \\
\hline
$400$  &  $0.214$   & $0.000156$  &  $0.214$ &  \\
$450$  &  $0.225$   & $0.000248$  &  $0.225$ &  $ 0.235 \pm 0.026 $ \cite{Arneodo:1994sh} \\
$500$  &  $0.236$   & $0.000356$  &  $0.237$ & \\
\hline
\end{tabular}
\end{center}
\end{table}
In contrast to the isoscalar unpolarized structure function, the isovector part does not undergo regularization. 
Such an alternating behavior between (un)regularized quantities is well-known for static 
properties \cite{Alkofer:1994ph,Christov:1995vm} but it is interesting to see that it also holds for structure 
functions. Of course, that is a prediction of the formalism.

\subsection{Polarized structure functions}

\begin{figure}[t]
\centerline{
\includegraphics[width=14cm,height=5cm]{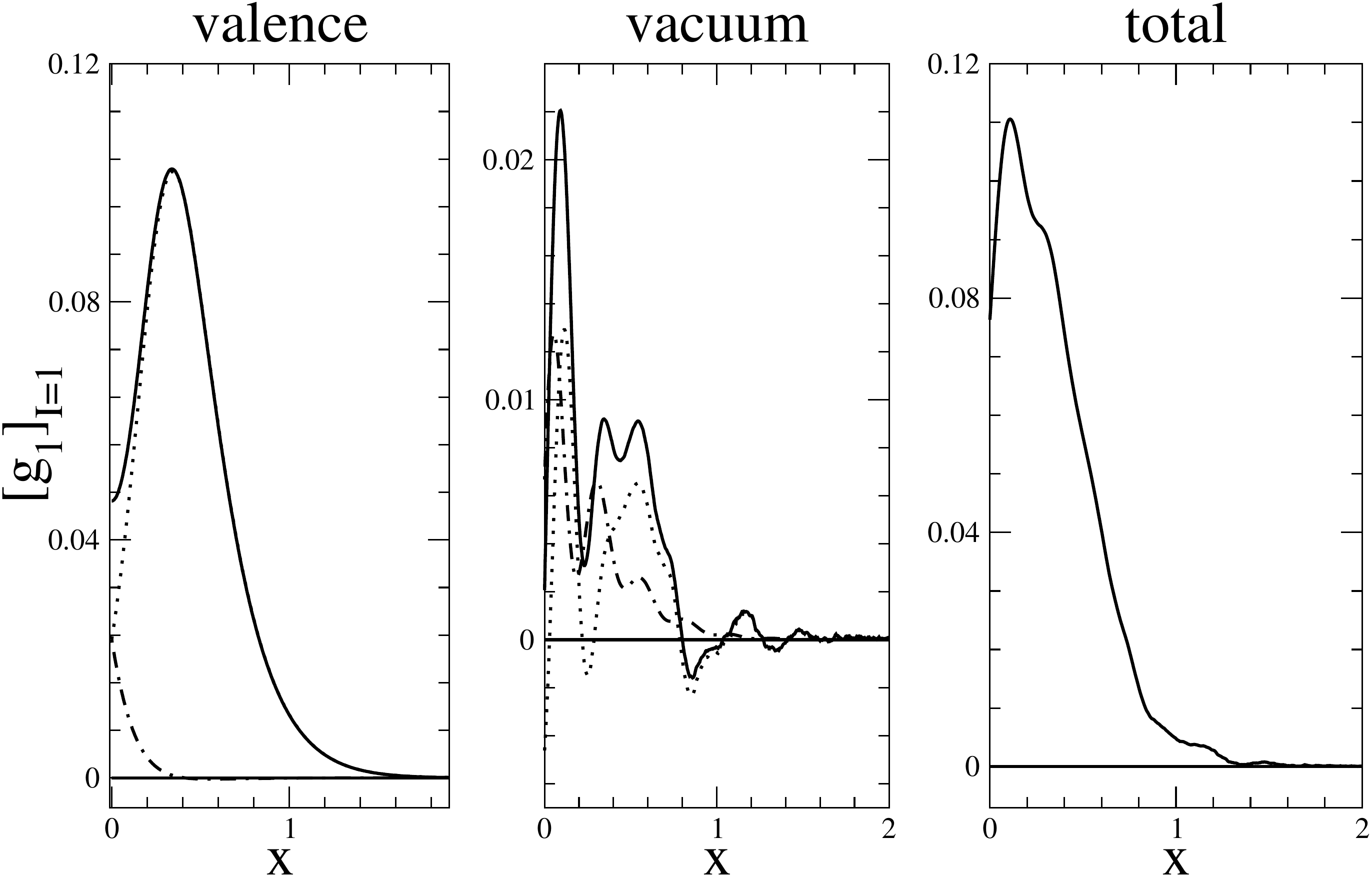}}
\caption{Model prediction ($m=400 {\rm MeV}$) for the isovector longitudinal polarized structure
functions. For the valence and vacuum contributions we separately display the positive (dotted) 
and negative (dotted-dashed) frequency contributions.}
\label{f3a}
\bigskip
\end{figure}

\begin{figure}[b]
\centerline{
\includegraphics[width=14cm,height=5cm]{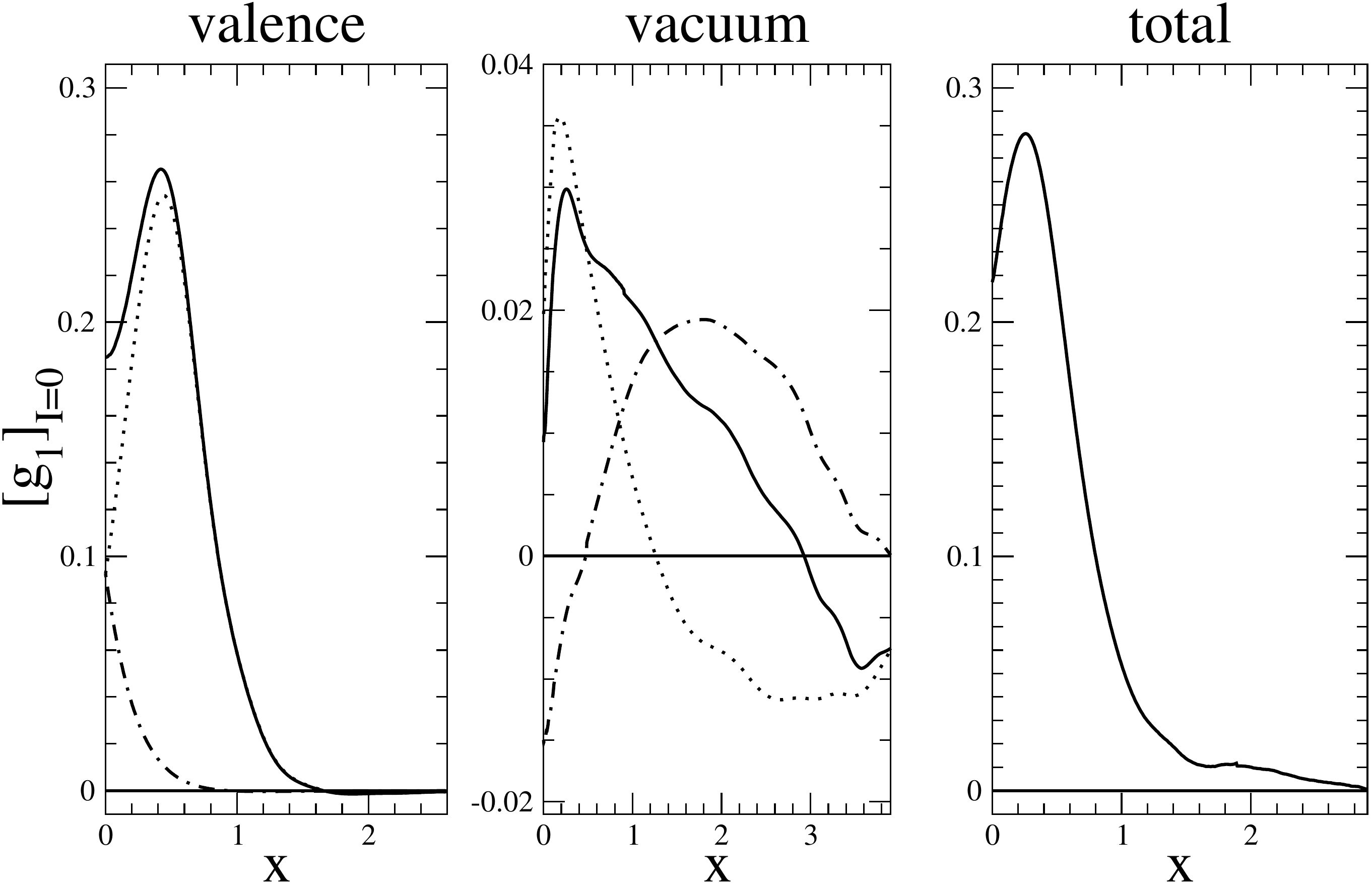}}
\caption{Same as Fig.~\ref{f3a} for the isoscalar longitudinal polarized structure functions.}
\label{f3b}
\end{figure}

For the polarized structure functions we will only list explicit formulas for the 
isovector longitudinal piece which is leading in $\frac{1}{N_C}$. Essentially this is the 
Fourier transform of Eq.~(\ref{g1x}). The vacuum contribution reads
\begin{align}
\left[g_{1}^{s}(x)\right]_{I=1}^{\mp}&= 
- \frac{M_N N_{c}}{36\pi} \left\langle N | I_{3} | N \right\rangle \sum_{\alpha} \sum_{i=0}^{2} c_i  
\Biggl\{ \mp \int_{\vert M_N x_{\alpha}^{\pm } \vert}^{\infty} 
d p\,M_N x_{\alpha}^{\pm} \int d \Omega_{p}   
\widetilde{\Psi}_{\alpha}^{\dagger} (\vec{p}) \hat{p} \cdot \vec{\tau} 
\gamma_{5} \widetilde{\Psi}_{\alpha} (\vec{p}) 
\label{vac_g1x} \\
&  - \frac{\epsilon_{\alpha}}{\sqrt{\epsilon_\alpha^2+\Lambda_i^2}}
\int_{\vert M_N x_{\alpha}^{\pm} \vert}^{\infty} d p\,p^{2} 
\Biggl[ A_{\pm} \int d \Omega_{p}  
\widetilde{\Psi}_{\alpha}^{\dagger} (\vec{p}) \vec{\tau} 
\cdot \vec{\sigma} \widetilde{\Psi}_{\alpha} (\vec{p})   
+  B_{\pm}  \int d \Omega_{p} \widetilde{\Psi}_{\alpha}^{\dagger} (\vec{p}) 
\hat{p} \cdot \vec{\tau} \hat{p} \cdot \vec{\sigma}\widetilde{\Psi}_{\alpha} 
(\vec{p}) \Biggr] \Biggr\}\,, 
\nonumber 
\end{align}
where we have introduced the abbreviations, see also Eq.~\eqref{def_of_Mx},
\begin{align}
A_{\pm} & = \frac{1}{2p} \left( 1 - \frac{(M_N x_{\alpha}^{\pm})^{2}}{p^{2}}\right),
\quad  B_{\pm}  = \frac{1}{2p} \left( 3\frac{(M_N x_{\alpha}^{\pm})^{2}}{p^{2}} -1 \right).
\label{long_multiple_factors}
\end{align}
As before, the superscripts denote the positive and negative frequency components. The total Dirac sea 
contribution to $g_1(x)$ again is the sum of the positive $(+)$ and negative $(-)$ frequency components. 
The valence quark contribution to the isovector longitudinal polarized structure function reads
\begin{align}
\left[g_{1}^{\rm v}(x)\right]_{I=1}^{\mp}& = \frac{M_N N_{c}}{36\pi} 
\left[1+{\rm sign}(\epsilon_{\rm v})\right]\left\langle N | I_{3} | N \right\rangle  
\Biggl\{\mp \int_{\vert M_N x^{\pm} \vert}^{\infty} d p\,M_N x_{\rm v}^{\pm} 
\int d \Omega_{p} \widetilde{\Psi}_{\rm v}^{\dagger} (\vec{p})\hat{p}\cdot \vec{\tau} 
\gamma_{5} \widetilde{\Psi}_{\rm v} (\vec{p}) \cr 
& -\int_{\vert M_N x_{\rm v}^{\pm} \vert}^{\infty} d p\,p^{2} 
\Biggl[ A_{\pm} \int d \Omega_{p} \widetilde{\Psi}_{\rm v}^{\dagger} 
(\vec{p}) \vec{\tau} \cdot \vec{\sigma} \widetilde{\Psi}_{\rm v} (\vec{p})  
+ B_{\pm} \int d \Omega_{p} \widetilde{\Psi}_{\rm v}^{\dagger} (\vec{p}) \hat{p} \cdot 
\vec{\tau} \hat{p} \cdot \vec{\sigma} \widetilde{\Psi}_{\rm v} (\vec{p}) \Biggr] \Biggr\}\,.
\label{val_g1x}
\end{align}
The numerical results are shown in Fig.~\ref{f3a}.
The isoscalar counterpart is subleading in $\frac{1}{N_C}$ and we display a typical model prediction
in Fig.~\ref{f3b}.
We have already discussed the Bjorken sum rule for the isovector piece. Its verification serves as
a test for the accuracy of the numerical simulation. The isoscalar combination also has a sum rule
which gives the matrix element of the isoscalar axial current $\overline{\Psi}\gamma_\mu\gamma_5\Psi$. 
As mentioned, its empirical determination
has triggered much of the research on structure functions. Our results for both sum rules are 
shown in Tab.~\ref{tab_3}. We note that the isoscalar axial charge is significantly less than 
one in agreement with phenomenology of the proton spin puzzle \cite{Ashman:1987hv}.

\begin{table}[t]
\vskip-0.3cm
\caption{Axial isovector and isoscalar charges for various values of the constituent quark mass $m$
from integrating the longitudinal structure functions. Subscripts are as in Tab.~\ref{tab_2}. Data in 
parenthesis give the numerical results as obtained from 
the coordinate space representation, {\it cf.} Eq.~(\ref{eq:ga7}) and Tab.~\ref{tab:model}.}
\label{tab_3}
\begin{center}
\begin{tabular}{|c| c| c| c| c||c| c| c| c| c|}
\hline
$m\,[{\rm MeV}]$  & $[ g_{A}]_{{\rm v}}$ & $ [ g_{A}]_{{\rm s}}$  & $ g_{A}$
& emp.\@ value & $[ g_{A}^{0}]_{{\rm v}}$ & $[ g_{A}^{0}]_{\rm s}$ & $ g_{A}^{0}$
& emp.\@ value  \\
\hline
$400$  &  $0.734$  & $0.065$ &  $0.799$  ($0.800$) & $1.2601$ &
$0.344$  & $0.0016$  &  $0.345$  ($0.350$) & \\
$450$  &  $0.715$  & $0.051$ &  $0.766$  ($0.765$) & $\pm0.0025$  &
$0.327$  & $0.0021$  &  $0.329$  ($0.332$) & $0.33 \pm 0.06$ \\
$500$  &  $0.704$  & $0.029$ &  $0.733$  ($0.733$) & \cite{Barnett:1996hr} &
$0.316$  & $0.0028$  &  $0.318$  ($0.323$) &  \cite{Alexakhin:2006oza} \\
\hline
\end{tabular}
\end{center}
\vskip-0.2cm
\end{table}

\begin{figure}[t]
\centerline{
\includegraphics[width=14cm,height=5cm]{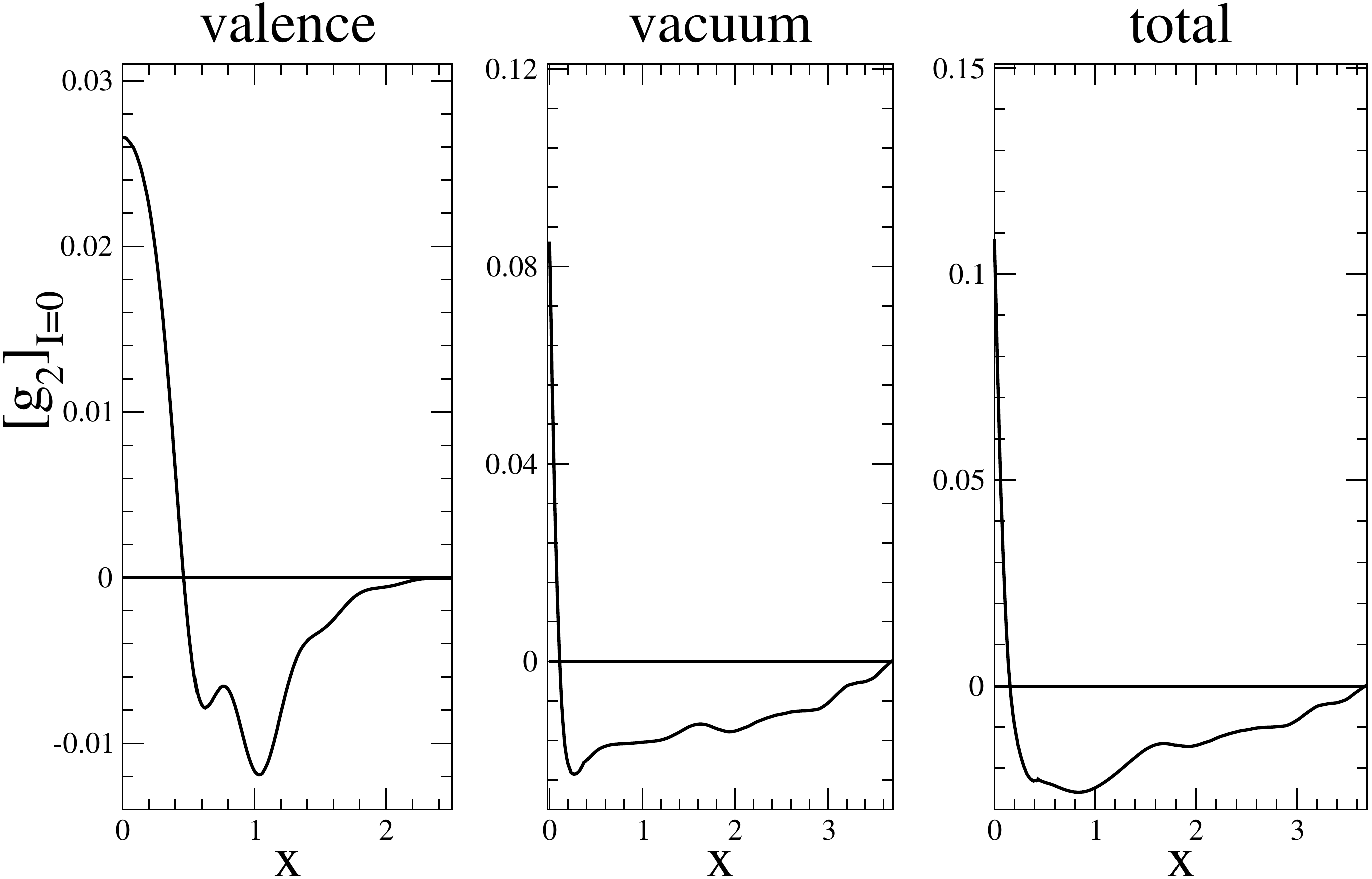}}
\caption{Model prediction of the isoscalar structure function, $g_2$, 
for the constituent quark mass of $m=400 {\rm MeV}$.}
\label{f4a}
\bigskip
\end{figure}

\begin{figure}[b]
\centerline{
\includegraphics[width=14cm,height=5cm]{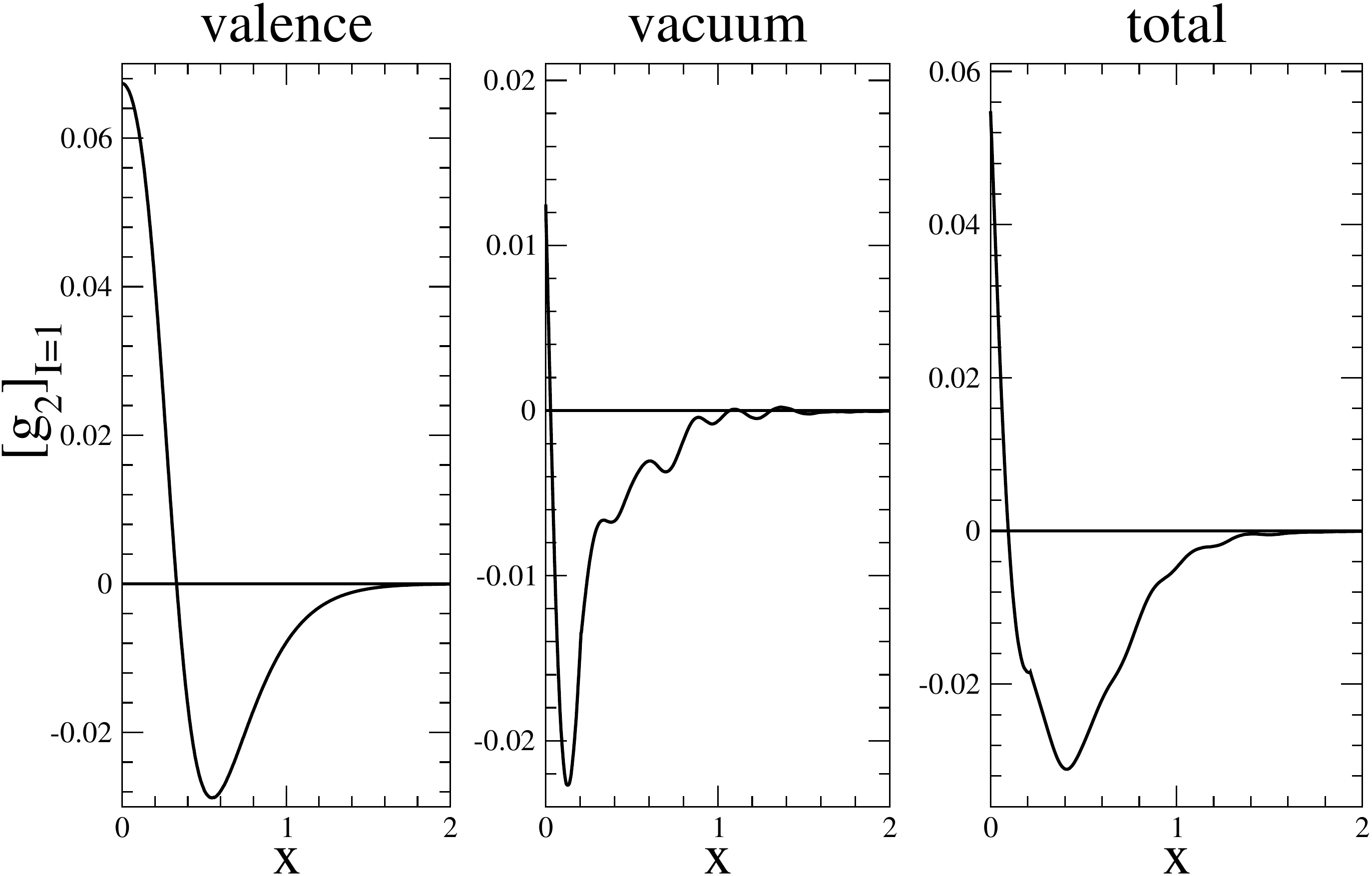}}
\caption{Model prediction of the isovector polarized structure functions, $g_2$, 
frame for the constituent quark mass of $m=400 {\rm MeV}$.}
\label{f4b}
\end{figure}

According to the projectors listed in Tab.~\ref{tab_1} the transverse polarized structure 
function $g_T(x)$ has matrix elements similar to those above. From its computation we subsequently
identify 
\begin{equation}
g_2(x)=g_1(x)-g_T(x)
\label{eq:g2x}
\end{equation}
for both the isoscalar and isoscalar combinations. Typical results are shown in Figs.~\ref{f4a}
and~\ref{f4b}. Here it occurs that the vacuum piece dominates. However, that is mainly a consequence
of cancellations for the valence contribution via Eq.~(\ref{eq:g2x}).

\subsection{Boosting to the infinite momentum frame}
\label{ssec:boost}

It is customary to introduce light-cone coordinates 
$x^{\pm}=\left(x^0\pm\hat{n}\cdot\vec{x}\right)/\sqrt{2}$ to discuss structure functions in 
the context of the parton model. Using these coordinates the Bjorken limit is particularly 
transparent
$$
q^{-}\to\infty \qquad {\rm and}\qquad 
x=-\frac{q^{+}}{p^{+}}\,.
$$
As discussed before, the fermion propagator is free and massless in the Bjorken limit.
Massless fermions have the singular function 
$\left\{\Psi(\xi),\overline{\Psi}(0)\right\}=\frac{1}{2\pi}\dslash\delta(\xi^2)\epsilon(\xi^0)$. This 
can be used to turn the current-current correlator in the hadron tensor into a matrix element of 
bilocal bilinear fermion operators \cite{Jaffe:85pi}
(These are fundamental fermion operators, not the eigenfunctions of $h$ in Eq.~(\ref{hedgehog}).)
$$
f_1(x)=\frac{x}{4\pi}\int d\xi^{-}\,{\rm e}^{-\imu p^{+}\xi^{-}}
\langle N|\overline{\Psi}(\xi)\gamma^{+}\mathcal{Q}^2\Psi(0)
-\overline{\Psi}(0)\gamma^{+}\mathcal{Q}^2\Psi(\xi)|N\rangle_{\xi^{+}=0,\xi_\perp=0}\,.
$$
This singles out the coordinate along the photon momentum as the most relevant variable.
We will see shortly that this is indeed realized in the IMF, which also has $\xi^{+}=0$.

Assuming translational invariance and inserting a complete set of states with momenta $p_n$, the 
matrix elements of bilocal bilinear quark operators can be shown to be non-zero only when
$$
p_n^{+}-(1-x)p^{+}=0\,.
$$
In the vicinity of $x=1$ this can only be fulfilled when the masses of both the partons $n$ and
the nucleon are negligible small and/or $p^{+}$ becomes very large. The limit of large $p^{+}$ 
defines the IMF. It is therefore suggestive to consider the soliton model structure functions in 
the IMF as well. To boost the system to the IMF, the collective coordinate method of Eq.~(\ref{nuc1}) 
for any local object $\Gamma$ must be extended to 
\begin{align}
\Gamma(\vec{\xi},\xi^0)\,\longrightarrow\, 
S(\Lambda) \Gamma(\vec\xi^\prime-\vec{R}^\prime,\xi^{\prime0})S^{-1}(\Lambda)
\qquad {\rm where}\qquad
\xi^{\prime\mu}=\left(\Lambda^{-1}\right)^\mu_{\hspace{2mm}\nu}\xi^\nu\,.
\label{eq:LT1}
\end{align}
Here $\Lambda$ parameterizes a Lorentz transformation and $S(\Lambda)$ is the corresponding
generator for $\Gamma$. Subsequently, the collective coordinates $\vec{R}$ are averaged as 
in Eq.~(\ref{nuc1}). 

A Lorentz boost with rapidity $\Omega$ along the light cone transforms the RF coordinates as
\begin{equation}
p^{+}\,\longrightarrow \, \frac{M_N}{\sqrt{2}}\,{\rm e}^\Omega
\qquad {\rm and}\qquad 
p^{-}\,\longrightarrow \, \frac{M_N}{\sqrt{2}}\,{\rm e}^{-\Omega}
\label{eq:LT2}
\end{equation}
while the transverse components are left unchanged. The transformation to the IMF is thus characterized
by $\Omega\to\infty$ which also implies that $\Lambda^{-1}$ singles out $\xi^{\prime-}$ so that 
$\xi^{\prime+}\to0$. In Ref.~\cite{Gamberg:1997qk} this transformation was applied together with
the collective coordinate average for bilocal bilinear quark composites like those in Eq.~(\ref{nuc5}). 
Essentially that study adapted a two-dimensional MIT bag model calculation~\cite{Jaffe:1980qx} to the 
soliton model by ignoring effects on the transverse coordinates as Lorentz covariance is only restored 
along $\hat{n}$. The result is a simple transformation prescription for the structure functions:
\begin{equation}
f_{\rm IMF}(x) = \frac{\Theta \left(1-x\right)}{1-x} f_{\rm RF} \left(-\ln (1-x) \right)\,,
\label{Infinite_momentum_frame} 
\end{equation}
where $f_{\rm RF}$ is any of the structure functions like that in Eq.~(\ref{g1x}) which are obtained from the 
hadron tensor in the RF according to the calculations in the previous section. Obviously this prescription 
ensures that the transformed structure functions have support only in the kinematically allowed interval 
$0\leq x\leq1$. Thus the structure function $f_{\rm IMF}(x)$ is a suitable input for the DGLAP evolution program. 
In what follows we will omit the label IMF for the boosted structure functions.

\subsection{DGLAP evolution}

Of course, we wish to compare our model predictions with data. In this section we describe the 
remaining step with focus on the polarized structure functions. All model results presented in this
Subsection have been obtained for the constitutent quark mass $m=400{\rm MeV}$.

So far, we have computed the structure functions within the NJL soliton model, which (at best) 
approximates QCD at a low mass scale, $\mu^{2}=Q_0^2$ which is thus an adjustable hidden parameter 
in the approach and can be thought of as the {\it identification scale} with QCD. This low mass 
scale is different from the high energy scales, $Q^2$ at which DIS data are available. To compare with 
the DIS data, we adopt the leading order Altarelli-Parisi (DGLAP) equations~\cite{Gribov:1972ri} for 
parton distributions to evolve the model structure functions. 
To apply this formalism we, unfortunately, have to identify the model structure functions with QCD 
distribution functions of quarks since the chiral model is not renormalizable and does not have 
a renormalization group equation to sum the {\it leading logs}. 

Let $h^{(I=1)}(x,t)$ be the isovector combination of any twist-2 
distribution with $t = \ln \left( \frac{Q^2}{\Lambda^2_{QCD}} \right)$. The change in momentum 
scale is governed by the differential equation  
\begin{equation}
\frac{d h^{(I=1)}(x,t)}{d t}=\frac{\alpha_s(t)}{2 \pi}C_{R}(F)\int_{x}^{1} \frac{d y}{y}
P_{qq}(y) h^{(I=1)}\left(\frac{x}{y},t\right):=
\frac{\alpha_s(t)}{2 \pi}C_{R}(F) P_{qq}\otimes h^{(I=1)}(x,t) \,.
\label{evol_integ_diff_equa_isovec}
\end{equation}
Here $\alpha_s(t)=\frac{4\pi}{\beta_{0} t}$, is the running coupling constant of QCD, in which
$\beta_{0}=\frac{11}{3}N_{C}- \frac{2}{3}N_{f}$ and $C_{R}(F)=\frac{N_{f}^{2}-1}{2N_{f}}$ are
combinatoric factors in the QCD renormalization group equation for $N_{f}$ flavors. Most importantly 
$P_{qq}(y)$ is the splitting function that describes the probability of a quark emitting a gluon 
and a quark with momentum fraction $y$. This splitting function and those for the isoscalar 
combination to be discussed below are given in Ref.~\cite{Gribov:1972ri}. The right-hand-side of 
Eq.~(\ref{evol_integ_diff_equa_isovec}) serves as the definition of the evolution product "$\otimes$".
As initial condition, $h^{(I=1)}(x,t(\mu^2))$, to integrate this differential equation we take 
the distributions identified from the boosted structure functions in the IMF. The endpoint of
integration is the scale $Q^2$ at which data from experiment are available. We attempt to tune $\mu^2$ 
to optimize the agreement with these data and take the very same identification scale for all evolution 
calculations.

The isoscalar combinations, $h^{(I=0)}(x,t)$ are more complicated. By the pure nature of the quantum numbers 
$h^{(I=0)}(x,t)$ mixes with the gluon distribution $g(x,t)$ and the evolution equations are coupled 
differential equations
\begin{align}
\frac{d h^{(I=0)}(x,t)}{d t}&=\frac{\alpha_s(t)}{2\pi}C_{R}(F)
\left[ P_{qq}\otimes h^{(I=0)}(x,t)+P_{qg}\otimes g(x,t)\right]\cr
\frac{d g(x,t)}{d t}&=\frac{\alpha_s(t)}{2\pi}C_{R}(F)
\left[ P_{gq}\otimes h^{(I=0)}(x,t)+P_{gg}\otimes g(x,t)\right].
\label{evol_integ_diff_equa_isoscal}
\end{align}
The only sensible identification of the gluon distribution $g(x,t)$ is to have it vanish at $\mu^2$,
otherwise sum rules would be violated. This is again an unavoidable (and undesirable) identification of 
QCD degrees of freedom.

\begin{figure}
\centerline{
\includegraphics[width=6cm,height=3.5cm]{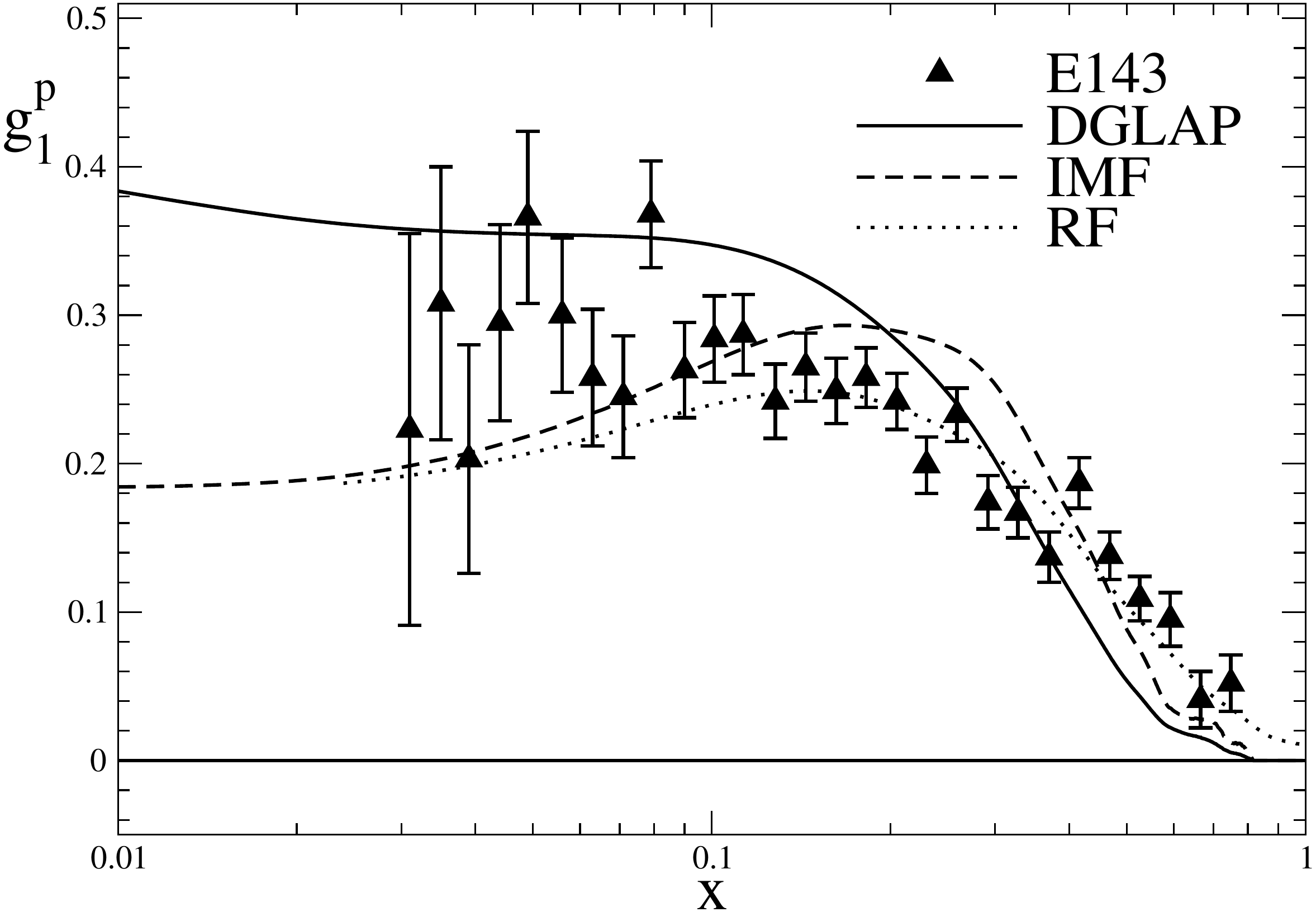}\hspace{1cm}
\includegraphics[width=6cm,height=3.5cm]{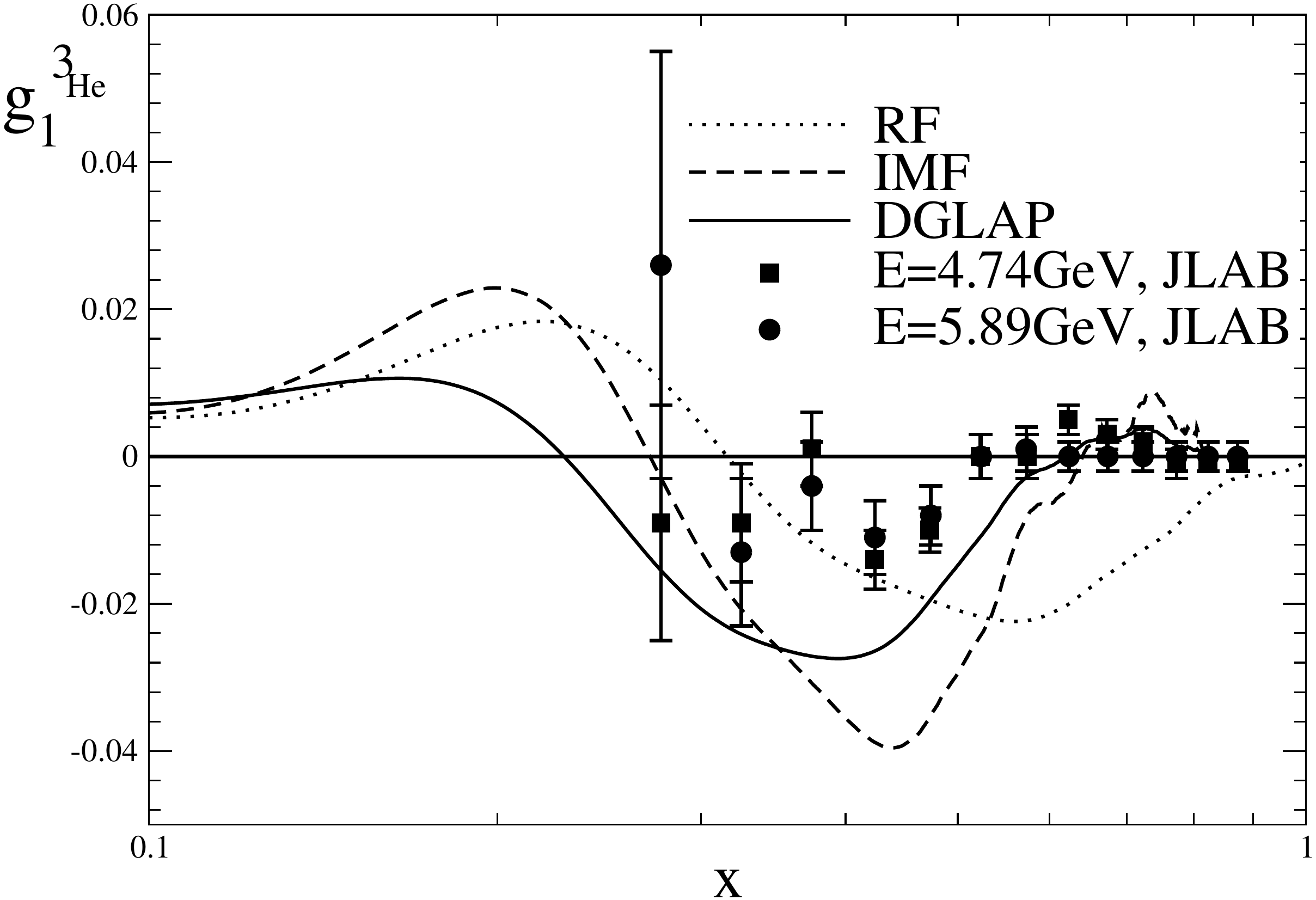}
}
\caption{Model prediction for the longitudinal polarized proton structure functions.
Left panel: $g_{1}^{p}(x)$ ; right panel: $g_{1}^{^3{\rm He}}(x)$.
These functions are "DGLAP'' evolved from $\mu^{2}=\mathrm{0.4\,GeV^{2}}$
to $Q^{2}=\mathrm{3\,GeV^{2}}$ after being projected to the IMF.
Data are from Refs. \cite{Abe:1994cp,Abe:1998wq} for the proton and
from Ref. \cite{Flay:2016wie} for helium. In the latter case $E$ refers to the electron energy.}
\label{f5}
\end{figure}

We are now in the position to confront the model prediction with data from experiment. For the longitudinal 
polarized structure function of the proton this is done in the left panel of Fig.~\ref{f5}.
We chose $\mu^2=0.4{\rm GeV}^2$ and get a reasonable (though not perfect) match with the data 
after evolving the boosted structure function to the scale of the experiment, $Q^2=3{\rm GeV}^2$.
Any further fine-tuning of $\mu^2$ has only marginal effects. The predictions are obviously in the 
right ballpark, but deviations clearly emerge in detail. Surprisingly, the RF result appears to 
match data best. This is an indication that the boost formalism overemphasizes the low $x$ regime.
For the neutron data are available in terms of the helium structure
function \cite{Flay:2016wie}\footnote{In Ref. \cite{Flay:2016wie} direct neutron
data are only given as the ratio $\mathsf{g}_1^{n}(x)/F_1(x)$.}
\begin{equation}
\mathsf{g}_1^{^3{\rm He}}(x)\approx P_n \mathsf{g}_1^{n}(x)
+P_p \mathsf{g}_1^{p}(x) -0.014\left[\mathsf{g}_1^{p}(x)-4\mathsf{g}_1^{n}(x)\right]\,,
\label{eq:g1he}
\end{equation}
with $P_n\approx 0.86$ and $P_p\approx-0.028$ arising from the nuclear model. Also from 
Fig.~\ref{f5} we see that in this case the DGLAP evolution indeed brings the model prediction
closer to data. At large $x$ we find the structure function to be small and positive while for
moderate $x$ the observed negative trough is present but somewhat too strong.

\begin{figure}
\centerline{
\includegraphics[width=9cm,height=3.5cm]{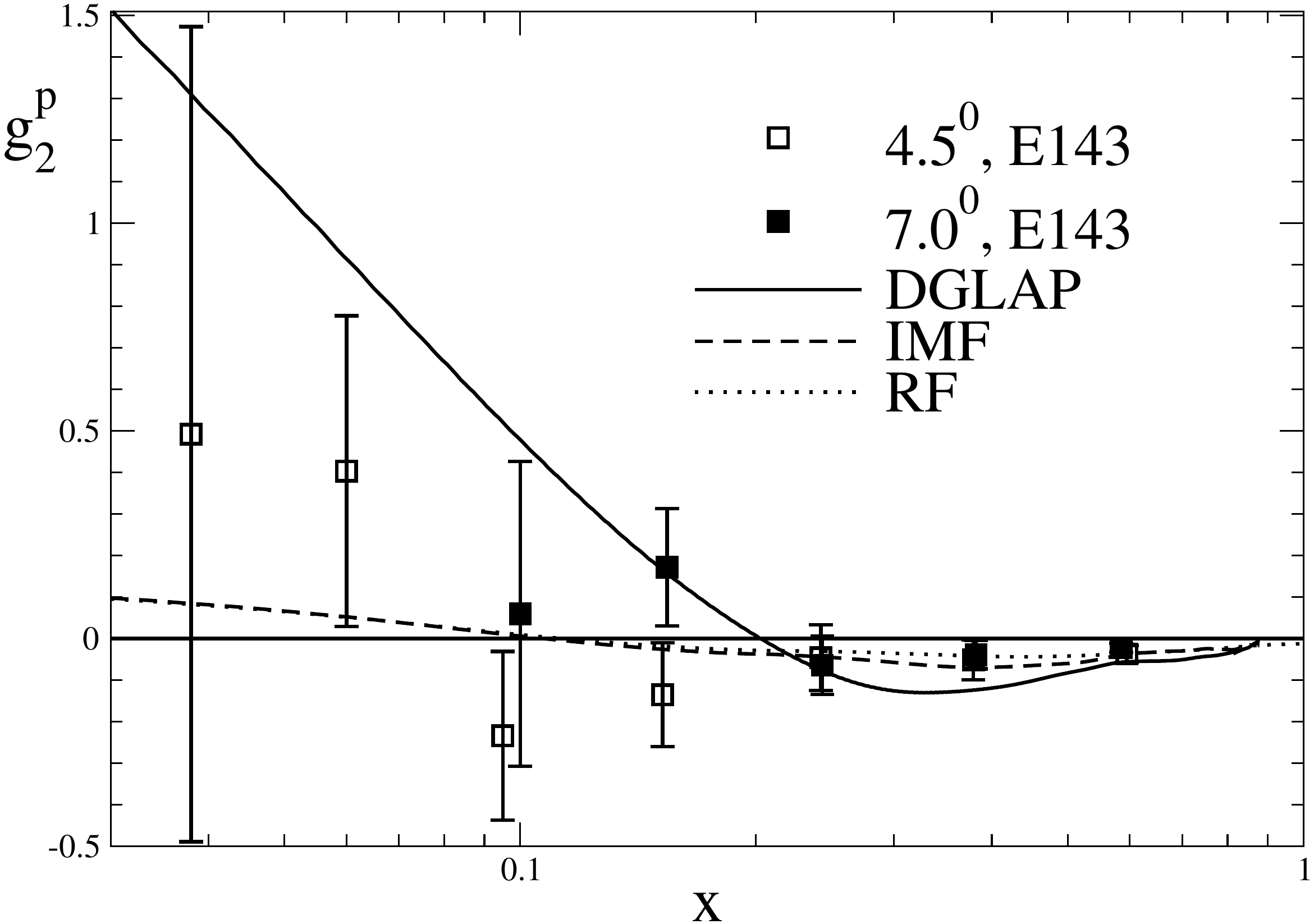}
}
\caption{Model prediction for the polarized proton structure functions
$g_{2}^{p}(x)$. This function is "DGLAP'' evolved from $\mu^{2}=\mathrm{0.4\,GeV^{2}}$ to
$Q^{2}=\mathrm{5\,GeV^{2}}$ after being projected to the IMF. Data are from Ref \cite{Abe:1995dc}.}
\label{f7}
\end{figure}

The evolution of the transverse polarized structure functions is even more complicated because
$g_2(x,t)$ is the sum of two terms. One has twist-2 \cite{Wandzura:1977qf}
\begin{equation}
g_{2}^{WW} (x,t)=-g_{1}(x,t) + \int_{0}^{1}dy\, \frac{1}{y} g_{1} (y,t)
\label{eq:g2t2}
\end{equation}
and the remainder, $\overline{g}_{2} (x,t)=g_{2}(x,t)-g_{2}^{WW} (x,t)$ is associated with twist-3.
The twist-2 part undergoes the DGLAP evolution described above. For the twist-3 piece we extract
Mellin moments 
\begin{equation}
M_j(Q^2)=\int_0^1 dx\, x^{j-1} \overline{g}_{2} (x,t)
\label{eq:momentsg2}
\end{equation}
that scale as
\begin{equation}
\frac{M_j(Q^2)}{M_j(\mu^2)}
=\left[\frac{\ln(\mu^2)}{\ln(Q^2)}\right]^{\frac{\gamma_{j-1}}{\beta_0}}\,.
\end{equation}
So far, only the leading large $N_C$ terms of $\gamma_{j-1}$ are known \cite{Jaffe:1989xx}.
At the initial scale $\mu^2$ we disentangle the twist components, evolve them separately to $Q^2$,
invert the Mellin transformation, and put the two components back together to build $g_2(x,t)$.
The result of this procedure for the proton channel is compared to available data in 
Fig.~\ref{f7}. Our estimate produces the main structure seen experimentally: $g_2^p(x,t)$ is 
negative and small in magnitude at large $x$ and increases substantially as $x$ decreases.

Twist-3 by itself is interesting as data have been recently reported \cite{Flay:2016wie}
for the second moment 
\begin{equation}
d_2^{(n)}(Q^2)=3\int_0^1 dx\, x^2\, \overline{g}^{{(n)}}_2(x,t)
\label{eq:d2n}
\end{equation}
in the neutron channel at two different transferred momenta: 
$d_2^{(n)}(3.21{\rm GeV}^2)=(-4.21\pm1.14)\times10^{-3}$
and $d_2^{(n)}(4.32{\rm GeV}^2)=(-0.35\pm1.04)\times10^{-3}$
(we added the reported errors in quadrature). Our model calculations for $m=400{\rm MeV}$ 
yield $-4.26\times10^{-3}$ and $-4.09\times10^{-3}$, respectively. While the lower $Q^2$ result 
matches the observed value, the higher one differs by about three standard deviations. The results 
indicates that the large $N_C$ approximation to evolve $\overline{g}_2$ requires improvement.

\begin{figure}
\centerline{
\includegraphics[width=9cm,height=3.5cm]{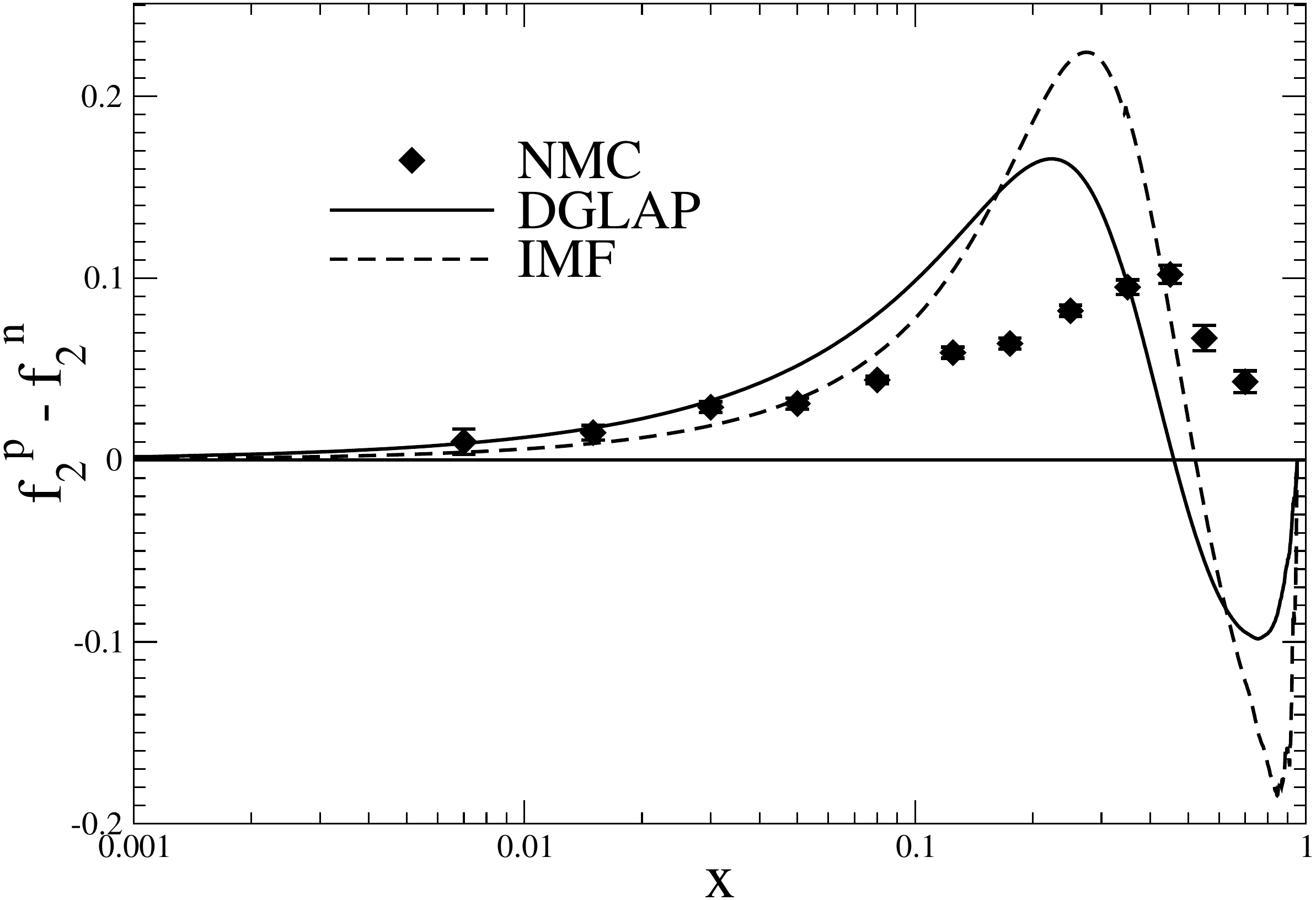}
}
\caption{Model prediction ($m=400{\rm MeV}$) for the unpolarized structure function that 
enters the Gottfried sum rule, Eq.~(\ref{eq:SG}). This function is "DGLAP'' evolved from 
$\mu^{2}=\mathrm{0.4\,GeV^{2}}$ to $Q^{2}=\mathrm{4\,GeV^{2}}$ after transformation to 
the IMF. Data are from Ref.~\cite{Arneodo:1994sh}.}
\label{f8}
\end{figure}

Finally we comment on the isovector unpolarized structure function that is compared to 
data in Fig.~\ref{f8}, see also Figs.~\ref{f1b} and \ref{f2}. Though the negative contribution to 
$f_1$ from the Dirac vacuum, cf. Fig.~\ref{f1b}, around $x=1$ is tiny in
the RF, it becomes relatively large when (i) multiplied by $x$ to obtain $f_2$ and (ii) when
transformed to the IMF because of the Jacobian factor $1/(1-x)$ thereby worsening the 
agreement with the experimental data from NMC \cite{Arneodo:1994sh}.
To some extend, this dilutes the perfect agreement between the model prediction and data for the
Gottfried sum rule, Eq.~(\ref{eq:SG}), discussed earlier . Under that integral the model result arises 
from cancellations not seen in the empirical structure function~\cite{Arneodo:1994sh}. 

\section{Related approaches}
\label{sec:others}

One of the major obstacles when computing structure functions within chiral quark soliton models is the 
consistent implementation of the regularization prescription. Various approaches have been undertaken. 
The numerical results do not differ significantly as the dominant contribution to the structure 
functions arises from the explicitly occupied valence level (in particular when $m\lesssim400{\rm MeV}$)
and this contribution is not subject to regularization. Even though the discrepancies among the various 
approaches to structure functions in chiral soliton models are presumably smaller than their systematic 
uncertainties we will nevertheless comment on alternative approaches in this Section.

A simple-minded but not too unrealistic procedure to avoid that problem is too simply ignore the 
vacuum contribution and compute structure functions in the so-called valence level only
approximation. This is guided by the observation that the most important role of the vacuum 
contribution in chiral quark soliton models is to stabilize the soliton but it is of lesser importance 
for the predictions of static nucleon properties \cite{Alkofer:1994ph,Christov:1995vm}. 
For example, for $m=400{\rm MeV}$ the valence level contributes almost 80\% to the moment of inertia 
in Eq.~(\ref{mominert}). This avenue for the structure functions was taken in the early works 
reported in Refs.~\cite{Weigel:1996kw,Weigel:1996jh}. It also allows 
for a sensitive estimate of $\frac{1}{N_C}$ effects and the separation of isoscalar and -vector 
components without encountering complicated expressions like those in Eqs.~(\ref{eq:wten3}) 
and~(\ref{blsfct}). The results from Sect.~\ref{sec:NR} show that this is indeed a reasonable
approximation for the polarized structure functions; maybe to a lesser extend for the unpolarized 
structure functions.

From Eq.~(\ref{eq:wten2}) we see that the regularized hadron tensor at leading order in 
$\frac{1}{N_C}$ is a sum of four terms, while the unregularized version only has two. Similarly,
when acting with the projection operators from Tab.~\ref{tab_1} to extract a certain structure
function, the spectral functions $f^{(\pm)}_\alpha(\omega)$ combine to reduce the number of terms 
that contribute to the hadron tensor to two as well. We have seen that explicitly for the longitudinal 
polarized structure function $g_1(x)$ in Eq.~(\ref{g1x}). These two terms are formally distribution
functions that take the fermions forward and backward in space time along the direction of the
virtual photon momentum. The (formal) appearance of such distributions is general to all 
fermion models in the Bjorken limit. It is therefore suggestive to consider
distribution functions in such models regardless of whether or not other peculiarities in 
the model, like regularization, require more detailed consideration. In this context the
authors of Ref.~\cite{Diakonov:1996sr} derived two equivalent\footnote{Equivalent expressions 
arise from trace identities. {\it E.g.}, $\sum_\alpha \epsilon_\alpha=0$ allows to write 
(for $\epsilon_{\rm v}>0$): $\epsilon_{\rm v}+\frac{1}{2}\sum_\alpha |\epsilon_\alpha|=
\sum_{\alpha,\rm occ.} \epsilon_\alpha$. Whether or not such identities hold depends on the 
particular regularization prescription.} expressions for unregularized quark distribution 
functions (coefficients adjusted to comply with Eq.~(\ref{ftrans}))
\begin{align}
D^{(1)}(x)&=\frac{2}{\pi}N_CM_N\sum_{\alpha,\rm occ.} \int d^3p\,
\widetilde{\overline\Psi_\alpha}(\vec{p})\nslash \Gamma \widetilde{\Psi}_\alpha(\vec{p})
\delta(p^3+\epsilon_\alpha-M_Nx)\cr
D^{(2)}(x)&=-\frac{2}{\pi}N_CM_N\sum_{\alpha,\rm non-occ.} \int d^3p\,
\widetilde{\overline\Psi}_\alpha(\vec{p})\nslash \Gamma \widetilde{\Psi}_\alpha(\vec{p})
\delta(p^3+\epsilon_\alpha-M_Nx)\,,
\label{eq:diak1}
\end{align}
in the large $N_C$ limit. Obviously these expression combine to
\begin{equation}
D(x)=\frac{1}{2}\left[D^{(1)}(x)+D^{(2)}(x)\right]
=-\frac{1}{\pi}N_CM_N\sum_{\alpha} {\rm sign}(\epsilon_\alpha) \int d^3p\,
\widetilde{\overline\Psi_\alpha}(\vec{p})\nslash \Gamma \widetilde{\Psi}_\alpha(\vec{p})
\delta(p^3+\epsilon_\alpha-M_Nx)\,,
\label{eq:diak2}
\end{equation}
for the Dirac sea contribution which, by definition, has the negative energy levels occupied (occ) 
and the positive energy levels empty (non-occ). Similarly anti-quark distribution functions are 
obtained with $\overline{D}(x)=-D(-x)$. Using $D(x)$, $\overline{D}(x)$ and suitable linear 
combinations of the  spin flavor matrices~$\Gamma$ the authors would then compute the structure 
functions. Considering, for example, the unregularized version of of the first term within the 
square brackets in Eq.~(\ref{g1x}) and noticing that
\begin{equation}
\left[\omega+\epsilon_\alpha\right]\delta(\omega^2-\epsilon_\alpha^2)
={\rm sign}(\epsilon_\alpha)\delta(\omega-\epsilon_\alpha)
\label{eq:diak3}
\end{equation}
we observe the very same structure as in $D(x)$. In the notation of Ref.~\cite{Diakonov:1996sr} the 
second term in Eq.~(\ref{g1x}) represents the antiquark distribution $\overline{D}(x)$ that must be added 
to complete the structure function. Not unexpectedly, without regularization these approaches are thus 
equivalent. Ref.~\cite{Diakonov:1996sr} performs a two step regularization for the distributions, first a 
smoothing function is multiplied in the level sum in Eq.~(\ref{eq:diak2}) with a scale $E_{\rm max}$. 
Then the calculation is repeated with a second, larger constituent quark mass and the difference is 
extrapolated to $E_{\rm max}\to\infty$. That second constituent quark mass conceptually is a 
Pauli-Villars mass, $M_{\rm PV}$. Its numerical value is determined from the pion decay constant $f_\pi$
as follows: compute the unregularized polarization functions, Eq.~(\ref{specfct}), that enter $f_\pi$ for 
both $m$ and $M_{\rm PV}$, multiply both polarization functions by $m$ and $M_{\rm PV}$, respectively and 
tune $M_{\rm PV}$ such that the difference is $f_\pi/4N_C$, with $f_\pi=93{\rm MeV}$. Though the procedure 
seems plausible, it is not rigorous\footnote{The caption to Fig.~1 in the second of Ref.~\cite{Diakonov:1996sr} 
suggests that the valence level contribution would also undergo this Pauli-Villars type subtraction. If 
correctly interpreted, that seems in contradiction to unit baryon number.}. 
Among other questions one might ask why should the second calculation have the same 
smoothing scale $E_{\rm max}$; and if different, what is the effect? We also note that a single 
subtraction does not produce a finite gap equation\footnote{In Ref.~\cite{Diakonov:1996sr} this 
problem is bypassed by postulating a non-zero constituent quark mass in $\bD^{(\pi)}$ and
define the model by that operator.}, Eq.~(\ref{gap}), and further obstacles may occur away from 
the chiral limit when quadratic divergences may occur. In the onset we have distinguished between 
regularized and non-regularized parts in the action, Eq.~(\ref{act1}). 
Any kind of {\it a posteriori} regularization faces the dilemma that
such a distinction is difficult to implement. There are combinations of distributions that are 
ultraviolet finite even without regularization. Must they nevertheless undergo regularization? In 
this context refer to the discussion on the Gottfried sum rule in Section \ref{sec:NR}.
We also note that the restriction to the leading $\frac{1}{N_C}$ terms 
does not distinguish between isoscalar and -vector components.

Most of those early distribution function calculations did not attempt the DGLAP evolution, rather 
compared the results with empirical distributions at a low renormalization point, that 
result from applying the inverse of the DGLAP evolution to data \cite{Gluck:1994uf}.

In Refs.~\cite{Pobylitsa:1998tk,Wakamatsu:1997en} similar
calculations for the $\frac{1}{N_C}$ corrections to the unpolarized distributions have been
performed while Ref.~\cite{Wakamatsu:1998rx} discusses the polarized distributions with
subleading $\frac{1}{N_C}$ terms included and also implements the DGLAP evolution program.
Similar to our calculations those authors observe that the Dirac sea contribution to the 
polarized structure functions is almost negligibly small.

The extension to three light flavors has also been addressed. These studies were first performed in the 
valence level only approximation for the hadron tensor \cite{Schroeder:1999fr} and soon after by formulating 
distributions incorporating the {\it a posteriori} regularization \cite{Wakamatsu:2003wg} with a Pauli-Villars 
mass as described above. Technically the main difference is that the collective coordinates are from $SU(3)$
and that there are eight instead of three angular velocities. Furthermore flavor symmetry breaking must be 
included because the strange quark mass (represented by the pseudoscalar kaon) is much larger than that of 
the up and down quarks. Of course, that extension allows a closer look at strangeness in the nucleon. 
In this regard the numerical results of that model calculation agree with data \cite{Bazarko:1994tt}, at 
least qualitatively.

Once the identification of distribution functions is accepted, other processes than DIS, that in QCD are 
described by various bilocal bilinear quark operators, can also be explored within chiral quark soliton 
models.  Let us mention two examples. Transversity distributions complete the description of the nucleon 
spin \cite{Ralston:1979ys} and are relevant for the Drell-Yan process \cite{Jaffe:1991ra}. There are two of 
them which are similar to the two polarized structure functions: the transverse $h_T(x)$ and longitudinal 
$h_L(x)$. In the language of distributions the relevant bilocal bilinear quark operators are similar 
to those for the polarized ones, except for different Dirac matrices. Again, transverse and longitudinal
refers to the alignment of spin and external momentum. In the soliton model these distributions
were first estimated in the large $N_C$ limit and valence level only approximation \cite{Pobylitsa:1996rs}.
Subsequently $\frac{1}{N_C}$ corrections were included \cite{Gamberg:1998vg} and finally Dirac
sea contributions were considered in Refs.~\cite{Wakamatsu:1998rx,Schweitzer:2001sr}. Transversity
distributions have sum rules with tensor charges, $\langle N|\Sigma_3(\tau_3)|N\rangle$. These charges
can be directly computed in the chiral soliton model without any ambiguity from regularizing the Dirac
sea component. That component was found to be very small \cite{Gamberg:1998vg} suggesting that the
valence level only approximation is reliable for these distributions. This was later confirmed by the
computation with the {\it a posteriori} regularization prescription \cite{Wakamatsu:1998rx}. The twist-3 
distribution $e(x)$, which has a sum rule with the $\pi N\,-\sigma$-term, has been 
considered in Refs.~\cite{Ohnishi:2003mf}. Again, this distribution is not a structure function
accessible in DIS but can be extracted from pion photoproduction \cite{Avakian:2003pk}. Even though 
the relevant bilocal bilinear quark operator is as simple as $\overline{\Psi}(0)(\tau_3)\Psi(\lambda n)$, 
the actual computation is quite intricate because of a potential $\delta$-function behavior of the isoscalar 
combination at $x=0$. The sum rule is only satisfied with the inclusion of such a behavior \cite{Burkardt:2001iy}.
The model calculation of Ref.~\cite{Ohnishi:2003mf} indeed confirms that singular structure.

Let us also briefly comment on the historical development.
Ref.~\cite{Diakonov:1996sr} mentions that some preliminary results on structure functions had been
''announced'' in Ref.~\cite{Diakonov:1996jx}. But that reference only states that these calculations are
in progress pointing to \cite{Diakonov:1996sr}. So it seems fair to state that the first results for
structure functions in a soliton model were presented in Ref.~\cite{Weigel:1996kw} according to the
journal received dates though there was some delay of the actual publication.

We have seen that, modulo regularization, the matrix elements to be computed are formally the same as if the 
bilocal bilinear quark operators were directly transferred from QCD to the model. That is, rather than merely 
taking the NJL model as one for some of the QCD symmetries, it is considered a model for QCD degrees of 
freedom.  This is a frequently adopted point of view, not only for the NJL model, but also, {\it e.g.} the 
MIT bag model \cite{Chodos:1974je}, in particular in the context of structure functions~\cite{Jaffe:1974nj}.
Furthermore it opens the door to explore quark distributions others than those parameterizing 
(electromagnetic) DIS.

\section{Conclusions}
\label{sec:concl}
The standard model for elementary particles is a gauge theory for leptons, quarks and gauge bosons. To make 
contact with the world of mesons and baryons, knowledge about their composition in terms of quarks (and gluons) 
is inevitable. The binding of the fundamental constituents to mesons and baryons, known as color-confinement in 
QCD, is a non-perturbative effect. Distribution functions that combine to structure functions parameterize this 
non-perturbative composition of mesons and baryons. These structure functions cannot be computed from first 
principles in QCD but are either extracted from empirical data, computed in lattice simulations or obtained from 
some model estimates. The chiral soliton model is one of the many popular and successful models for baryons.
Here meson fields are the model degrees of freedom and solitons are (static) solutions to the 
respective, non-linear fields equations.

The calculation of nucleon structure functions in chiral soliton models has been a long issue. Traditional 
soliton models like the Skyrme model and its extensions by incorporating vector mesons in addition to the 
pions face the problem of only representing local quark bilinear combinations. On the other hand, models
that carry through the bosonization, that transforms the quark into a meson theory, are plagued by the need 
for regularization. Here we have reviewed a method that takes regularization seriously from the initial 
formulation of the action for the quark model rather than empirically implementing regularization for 
distribution functions that linearly combine to structure functions. The formal relation between structure 
functions and quark distributions is no longer obvious when regularization is required. We stress that this 
formulation only identifies chiral symmetries of QCD with no statement on how the model and QCD quarks relate. 
Yet all the sum rules that relate integrals of the structure functions to observables like hadron masses, 
isospin etc.\@ and that are commonly derived from the probability interpretation of distribution functions 
are also valid in this approach. The project should thus be considered more like a {\it proof of concept} 
rather than attempting precise predictions for the structure functions.

The point of departure is a self-interacting chirally symmetric quark model. It is particularly formulated
to make feasible the full process of bosonization. At each step of this calculation regularization is 
carefully traced resulting in consistently regularized structure functions. The model is defined such that 
only one part of the bosonized action is regularized in order to maintain the chiral anomaly. Hence it is 
suggestive that only some of the structure functions will be subject to regularization. The treatment 
reviewed here is constructive in the sense that it dictates for which structure function regularization 
must be implemented and for which this is not the case. This goes beyond analyzing whether or not the 
particular structure function is ultraviolet convergent. The method is also predictive in case the 
structure function does not have a sum rule that is related to a static property with an established 
regularization prescription.

Regularization, of course, only concerns the vacuum (Dirac sea) contribution to any observable. In 
addition there is always the contribution from the valence level (strongly) bound by the self-consistent 
soliton. This level contribution must be included to deal with a unit baryon number object. We have 
actually seen that this level contribution is dominant for almost all structure functions except the 
isoscalar unpolarized combination. For this combination we see a strong enhancement at small 
Bjorken~$x$. This was also seen in the numerical simulation of Ref.~\cite{Diakonov:1996sr}, though 
not quite as drastic as in our case. We recall that the unpolarized isoscalar structure function has a 
sum rule with the energy, which in soliton models is the classical soliton energy. The standard definition
of this energy subtracts the zero soliton vacuum counterpart to get a finite result for the soliton 
energy and therefore this structure function should undergo an analog subtraction. It is important to note that 
this energy subtraction has no dynamic effect, {\it i.e.} it does not enter the field equation for the 
soliton. Any additional (finite) subtraction would be possible in a renormalizable theory. Hence this 
piece is not without ambiguity. Of course, it is very suggestive to subtract the zero soliton energy to 
determine the binding of the soliton. But that is only a (regularization) condition for the integrated 
structure function. Formally, however, the subtraction is obtained from a different action functional. 
The result that the zero soliton vacuum structure function is not a constant is kind of surprising as 
it suggests that the trivial vacuum has structure. One may also speculate whether this unexpected result 
is related to the numerical treatment of discretizing wave-functions with box boundary conditions. In close 
proximity to the boundary, completeness of the wave-functions is not guaranteed \cite{Reinhardt:2012xs}. 
We are currently exploring this speculation.

The biggest conceptual problem unsolved so far is the fact that the computed structure functions 
have support for $|x|>1$ resulting from the soliton not being translationally invariant and that
the collective coordinate approach to restore this symmetry is merely an approximation. The computed
structure functions are small (or even tiny) for $x>1$ but not exactly zero. In order to
apply the DGLAP evolution formalism, the support must be restricted to $|x|\le1$. We have adopted
a procedure from the $D=1+1$ MIT bag model boosting the rest frame structure functions to the 
infinite momentum frame. This may overemphasize the small $x$ regime and also interfere with the
rotational $\frac{1}{N_C}$ corrections but it maintains the sum rules. Other approaches, that
merely omit the $|x|>1$ piece, multiply a differentiable function that models a function with 
a step at $x=1$ \cite{Wakamatsu:1998rx,Wakamatsu:2003wg}. Though that may better reproduce the 
data in the small $x$ regime, this {\it ad hoc} approach violates the sum rules, at least formally.

A possible extension of the approach reviewed here would be the consideration of inelastic scattering
with neutrino induced interactions. That would bring in a generalization of the Compton tensor that
would also include couplings to the axial current as governed by the weak component of the standard model
and thus form factors and structure functions that are not disallowed by parity conservation.
\bigskip

\authorcontributions{The authors mutually agree that their contributions warrant co-authorship. That must be enough
information for the public readership!}

\funding{H.\@ W.\@ is supported in part by the National Research Foundation of
South Africa (NRF) by grant~109497.}

\conflictsofinterest{The authors declare no conflict of interest.}
%
%
%
%
\reftitle{References}
%
%

\end{document}